%% file: UQ_correlations_EOS.tex
\documentclass[10pt,aps,prc,floatfix,twocolumn,nofootinbib,superscriptaddress]{revtex4-1}

\usepackage[labelfont={small},subrefformat=parens,caption=false]{subfig}
\usepackage{graphicx,amsmath,amssymb,bm}
\usepackage{amsfonts}
\usepackage[utf8]{inputenc}
\usepackage{verbatim}
\usepackage{float}
\usepackage{cancel}
\usepackage{multirow}
\usepackage{array}
\usepackage{xparse,xspace}
\usepackage{physics}
\usepackage{color}
\usepackage{dsfont}
\usepackage[pdfpagelabels, pdfencoding=auto, psdextra]{hyperref}
\usepackage[overload]{textcase}

\hypersetup{%
    pdfsubject=Paper,
    pdfkeywords={nuclear physics} {Bayesian} {chiral EFT} {nuclear matter},
    unicode = true,
    breaklinks = true,
    colorlinks = true,
    linkcolor = blue,
    menucolor = blue,
    citecolor = blue,
    urlcolor = blue
}

\usepackage{cellspace}
\usepackage{dcolumn}
\newcolumntype{d}[1]{D{.}{.}{#1}}

\graphicspath{{./figures/}}

\usepackage{silence}
\usepackage[T1]{fontenc}

\newcommand{\phantomsublabel}[3]{%
\unitlength=1in%
\put(#1,#2){%
 \subfloat[]{%
 \label{#3}%
 }}%
}

\input{buqeye_macros}

\newcommand{\meangrad}{m^{\!\nabla\!}}
\newcommand{\kernelgrad}{\kappa^{\!\nabla\!}}
\newcommand{\eg}{\textit{e.g.}\xspace}
\newcommand{\ie}{\textit{i.e.}\xspace}

\begin{document}

% Overall checks and general comments
\title{Quantifying uncertainties and correlations in the nuclear-matter equation of state}

\author{C. Drischler}
\email{cdrischler@berkeley.edu}
\affiliation{Department of Physics, University of California, Berkeley, California 94720, USA}
\affiliation{Nuclear Science Division, Lawrence Berkeley National Laboratory, Berkeley, California 94720, USA}

\author{J.~A. Melendez}
\email{melendez.27@osu.edu}
\affiliation{Department of Physics, The Ohio State University, Columbus, Ohio 43210, USA}

\author{R.~J. Furnstahl}
\email{furnstahl.1@osu.edu}
\affiliation{Department of Physics, The Ohio State University, Columbus, Ohio 43210, USA}

\author{D.~R. Phillips}
\email{phillid1@ohio.edu}
\affiliation{Department of Physics and Astronomy and Institute of Nuclear and Particle Physics, Ohio University, Athens, Ohio 45701, USA}

\date{\today}

%%%%%%%%%%%%%%%%%%%%%%%%%%%%%%%%%%%%%%%%%%%%%%%%%%%%%%%%%%%%%%%
\begin{abstract}

We perform statistically rigorous uncertainty quantification (UQ) for chiral
effective field theory ($\chi$EFT) applied to infinite nuclear matter up to twice
nuclear saturation density. The equation of state (EOS) is based on high-order
many-body perturbation theory calculations with nucleon-nucleon and
three-nucleon interactions up to fourth order in the $\chi$EFT expansion. From
these calculations our newly developed Bayesian machine-learning
approach extracts the size and smoothness properties of the correlated EFT truncation
error. We then propose a novel extension that uses multitask machine learning to
reveal correlations between the EOS at different proton fractions.
The inferred in-medium $\chi$EFT breakdown scale in pure neutron matter and
symmetric nuclear matter is consistent with that from free-space
nucleon-nucleon scattering. These significant advances allow us to provide posterior distributions
for the nuclear saturation point and propagate theoretical uncertainties to
derived quantities: the pressure and incompressibility of symmetric nuclear
matter, the nuclear symmetry energy, and its derivative. Our results, which are validated by statistical diagnostics, demonstrate
that an understanding of truncation-error correlations between different
densities and different observables is crucial for reliable UQ\@. The methods
developed here are publicly available as annotated Jupyter notebooks.

\end{abstract}
%%%%%%%%%%%%%%%%%%%%%%%%%%%%%%%%%%%%%%%%%%%%%%%%%%%%%%%%%%%%%%%

%%%%%%%%%%%%%%%%%%%%%%%%%%%%%%%%%%%%%%%%%%%%%%%%%%%%%%%%%%%%%%%
\maketitle
%%%%%%%%%%%%%%%%%%%%%%%%%%%%%%%%%%%%%%%%%%%%%%%%%%%%%%%%%%%%%%%

%%%%%%%%%%%%%%%%%%%%%%%%%%%%%%%%%%%%%%%%%%%%%%%%%%%%%%%%%%%%%%%
\section{Introduction}
\label{sec:introduction_matter_uq}

Calculations of observables in chiral effective field theory (\chiEFT)~\cite{Epel09RMP, Mach11PR,
Hammer:2019poc, Tews:2020hgp} are
truncated at a finite order in the EFT expansion, leaving a residual error that
should be quantified to enable robust comparisons to experiment and competing theories~\cite{PhysRevA.83.040001}. 
While \chiEFT\ is widely used to predict the nuclear-matter equation of state (EOS)~\cite{Tews13N3LO, Hebe13ApJ,Baar13CCinf,Drischler:2013iza, Hagen:2013yba,Carbone:2013rca,Coraggio:2014nva, Well14nmtherm, Rogg14QMC, Holt:2016pjb,Drischler:2016djf,Ekstrom:2017koy, Drischler:2017wtt, Lonardoni:2019ypg, Piarulli:2019pfq} (see also
Refs.~\cite{Hebe15ARNPS, Drischler:2019xuo, Sammarruca:2019ncy} for recent reviews),
a proper statistical analysis of the \chiEFT\ truncation errors for the EOS and associated observables is lacking.
This work fills that gap. Implications for neutron-rich matter are elucidated in a companion paper~\cite{Drischler:2020preparationAstro}. 

In a recent paper, a Bayesian model for
EFT truncation errors was developed that included their correlation across
continuous independent variables, such as energy or scattering
angle~\cite{Melendez:2019izc}. The model uses machine learning (ML) to
determine the convergence and correlation pattern of the \chiEFT\ expansion by calibrating Gaussian processes (GPs) to the computed orders; this leads to statistical estimates of
the omitted higher orders. Here we extend this method and apply it to
many-body observables in infinite (nuclear) matter computed using \chiEFT\ nucleon-nucleon (NN) and three-nucleon (3N) interactions up to
next-to-next-to-next-to-leading order (\NNNLO); specifically, to the EOS 
in the limits of pure neutron
matter (PNM) and symmetric nuclear matter (SNM). 

Our Bayesian model incorporates two types of correlations in the \chiEFT\ truncation error for different quantities. Given an observable $y(x)$, we diagnose and assess the impact of
\begin{itemize}
    \item[(1)] Type $x$: Correlations between $y(x)$ and $y(x')$. In infinite
    matter, the input points $x$ could be Fermi momentum $\kf$ or the density
    $n$. Type-$x$ correlations are quantified and propagated via the ML
    truncation-error model proposed in Ref.~\cite{Melendez:2019izc}. 

    \item[(2)] Type $y$: Correlations between discrete observables $y_i(x)$ and
    $y_j(x')$, \eg, between the EOS of PNM and SNM\@. This can include type-$x$
    correlations if the observables are considered at different input locations.
    Type-$y$ correlations also include correlations between an observable and
    its derivatives~\cite{Rasmussen:2003gphmc, Solak:2003dgpds,
    Eriksson:2018scaling,Chilenski_2015_gptools}. We expand upon our ML framework to incorporate these
    novel correlations using \emph{multitask} ML
    algorithms~\cite{alvarez2012kernels, Melkumyan:2011multikernel,
    Caruana:1997MultitaskLearning, Zhang:2018MultitaskLearning}.
\end{itemize}
Both types are not only crucial to a robust uncertainty quantification
(UQ) in infinite matter but also reveal physics about the system. In particular, type-$x$ correlations tell us the persistence of information in \chiEFT\ across different densities; building them into our error model facilitates the reliable computation of derivatives of $y(x)$. 

Our previous applications of Bayesian analysis to \chiEFT\ truncation errors focused on NN observables
calculated using Weinberg power counting~\cite{Furnstahl:2015rha,
Melendez:2017phj, Wesolowski:2018lzj, Melendez:2019izc}. Reference~\cite{Melendez:2019izc} developed both the truncation-error model with correlations and statistical model-checking diagnostics (see also Ref.~\cite{BastosDiagnosticsGaussianProcess2009}) to check the validity of the model's postulated order-by-order convergence pattern and correlation structure.
For some NN potentials we
found a validation of basic EFT convergence expectations, with a breakdown scale $\Lambda_b$ that is consistent with $600 \MeV$, despite the lack of
renormalization-group (RG) invariance with this power-counting scheme. For other
potentials our diagnostics clearly show deviations from these expectations,
which were attributed to regulator artifacts. Thus, the statistical tools offer
not only theoretical uncertainty bands, but also an alternative to RG diagnostics of
whether an EFT is performing as advertised.

The stage is set for more wide-ranging applications and tests of these tools.
There are three main reasons why infinite matter provides an attractive laboratory for studying their use and what they tell us about the convergence pattern of \chiEFT\@.
First, translational invariance provides many simplifications in characterizing the system: it permits clear focus on the bulk properties of nuclear matter without confusion from surface effects that might complicate observables' convergence pattern in finite nuclei.
Second, the densities of relevance for infinite matter are higher than for light nuclei, which provides a different perspective on the convergence of \chiEFT\@.
Also, the controlled specification of density can help to illuminate the nature of that convergence, which can be obscured in applications to observables in finite nuclei.
Third, 3N interactions drive nuclear saturation in SNM\@.
They are quantitatively important at every proton fraction at nuclear saturation density $n_0 \approx 0.16 \fmiq$ ($\rho_0 \approx 2.7 \times 10^{14} \, \text{g\,cm}^{-3}$) and above.

Calculations of infinite matter with the \NNNLO \chiEFT\ interactions we use for our study are feasible because 
many-body perturbation theory (MBPT) is available as a controlled and computationally efficient many-body method.
The technology from Refs.~\cite{Drischler:2016djf,Drischler:2017wtt} extends to high enough order in MBPT to ensure adequate many-body convergence for these \chiEFT\ NN and 3N potentials.
Furthermore, the ability of MBPT to isolate contributions from different classes of diagrams can provide new insights into what determines the
convergence pattern: calculations with and without 3N forces can be directly compared.

Given the marked difference in density between the NN system and infinite matter the reader may doubt that the same \chiEFT\ convergence pattern prevails in the two systems. 
3N forces are present in infinite matter but, tautologically, not in the NN system.
Many-body effects such as Pauli blocking may also affect the way that different \chiEFT\ orders contribute to observables.
The diagnostics used to check that truncation errors were behaving as advertised in Refs.~\cite{Furnstahl:2015rha, Melendez:2017phj, Melendez:2019izc} can also be applied to \chiEFT\ calculations of the EOS\@.
They allow us to determine whether the EFT breakdown scale in infinte matter is consistent with that found from analyses of few-nucleon observables.

The paper is organized as follows. The statistical model of 
Ref.~\cite{Melendez:2019izc} is briefly reviewed in Sec.~\ref{sec:the_model_nuclear_matter}, including an explanation of the model's treatment of type-$x$ correlations. In 
Sec.~\ref{sec:results_for_pnm_and_snm} we
present the EOS for recent \chiEFT\ NN and 3N interactions up to \NNNLO, analyze the corresponding order-by-order EFT coefficients, obtain a Bayesian posterior for the EFT breakdown scale, provide error bands for the EOS in the limit of SNM and PNM, and present error ellipses for the predicted nuclear saturation point. We then study type-$y$
correlations (those between observables) in Sec.~\ref{sec:derivative_quantities} and thereby derive uncertainty bands for the symmetry energy as a function of density. This is followed by new results for derivatives of the SNM EOS:  order-by-order uncertainty
estimates for the pressure and incompressibility. Section~\ref{sec:summary_matter_uq} has our summary and outlook.
Further details---regarding our statistical model-checking diagnostics, the order-by-order values of the SNM and PNM EOS, and derivatives of GPs and multitask GPs---are presented in
Appendices~\ref{sec:model_checking_infinite_matter}--\ref{sec:gp_details_infinite_matter}. Additional figures are given in the Supplemental Material~\cite{SuppMatLong}.  

Although some results for PNM are presented here, this paper focuses on SNM\@.
A companion paper~\cite{Drischler:2020preparationAstro} discusses the parallel analysis for PNM,
and provides error estimates for the symmetry energy and its slope parameter that include the impact of both type-$x$ and type-$y$ correlations.
The results presented in these two papers do not constitute an exhaustive study; rather we identify trends and issues and anticipate future 
refinements of our approach.

%%%%%%%%%%%%%%%%%%%%%%%%%%%%%%%%%%%%%%%%%%%%%%%%%%%%%%%%%%%%%%%
\section{A model of EFT truncation errors that includes correlations}
\label{sec:the_model_nuclear_matter}

%%%%%%%%%%%%%%%%%%%%%%%%%%%%%%%%%%%%%%%%%%%%%%%%%%%%%%%%%%%%%%%
\subsection{Previous work}

In previous works we proposed a pointwise Bayesian statistical model to
estimate EFT truncation errors for predicted observables
$\genobs$~\cite{Furnstahl:2015rha,Melendez:2017phj}. This model formalizes the
notion of convergence of $\genobs$ at a single sampling point to allow one to
credibly assess whether experimental data are consistent with theory. The
convergence pattern should be a consequence of the EFT power counting. We
incorporate this ``expert knowledge'' of the convergence pattern in our prior
probability distributions. Those beliefs are subsequently updated using the
actual order-by-order EFT results
$\{\genobs_0,\,\genobs_1, \,\genobs_2, \dotsc, \,\genobs_k
\}$~\cite{Cacciari:2011ze}. Note that $\genobs_1 \equiv 0$ in \chiEFT\@.

The authors of Ref.~\cite{Hu:2019zwa} applied the simplest version of the
pointwise model from Ref.~\cite{Furnstahl:2015rha} to Brueckner-Hartree-Fock
calculations of infinite matter based on an NN-only \chiEFT\ potential. They used
$Q = \kf/\Lambda_b$ as the expansion parameter and found credibility intervals
for PNM and SNM\@. They then followed Ref.~\cite{Furnstahl:2015rha} and applied
a consistency check on the empirical coverage to validate their choice of
$\Lambda_b$. The present work goes significantly beyond that of
Ref.~\cite{Hu:2019zwa} in using a much more accurate many-body method, including
3N forces, analyzing and accounting for correlations within and between
observables, applying a suite of model-checking diagnostics,
and deriving posteriors for $\Lambda_b$.

%%%%%%%%%%%%%%%%%%%%%%%%%%%%%%%%%%%%%%%%%%%%%%%%%%%%%%%%%%%%%%%
\subsection{Including correlations}

To include the effects of correlations between EFT predictions at different
values of independent variables, generically denoted as $\kinparvec$, we extended
the pointwise model to functions $\genobs(\kinparvec)$, encoding the idea of
curvewise convergence for observables via GPs~\cite{Melendez:2019izc}.
GPs are powerful tools for both regression and classification, and have become
popular in many fields, including statistics, physics, and applied
mathematics~\cite{sacks1989design,cressie1992statistics,rasmussen2006gaussian}.
The GP parameters are interpretable from an EFT convergence standpoint, and can
be easily calibrated against known order-by-order predictions.

We give here a brief overview of the statistical model and refer the reader to
Ref.~\cite{Melendez:2019izc} for details and examples (including a Jupyter
notebook that reproduces the figures in that paper). The model for the
truncation error $\delta\genobs_k(\kinparvec)$ at order $k$ in the EFT expansion
(\eg, $k=4$ at \NNNLO in \chiEFT) is based on the decomposition
\begin{align} \label{eq:discrepancy_k_definition}
    \delta\genobs_k(\kinparvec) = \genobsref(\kinparvec) \sum_{n=k+1}^\infty c_n(\kinparvec) Q^n(\kinparvec) \,,
\end{align}
where $\genobsref(\kinparvec)$ is a dimensionful quantity that sets the
reference scale of variation with $\kinparvec$, $Q(\kinparvec)$ is a
dimensionless expansion parameter, and the $c_n(\kinparvec)$ are dimensionless
coefficients. The observable $\genobs_k(\kinparvec)$ itself at order $k$ is
decomposed as
\begin{align} \label{eq:Xk}
    \genobs_k(\kinparvec) = \genobsref(\kinparvec) \sum_{n=0}^k c_n(\kinparvec) Q^n(\kinparvec) \,,
\end{align}
where the observable coefficients $c_n(\kinparvec)$ are extracted from the
order-by-order calculations, given $\genobsref(\kinparvec)$ and $Q(\kinparvec)$,
using
\begin{align} \label{eq:constraints}
    \genobs_0(\kinparvec) & \equiv \genobsref(\kinparvec) c_0(\kinparvec) \,, \\
    \label{eq:Delta_y}
    \Delta\genobs_n(\kinparvec) & \equiv \genobsref(\kinparvec) c_n(\kinparvec) Q^n(\kinparvec) \,.
\end{align}
Here, $\Delta\genobs_n(\kinparvec)$ is the order-$n$ correction to the
observable. Since all scales have been factored into $\genobsref(x)$ and $Q(x)$,
the $c_n(x)$ are expected to be \emph{natural}, or, in other words, of order 1,
assuming there are no systematic cancellations (\eg, fine tuning) that would
make the coefficients much smaller than the reference size.

We postulate that the properties of the unobserved $c_{n > k}(x)$ are the
same as the observed $c_{n\leq k}(x)$. Specifically, our model assumes that all the
$c_n(x)$ are independent and identically distributed (\iid) random curves. We
formalize the EFT convergence assumptions by modeling the coefficients $c_n(x)$
as independent draws from a single underlying GP\@.
A brief introduction to GPs in this context is given in Ref.~\cite{Melendez:2019izc}, whose notation is
$\GP[m(x),\, \kernel(x,\, x')]$ for some mean function $m(x)$ and
positive-semidefinite covariance function (also called kernel) $\kernel(x,\,x')$.
For more in-depth discussions, see Refs.~\cite{rasmussen2006gaussian,
Mackay:1998introduction, Mackay:2003information}.
We adopt  $m(\kinparvec) = 0$ since corrections are just as likely to be positive as they are to be negative, and
$\kernel(\kinparvec,\, \kinparvec';\,\sdth,\, \ell) = \sdth^2 \, r(\kinparvec,\, \kinparvec'; \, \ell)$,
so that\footnote{We use the common shorthand notation
in statistics, in which $z \sim \cdots$ reads as ``the variable $z$ is
distributed as $\cdots$.'' Some authors also use $\pr(z) = \cdots$. See also
Ref.~\cite{Melendez:2019izc}. The ``iid'' above the $\sim$ indicates that the $c_n$s are a set of \iid\ random curves.}
\begin{align}
    c_n(\kinparvec) \given \sdth^2, \ell & \overset{\text{\tiny iid}}{\sim} \GP[0,\, \sdth^2r(\kinparvec,\, \kinparvec';\ell)] \,.
    \label{eq:cn_iid} %\\
%   \param & \equiv \{\sdth^2, \ell\} \,. \label{eq:param}
\end{align}

We choose the correlation function $r(\kinparvec,\,\kinparvec';\,\ell)$ to be a
radial basis function (RBF), which ensures that the $c_n(\kinparvec)$ are very
smooth functions (up to numerical noise that is handled by a white-noise term),
\begin{align} \label{eq:rbf_kernel_nuclear_matter}
    r(\kinparvec, \,\kinparvec'; \,\ell) = \exp[-\frac{(x-x')^2}{2\ell^2}] \,.
\end{align}
The length scale $\ell$ controls how quickly the $c_n(\kinparvec)$ vary as a
function of $x$; a small length scale implies that the $c_n(\kinparvec)$ vary
quickly, whereas a large length scale implies the opposite. Importantly, this
kernel is \emph{stationary}, meaning that it only depends on the absolute
difference $|x-x'|$. Stationarity implies that the $c_n(x)$ should share similar
correlation properties across all $x$, up to fluctuations. For example, the
curves should not vary rapidly at small $x$ and flatten out at large $x$.

To update the $\cbar^2$ based on observed $c_n(\kinparvec)$ requires a prior. We
choose the scaled inverse-chi-squared distribution~\cite{Melendez:2019izc} with
$\nu=10$ degrees of freedom and scale parameter $\tau^2 = (\nu - 2) / \nu$. This
informative prior on $\cbar^2$ has a mean value of 1, which builds in our
assumption that the coefficients should be naturally sized.

Thus, in summary, the $c_n(\kinparvec)$ are modeled as random functions. These
functions share a common variance $\sdth^2$, and should each look like random
draws from a GP with an RBF kernel $\kernel(x,
x';\ell)$.\footnote{Generalizations of this model are discussed in
Ref.~\cite{Melendez:2019izc};
Appendix~\ref{sec:model_checking_infinite_matter} considers
relaxing the assumption of a single GP in favor of separate GPs for NN and 3N
contributions.} The hyperparameters $\cbar^2$ and $\ell$ can be learned from the
order-by-order \chiEFT\ predictions, see Ref.~\cite{Melendez:2019izc}.

%%%%%%%%%%%%%%%%%%%%%%%%%%%%%%%%%%%%%%%%%%%%%%%%%%%%%%%%%%%%%%%
\begin{figure*}[tb]
  \includegraphics{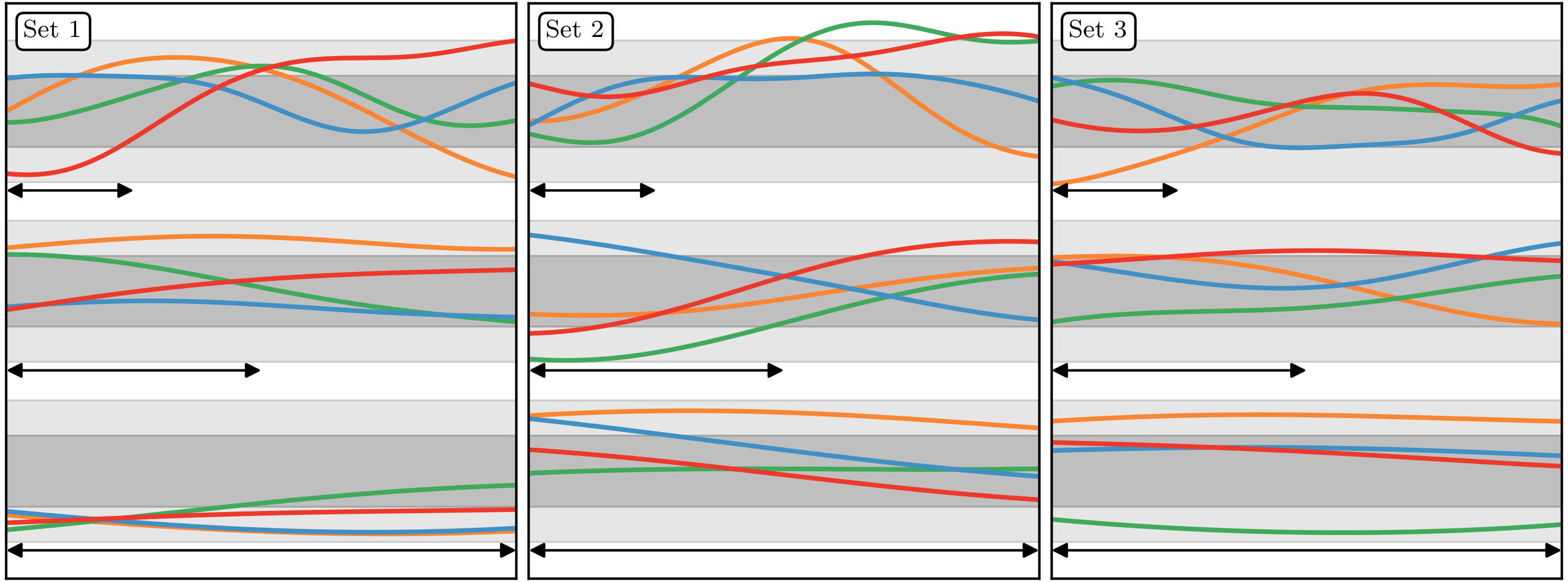}
  \caption{
  The three columns show three sets of random functions. 
  For each column the three rows show the output of GPs that have
  the same (arbitrary) mean and marginal variance, but  differ in their radial basis function (RBF) kernel: a different length scale is used in each of the three rows. The length scales are indicated by the
  double-headed arrow at the bottom of each panel
  Each panel then contains four
  functions, \ie, four draws from the GP. The $x$-axis could
  represent an independent variable such as energy, angle, or density and the $y$-axis can be thought of as the order-by-order EFT coefficients for an observable of interest. The dark
  (light) shaded bands represent one (two) standard deviations from the mean.
  }\label{fig:gp_draws_lengthscales}
\end{figure*}
%%%%%%%%%%%%%%%%%%%%%%%%%%%%%%%%%%%%%%%%%%%%%%%%%%%%%%%%%%%%%%%

Some intuition about the nature of the GPs used to model $c_n(\kinparvec)$ in this work
can be gleaned from Fig.~\ref{fig:gp_draws_lengthscales}. In each subplot are
four random draws from a GP with an RBF kernel.
The GPs have an arbitrary mean and marginal variance that are the same in all panels.
Each row depicts a different GP, with increasing length scale $\ell$ from the top row to the bottom.
Comparing different columns illustrates the
nature of the fluctuations one should expect when working with only a few
samples.
The smoothness of the coefficient functions $c_n(\kinparvec)$---a feature of the RBF kernel---is consistent with expectations and
observations from EFTs; \eg, compare to the real $c_n(\kinparvec)$s from NN scattering in Fig.~10\ of Ref.~\cite{Melendez:2017phj}
and from nuclear matter in Figs.~\ref{fig:energy_per_neutron_coefficients_lambda-500}
and~\ref{fig:energy_per_particle_coefficients_lambda-500}.
We contend that the modeling of $c_n(\kinparvec)$ as draws from GPs is supported
by the examples shown in Fig.~\ref{fig:gp_draws_lengthscales}. However, it is easy to be fooled by visual
evidence. We therefore rely on the model checking diagnostics in
Appendix~\ref{sec:model_checking_infinite_matter} to validate the GP hypothesis.

Once we make the inductive step that the higher-order coefficients (which we do not have) also obey Eq.~\eqref{eq:cn_iid}, it follows that
the truncation error $\delta\genobs_k(\kinparvec)$ defined by
Eq.~\eqref{eq:discrepancy_k_definition} is a geometric sum over independent normally
distributed variables. Its distribution is~\cite{Melendez:2019izc}
\begin{align} \label{eq:discr_k_prior}
    \delta\genobs_k(\kinparvec) \given \sdth^2, \, \ell, \, Q \sim \GP[0,\,  \sdth^2\discrcorr{k}(\kinparvec, \, \kinparvec'; \, \ell)] \,,
\end{align}
with
\begin{equation}
\begin{split}
    \discrcorr{k}(\kinparvec,\, \kinparvec';\,\ell) & \equiv \genobsref(\kinparvec)\genobsref(\kinparvec') \\
    & \quad \times \frac{[Q(\kinparvec)Q(\kinparvec')]^{k+1}}{1 - Q(\kinparvec)Q(\kinparvec')} \, r(\kinparvec, \, \kinparvec'; \, \ell)\,.
\end{split}
 \label{eq:truncation_corrfunc_matter}
\end{equation}
The marginal variance of Eq.~\eqref{eq:discr_k_prior} is $\kinparvec$ dependent
in general and equal to $\sdth^2\discrcorr{k}(\kinparvec, \, \kinparvec;\,
\ell)$~\cite{Melendez:2019izc}.

Note that one can embed our truncation error model within a Bayesian parameter
estimation framework to find posteriors for the low-energy constants (LECs) of
\chiEFT\  interactions~\cite{Wesolowski:2015fqa, Wesolowski:2018lzj}. Here we
will take the LECs as given for our analysis of infinite matter. An important
subject for future work is a complete Bayesian analysis that consistently
combines uncertainties on LECs from fitting nuclear interactions to data with
truncation errors in order to find the full uncertainty in \chiEFT\ predictions.

%%%%%%%%%%%%%%%%%%%%%%%%%%%%%%%%%%%%%%%%%%%%%%%%%%%%%%%%%%%%%%%
\section{Results for PNM and SNM} \label{sec:results_for_pnm_and_snm}

This section describes how we incorporate type-$x$ correlations in our model of EFT convergence, \ie, account for the fact that the truncation
error varies smoothly with density. With the Bayesian framework described
in Sec.~\ref{sec:the_model_nuclear_matter}, it is straightforward to analyze the convergence
patterns of the energy per particle in PNM ($E/N$) and SNM ($E/A$) and
obtain the size and correlation structure of the truncation error. We then extract  first posterior
distributions for the predicted saturation point of SNM\@.

%%%%%%%%%%%%%%%%%%%%%%%%%%%%%%%%%%%%%%%%%%%%%%%%%%%%%%%%%%%%%%%
\subsection{Nuclear-matter equation of state}

Our analysis is based on the MBPT calculations of $E/N(n)$ and $E/A(n)$ up to $2n_0$ in Refs.~\cite{Drischler:2017wtt, Drischler:2020preparationAstro} and Refs.~\cite{Drischler:2017wtt, Leonhardt:2019fua}, respectively, equidistantly sampled at $n= 0.05,\, 0.06,\, \dotsc, \, 0.34 \fmiq$.
These high-order MBPT
calculations are driven by the novel Monte Carlo framework introduced in Ref.~\cite{Drischler:2017wtt}. In this framework, arbitrary interaction
and many-body diagrams can be efficiently evaluated using automatic code
generation, which enables calculations with controlled many-body uncertainties for the employed NN and 3N interactions
(see the references for details).

%%%%%%%%%%%%%%%%%%%%%%%%%%%%%%%%%%%%%%%%%%%%%%%%%%%%%%%%%%%%%%%
\begin{table}[tb]
    \centering
    \caption{
    NN and 3N interactions considered in this work~\cite{Drischler:2017wtt}. The interactions are based on the order-by-order NN potentials by
    Entem, Machleidt, and Nosyk~\cite{Entem:2017gor} (EMN) with momentum cutoffs $\Lambda = 450$
    and $500\MeV$ up to \NNNLO and 3N forces at the same order and cutoff.
Reference~\cite{Drischler:2017wtt} fit the two 3N LECs $c_D$ and $c_E$ to the triton binding energy and the empirical saturation point of SNM\@. Chiral 3N forces up to \NNNLO also depend on the NN LECs $C_S$ and $C_T$ as well as the $\pi$N LECs $c_1$, $c_3$, and $c_4$. Their values were taken from the associated NN potential. The applied LO and NLO potentials are NN only and therefore omitted. 
More details can be found in Ref.~\cite{Drischler:2017wtt} and its Supplemental Material.}
    \label{tab:drischler_potential_fits}
    \begin{ruledtabular}
    \begin{tabular}{ScScd{2.2}d{2.2}}
        Chiral order &	NN potential &	\multicolumn{1}{Sc}{$c_D$} &	\multicolumn{1}{Sc}{$c_E$} \\
        \hline
        %LO  &	EMN~$450\MeV$ &	&  \\
        %        LO  &	EMN~$500\MeV$&	&  \\
        %NLO &	EMN~$450\MeV$ &	- &	 \\
        %        NLO &	EMN~$500\MeV$&	&	 \\
        \NkLO{2}&	EMN~$450\MeV$&	2.25&	0.07 \\
        % 2&	\NkLO{2}&	450&	2.50&	0.10 \\
        % 3&	\NkLO{2}&	450&	2.75&	0.13 \\
        \NkLO{2}&	EMN~$500\MeV$ &	-1.75&	-0.64 \\
        % 5&	\NkLO{2}&	500&	-1.50&	-0.61 \\
        % 6&	\NkLO{2}&	500&	-1.25&	-0.59 \\
        %hline
        \NkLO{3}&	EMN~$450\MeV$ &	0.00&	-1.32 \\
        % 8&	\NkLO{3}&	450&	0.25&	-1.28 \\
        % 9&	\NkLO{3}&	450&	0.50&	-1.25 \\
        \NkLO{3}&	EMN~$500\MeV$ &	-3.00&	-2.22 \\
        % 11&	\NkLO{3}&	500&	-2.75&	-2.19 \\
        % 12&	\NkLO{3}&	500&	-2.50&	-2.15 \\

    \end{tabular}
    \end{ruledtabular}
\end{table}
%%%%%%%%%%%%%%%%%%%%%%%%%%%%%%%%%%%%%%%%%%%%%%%%%%%%%%%%%%%%%%%

The authors of Ref.~\cite{Drischler:2017wtt} combined NN potentials developed by
Entem, Machleidt, and Nosyk~\cite{Entem:2017gor} with 3N
forces at the same order and cutoff to construct a set of order-by-order NN and 3N interactions up to \NNNLO.
%\emph{Consistently applied} means here that these interactions are considered at the same chiral order and with equal momentum cutoffs~\cite{Drischler:2017wtt}. 
%For a discussion of inconsistencies that arise from na{\"i}vely assigning regulator functions to many-body forces derived in dimensional regularization beyond \NNLO, we refer the reader to Section~4.2 in Ref.~\cite{Epelbaum:2019kcf}.
%This definition of \emph{consistency} has been criticized in Ref.~\cite{Epelbaum:2019kcf}.
The two 3N LECs $c_D$ and $c_E$ that govern, respectively, the intermediate- and short-range 3N
contributions at N$^2$LO, were adjusted to the triton binding energy and the empirical saturation point of SNM\@. For the momentum cutoffs $\Lambda = 450$
and $500\MeV$, three 3N forces with different combinations of $c_D$ and $c_E$ and
reasonable saturation properties were obtained at \NNLO and \NNNLO. However,
the terms of the 3N forces proportional to $c_D$ and $c_E$ do not contribute to PNM with nonlocal
regulators~\cite{Hebeler:2010xb}, so there is only one neutron-matter EOS
determined for each momentum cutoff and chiral order. And even our results for SNM at a given cutoff do not differ
significantly for the different 3N fits.
We therefore restrict the discussion here to
one Hamiltonian for each cutoff as summarized in
Table~\ref{tab:drischler_potential_fits}. Additional figures focusing on the $450 \MeV$ potentials are given in the Supplemental Material~\cite{SuppMatLong}.

%%%%%%%%%%%%%%%%%%%%%%%%%%%%%%%%%%%%%%%%%%%%%%%%%%%%%%%%%%%%%%%
\subsection{Extracting observable coefficients}
\label{sec:extracting_coefficients_pnm_snm}

The observable coefficients $c_n(\kinparvec)$ form the backbone of the convergence model
and the training of the GP hyperparameters. To extract the
coefficients, we need to assign values to $\kinparvec$, $\genobsref(x)$, and
$Q(x)$ based on the system under consideration.

For the independent variable $x$ we have two clear choices: $\kf$ or the density
$n = g\, \kf^3/(6\pi^2)$ with the spin-isospin degeneracy $g = 2$ for PNM and $g = 4$ for SNM\@.
The choice of $x$ is important because we assume a GP kernel
for $c_n(\kinparvec)$ that is \emph{stationary}, \ie,
the variance and correlation length are the same across the entire space of the independent variable selected. In other words, we want the $c_n(x)$ to be
approximately as curvy at low density as at high density, and this will (presumably) be better satisfied for one of these two choices than for the other.
Since we do not have
strong theoretical arguments regarding stationarity of the EFT truncation error in either $\kf$ or $n$, we instead rely on
empirical evidence. When plotting the observable coefficients as functions of $n$,
they are more compressed at low density and stretched out at
high density; when plotting versus $\kf$, the coefficients appear
to have a slightly more homogeneous correlation structure. Therefore, although the evidence is only slight, we choose our GP input
space to be $\kf$. Because readers usually want to know the density dependence of results, we still display graphs in which $n$ varies linearly and is the main independent variable. But it should be borne in mind that the type-$x$ correlations are actually formulated in $\kf$. Predictions across a larger range of $\kf$
would be needed to provide more conclusive evidence for this choice of $x$.

Although we have chosen $\kf$ as the GP input space, we could, in fact, have defined the input space with $\kf$ replaced by $\gamma \kf$ for some constant $\gamma > 0$.
However, we emphasize that our analysis is actually independent of the choice of $\gamma$. This is because the RBF kernel~\eqref{eq:rbf_kernel_nuclear_matter} is stationary and we use a scale
invariant prior on $\ell$ for each system. Therefore, the posterior for $\cbar$ and $\ell$ contain the same information regardless of how we scale $\kf$. The only effect of the choice of $\gamma$ is a cosmetic one: the posterior in $\ell$ will be scaled by that factor too.
We return to this point when discussing the nuclear symmetry energy in
Sec.~\ref{sec:symmetry_energy}.

We choose the reference scale for nuclear-matter EOS to be
\begin{align} \label{eq:y_ref}
    \genobsref(\kf) = 16 \MeV \times \left(\frac{\kf}{\kfzero}\right)^2 \,,
\end{align}
where $\kfzero$ is the Fermi momentum associated with $n_0 =0.16 \fmiq$; \ie, ${\kfzero^\text{PNM}} = 1.680 \fmi$ and $\kfzero^\text{SNM} = 1.333 \fmi$. Our findings indicate that this is a good approximation to the
size of the LO predictions of $E/N(n)$ and $E/A(n)$, and sets a
reasonable scale for the convergence for higher \chiEFT\ orders.

A natural choice for the expansion parameter based on experience with free-space
NN scattering is $Q \propto \kf/\Lambda_b$. Conceptually, we might want to consider $Q = \gamma \kf/\Lambda_b$, where the constant prefactor $\gamma > 0$ arises, \eg, from an average of momenta over the
Fermi sea ($\gamma = \sqrt{3/5}$)~\cite{Drischler:2017wtt}.
In fact, because our statistical model only constrains $Q(\kf)$, different choices of $\gamma$ are implicitly considered if we take the simplest choice:
\begin{equation} \label{eq:Qkf}
    Q(\kf) = \frac{\kf}{\Lambda_b} \,.
\end{equation}
A soft scale $\gamma \kf$ with $\gamma \neq 1$ will give the same results we find with Eq.~\eqref{eq:Qkf} as long as the inferred value of $\Lambda_b$ is also rescaled to $\gamma \Lambda_b$. This means we cannot make a connection to NN scattering results, where $\Lambda_b \approx 600\MeV$ is favored, solely based  on the results of the statistical analysis. If $\Lambda_b=600 \MeV$ were asserted to be the breakdown scale we could infer $\gamma$. Conversely, if theoretical arguments for a particular $\gamma$ were adduced we could determine $\Lambda_b$. For this first study we take the simplest choice, Eq.~\eqref{eq:Qkf}, let $\Lambda_b$ adopt a value
learned from the data, and reserve speculation over the broader meaning of the result found.

However, the choice of the soft scale in the numerator of Eq.~\eqref{eq:Qkf} is not clear when computing a quantity, such as the symmetry energy, that is obtained from
both $E/N(n)$ and $E/A(n)$ at a specific density.
Then $\kf^\text{PNM}$ and $\kf^\text{SNM}$ differ,  and this would have to be accounted for when defining the $Q$ used to extract the coefficients. Ultimately this issue does not affect our results though: we deal with it through the use of \emph{multitask} GPs (see Sec.~\ref{sec:symmetry_energy}). 

%%%%%%%%%%%%%%%%%%%%%%%%%%%%%%%%%%%%%%%%%%%%%%%%%%%%%%%%%%%%%%%
\begin{figure}[tb]
    \centering
    \includegraphics{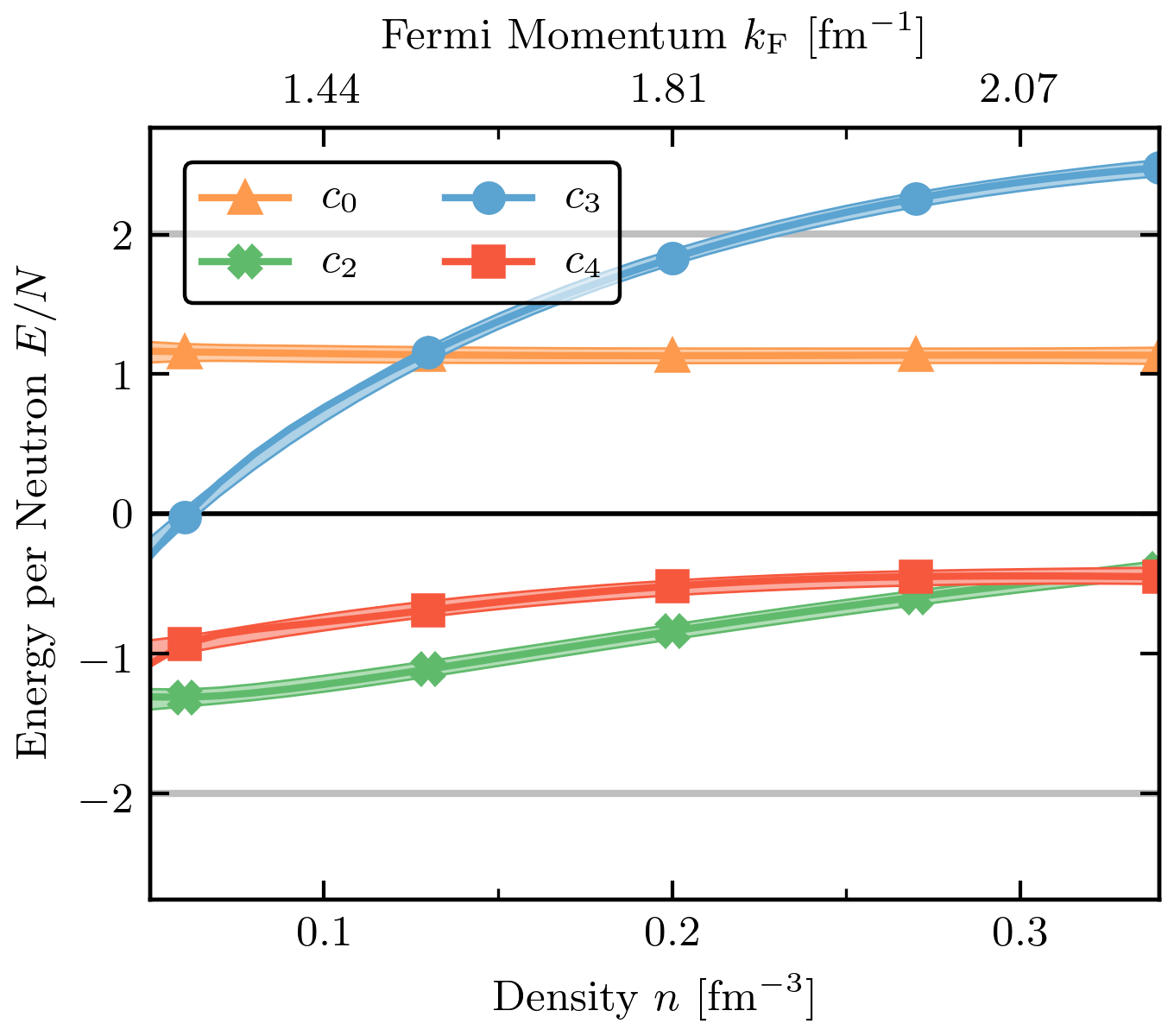}
    \caption{
    Observable coefficients, $c_n$, for $E/N(n)$ up to \NNNLO, as a function of density $n$, obtained using the $\Lambda = 500\MeV$ interactions in Table~\ref{tab:drischler_potential_fits}. Markers indicate training points, gray lines
    indicate $2\cbar$, and colored bands are 68\% credible intervals of the
    interpolating GPs.
    The estimated hyperparameters are given by $\cbar = 1.0$ and $\ell = 0.97\fmi$.
    Note that the secondary $x$-axis (at the top of the figure) is not linear in $\kf^\text{PNM}$.
    }
    \label{fig:energy_per_neutron_coefficients_lambda-500}
\end{figure}
%%%%%%%%%%%%%%%%%%%%%%%%%%%%%%%%%%%%%%%%%%%%%%%%%%%%%%%%%%%%%%%

%%%%%%%%%%%%%%%%%%%%%%%%%%%%%%%%%%%%%%%%%%%%%%%%%%%%%%%%%%%%%%%
\begin{figure}[tb]
    \centering
    \includegraphics{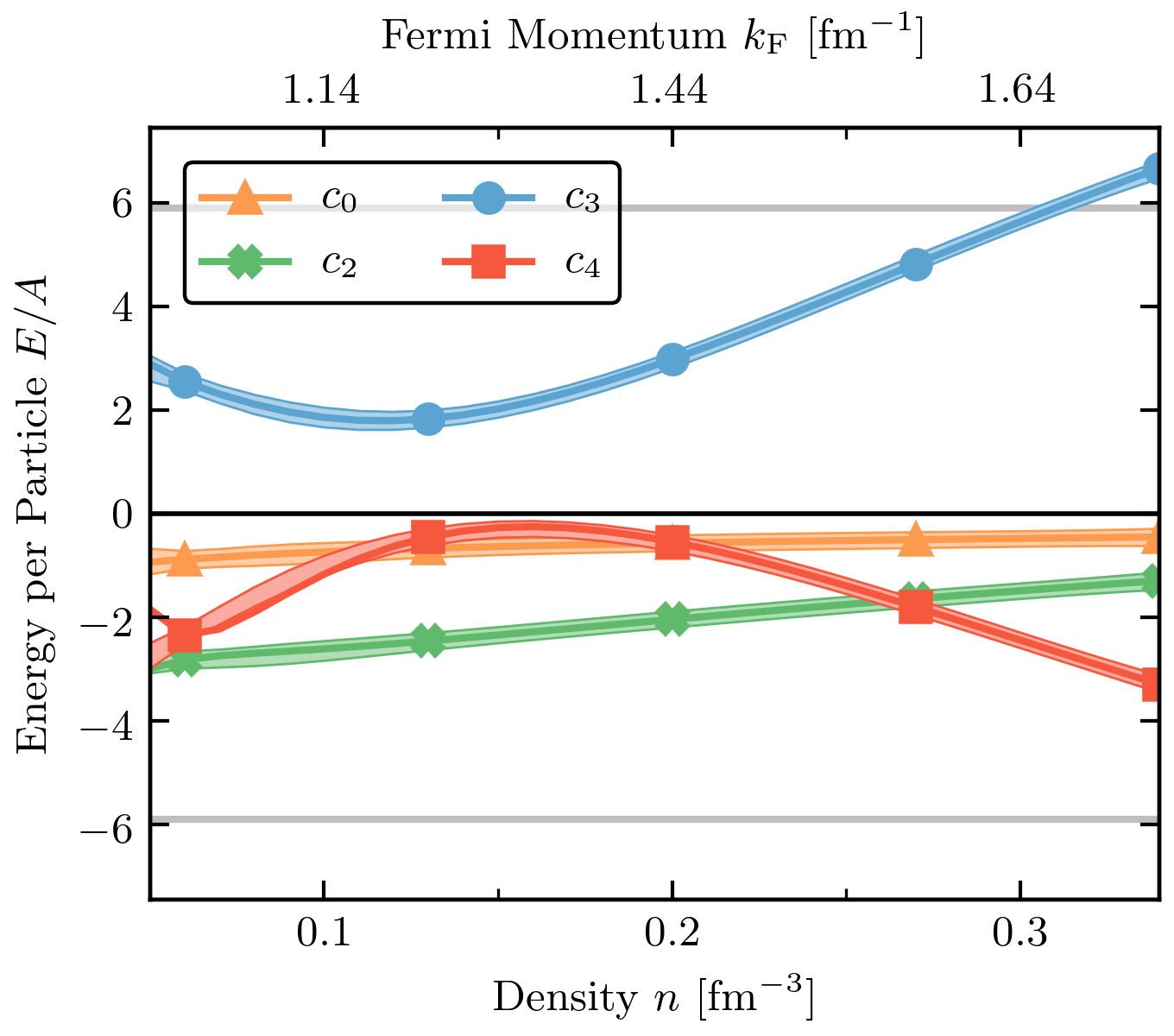}
    \caption{Observable coefficients $c_n$ for $E/A(n)$ up to \NNNLO, as a function of density $n$, using the $\Lambda = 500\MeV$ interactions in Table~\ref{tab:drischler_potential_fits}.
    See Fig.~\ref{fig:energy_per_neutron_coefficients_lambda-500} for the notation.
    The estimated hyperparameters are given by $\cbar = 2.9$ and $\ell = 0.48\fmi$.
        Note that the secondary $x$-axis (at the top of the figure) is not linear in $\kf^\text{SNM}$.
    }
    \label{fig:energy_per_particle_coefficients_lambda-500}
\end{figure}
%%%%%%%%%%%%%%%%%%%%%%%%%%%%%%%%%%%%%%%%%%%%%%%%%%%%%%%%%%%%%%%

The EFT expectation is then that each successive order
$\Delta\genobs_n(\kinparvec)$ should decrease by about a factor of $Q=
\kf/\Lambda_b \gtrsim 1/3$ (depending on density)---except for $\Delta y_2$, where the change from LO to NLO leads to $Q^2$
improvement. Tables~\ref{tab:energies_per_nucleon_450MeV} and
\ref{tab:energies_per_nucleon_500MeV} of Appendix~\ref{sec:tables_pnm_snm} show that the predicted energy per particle for PNM and SNM is consistent with this convergence pattern, as long as we allow for statistical variations of the
coefficients $c_2(\kf)$, $c_3(\kf)$, and $c_4(\kf)$ in Eq.~\eqref{eq:Delta_y}.

Reference~\cite{Drischler:2017wtt} showed that
the residual MBPT uncertainty is much smaller than the estimated \chiEFT\ truncation
error for the interactions considered here.
Nevertheless, to be conservative, we
assign an (uncorrelated) uncertainty of 0.1\% (or $\geqslant 20\keV$, whichever is greater) to the total
energy per particle in order to account for this residual uncertainty.
That is, when predicting the EOS with theory uncertainties given by Eq.~\eqref{eq:discr_k_prior}, we add a white noise term to the truncation error kernel, $\sdth^2 R_{\delta k}$, of Eq.~\eqref{eq:truncation_corrfunc_matter}.

But, before making those predictions, we need to obtain the GP hyperparameters from the convergence pattern of \chiEFT\@.
Because each $c_n(\kf)$ is extracted using a different power of $Q(\kf)$, even this small noise in the many-body calculations gets magnified as the EFT order $n$ grows. 
Our GP approach can straightforwardly account for such uncertainties in the training data. 
We smooth the EOS by fitting a GP at each order before 
computing the observable coefficients to obtain a noise level that is approximately constant in EFT order. 
This has no noticeable effect on the total energy per particle.
We then include a white noise term (called a nugget in this context) in the
RBF kernel [Eq.~\eqref{eq:rbf_kernel_nuclear_matter}] when training the GP hyperparameters.
This regularizes the matrix inversion of our GP framework.
The variance of the white noise is chosen to be $\sigma^2=5 \times 10^{-4}$.

With choices for $\kinparvec$, $\genobsref(x)$, and $Q(x)$ in hand, we can make the observations of the previous paragraph rigorous through a statistical analysis of
the convergence pattern of $E/N(n)$ and $E/A(n)$. The
corresponding observable coefficients $c_n(\kf)$ have been extracted  in
Figs.~\ref{fig:energy_per_neutron_coefficients_lambda-500}
and~\ref{fig:energy_per_particle_coefficients_lambda-500} using $\Lambda_b =
600\MeV$\@. (We return to the choice of $\Lambda_b$ shortly.) Each curve appears
to be naturally sized and relatively smooth across this range of density. Below
\NkLO{2}, where only NN forces are present in \chiEFT, the coefficients have
particularly large length scales, whereas the higher orders %clearly show signs
exhibit greater curvature because 3N contributions affect their density dependence. The GP hyperparameters, $\cbar$ and $\ell$, are trained on
all coefficients except $c_0$ (LO), and hence find values that represent the features of
all $c_n(\kf)$ simultaneously.
[The leading-order term is often disregarded when we perform this induction on the $c_n(\kf)$ because it informed the selection of  $\genobsref$.] 
As a test that these hyperparameters are
appropriate, we plot in Figs.~\ref{fig:energy_per_neutron_coefficients_lambda-500}
and~\ref{fig:energy_per_particle_coefficients_lambda-500} also the GP interpolants (colored $1\sigma$ bands) with these $\cbar$ and $\ell$ values.
These smoothly interpolate the density-dependence of each coefficient, including the untrained data.
%Each process follows the remainder of the curve given only the training data.
% Nevertheless, the GP hyperparameters, $\cbar$ and $\ell$, learned by fitting
% to all coefficients except leading order appear to describe all coefficients
% relatively well. We provide model checking diagnostics in
% Appendix~\ref{sec:model_checking} which support the claim that these
% coefficients can be treated as drawn from one underlying GP\@.

% Figures~\ref{fig:energy_per_neutron_coefficients_lambda-500}
% and~\ref{fig:energy_per_particle_coefficients_lambda-500} show the resulting
% $c_n$ and GPs tuned to them.
The hyperparameter $\cbar$ is easily updated due to its conjugate prior, whereas
$\ell$ is determined by optimizing the log-likelihood.
% The coefficients were found using a point estimate for $\Lambda_b$, and the
% GPs then found an optimal $\ell$ by maximizing the log likelihood, given this
% choice of $\Lambda_b$.
To support the use of $\Lambda_b = 600\MeV$ as a point estimate for the EFT
breakdown scale, along with the fit values of $\ell$, we provide posterior
distributions $\pr(\Lambda_b, \ell \given \data)$ trained on each $E/N(n)$,
$E/A(n)$, and both simultaneously (assuming that they are independent data). The
marginal posteriors $\pr(\Lambda_b \given \data)$ and $\pr(\ell \given \data)$
can then be obtained by integrating over $\ell$ and $\Lambda_b$, respectively. We
use a Gaussian prior $\pr(\Lambda_b) = 600\pm 150\MeV$ for the breakdown scale,
and a scale invariant prior $\pr(\ell) = 1 / \ell$ for the each length scale of
$E/N(n)$ and $E/A(n)$.

%%%%%%%%%%%%%%%%%%%%%%%%%%%%%%%%%%%%%%%%%%%%%%%%%%%%%%%%%%%%%%%
\begin{figure}[tb]
    \centering
    \includegraphics{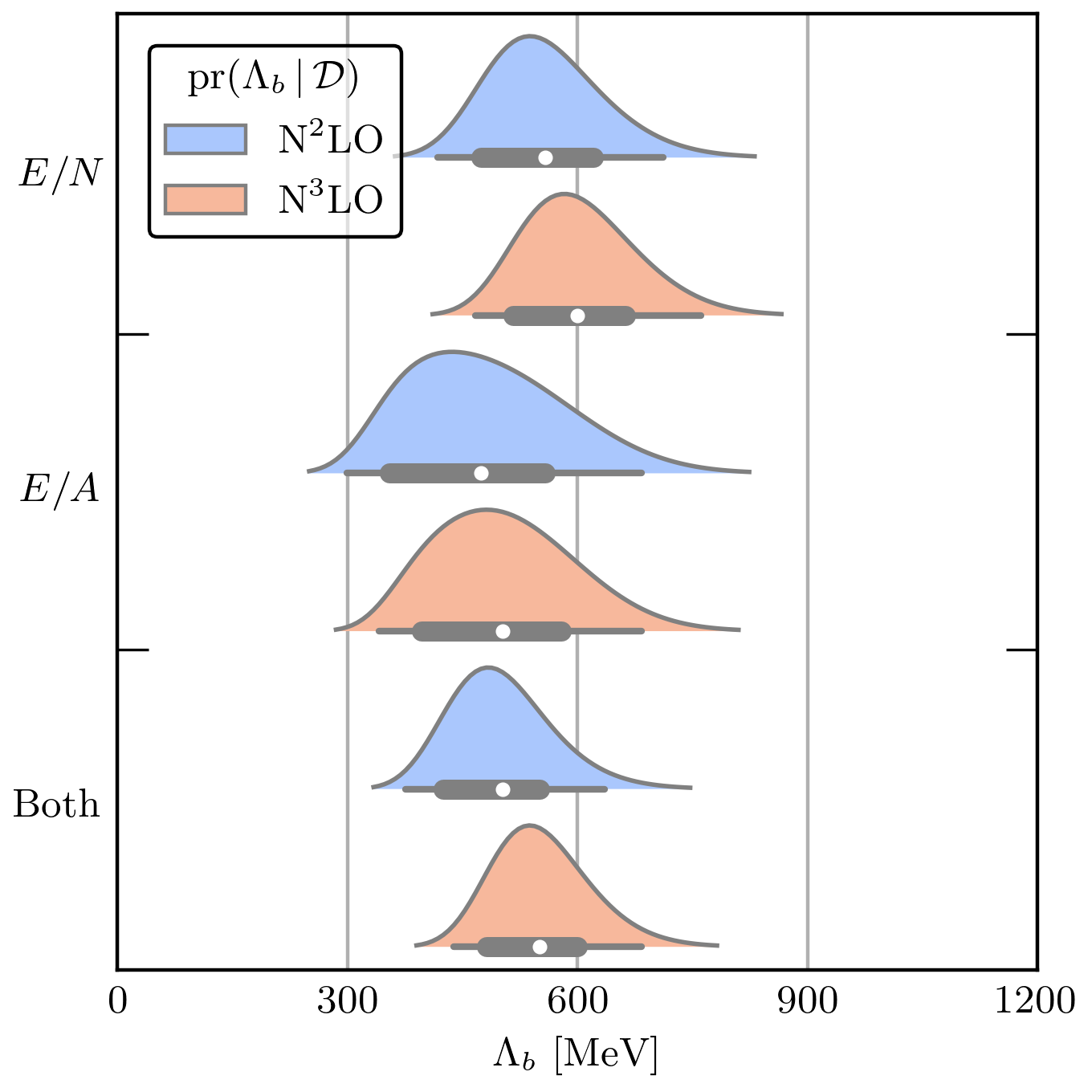}
    \caption{
    The posteriors for the EFT breakdown scale $\Lambda_b$ using orders through \NkLO{2}(blue bands) and \NkLO{3} (red bands) corresponding to
    Figs.~\ref{fig:energy_per_neutron_coefficients_lambda-500}
    and~\ref{fig:energy_per_particle_coefficients_lambda-500}. The upper pair of posteriors comes from analyzing $E/N$, the middle pair from $E/A$, and the bottom from a combined analysis. In all cases
    a Gaussian prior
    centered at $\Lambda_b = 600\pm 150\MeV$ is used.
    The combined \NkLO{3} posterior is consistent with the $\Lambda_b \approx 600\MeV$ found when considering free-space NN scattering observables~\cite{Melendez:2017phj}.
    } \label{fig:breakdown_posteriors_lambda-500}
\end{figure}

%%%%%%%%%%%%%%%%%%%%%%%%%%%%%%%%%%%%%%%%%%%%%%%%%%%%%%%%%%%%%%%

Figure~\ref{fig:breakdown_posteriors_lambda-500} shows the posteriors for $\Lambda_b$. We compute the posteriors using observable
coefficients up to \NkLO{2} and \NkLO{3} (neglecting LO) as a check of the
robustness of our analysis. It is clear that the results behave consistently
between orders and between observables, with $E/A(n)$ possibly preferring a
smaller breakdown scale than $E/N(n)$. When combined, the posterior has a median
around $\Lambda_b \approx 560\MeV$ with a $1\sigma$ spread of over $50\MeV$.
This is consistent with $\Lambda_b \approx 600\MeV$ as extracted from $\npr$ cross sections and angular observables of free-space NN scattering. Thus, we choose
$\Lambda_b = 600\MeV$
here for simplicity. We reiterate that, in reality, only the ratio $Q(\kf) = \kf/\Lambda_b$ is 
determined by this analysis, and hence extracting $\Lambda_b$ is contingent on
our choice of the numerator, $\kf$.
If, for example, $\kf^\text{PNM}$ were used
for $E/A(n)$, this would bring the $\Lambda_b$ posteriors to even better
agreement, with a median value $\Lambda \gtrsim 600\MeV$.
Because $\kf^\text{PNM} > \kf^\text{SNM}$ for a given density, and with the estimate $\Lambda_b \simeq 600\MeV$, the
truncation error of $E/A(n)$ would then grow accordingly.

\begin{figure}[tb]
    \centering
    \includegraphics{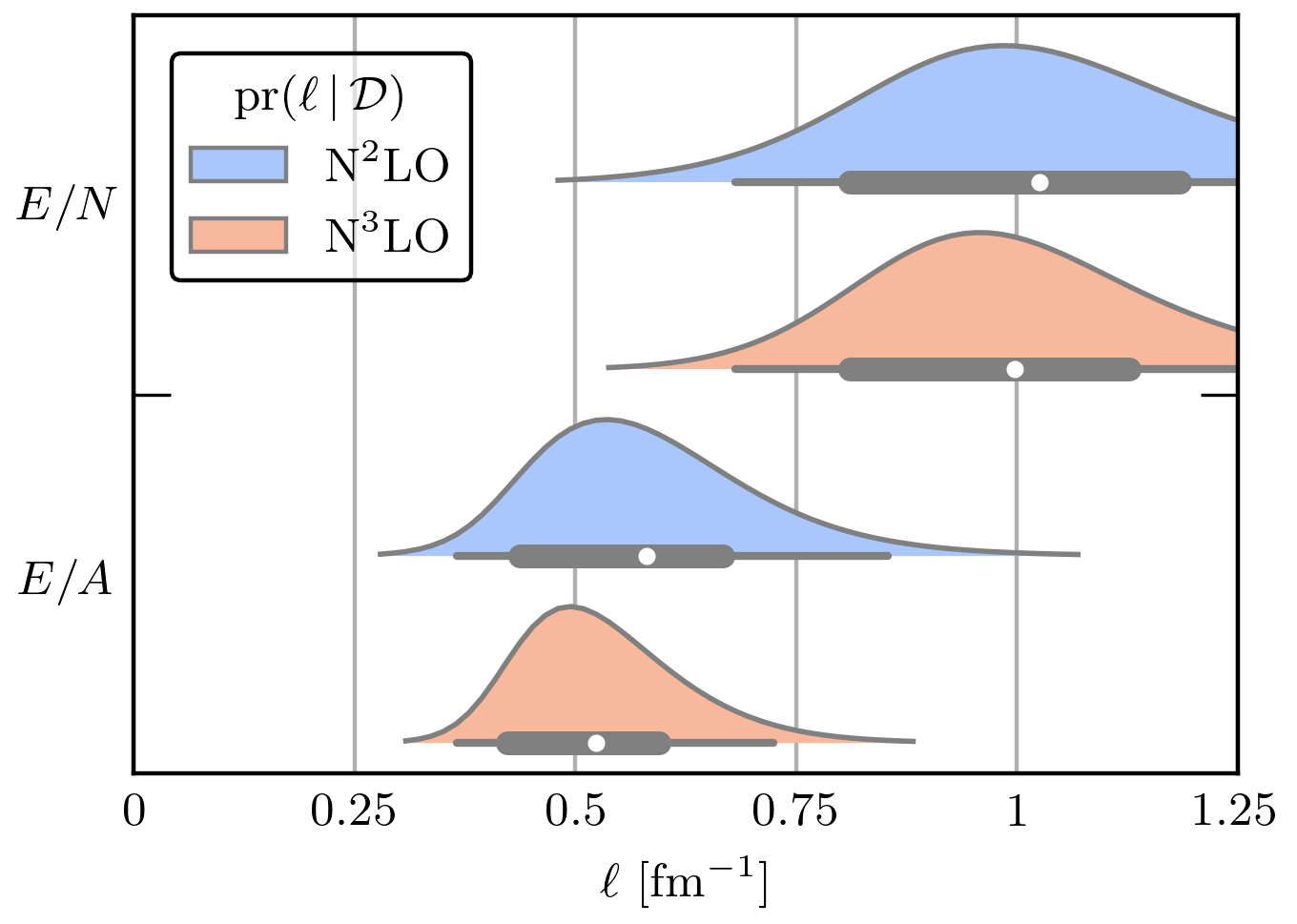}
    \caption{
    Length-scale posteriors organized similarly to Fig.~\ref{fig:breakdown_posteriors_lambda-500}. A scale-invariant
    prior proportional to $1/\ell$ is used, and each length scale is relative to
    the $\kf$ of each system.
    If we were to use a single
    $\kf$ prescription, the length scales in PNM and SNM, $\ell_\text{PNM}$ and $\ell_\text{SNM}$, respectively, would transform just as $\kf$, making the
    posteriors shift towards agreement. See the discussion of the input space in
    the main text.
    } \label{fig:length_scale_posteriors_lambda-500}
\end{figure}

Figure~\ref{fig:length_scale_posteriors_lambda-500} depicts the length scale
posteriors for the observable coefficients in PNM and SNM.
%Each $\ell$ is provided in units of the respective $\kf$ of the system.
If the length scales were put on a common
scale then the posteriors would become more aligned.
%, though we have no reason \emph{a priori} to expect that to be the case. 
The fact that the $c_n(\kf)$ from PNM and SNM could share a common $\ell$ in the same $\kf$ scale could
prove useful when modeling the correlations of the convergence patterns in
$E/N(n)$ and $E/A(n)$. We return to this in
Sec.~\ref{sec:symmetry_energy}. Both length scales are relatively
large given the range of $\kf$ used in this work. This implies that the
truncation error is a \emph{highly} correlated quantity, and hence that an
estimate of this correlation will prove crucial to a robust UQ in infinite
matter.

%%%%%%%%%%%%%%%%%%%%%%%%%%%%%%%%%%%%%%%%%%%%%%%%%%%%%%%%%%%%%%%
%%%%%%%%%%%%%%%%%%%%%%%%%%%%%%%%%%%%%%%%%%%%%%%%%%%%%%%%%%%%%%%

We have provided one reasonable implementation of a GP-based
EFT convergence model in this subsection. Other choices for $x$, $\genobsref(x)$, and $Q(x)$ could
be made. 
We provide an example of an alternative $\genobsref(x)$ in Appendix~\ref{sec:model_checking_infinite_matter}, which also
provides model checking diagnostics for the interested reader to verify our
convergence model for these systems. We find evidence that
$c_3(\kf)$ may be an outlier in terms of the large effect of the 3N
contributions that enter \chiEFT\ at \NNLO. If one does not believe that such large
corrections will continue, it may prove useful to leave $c_3(\kf)$ (\NNLO) out of our
inductive model for higher-order terms.

Additionally, the diagnostics point to
the possibility that the NN-only coefficients $c_0(\kf)$ (LO) and $c_2(\kf)$ (NLO) may have a different correlation structure than higher
orders. As noted above, this is suggested by a visual inspection of
Figs.~\ref{fig:energy_per_neutron_coefficients_lambda-500}
and~\ref{fig:energy_per_particle_coefficients_lambda-500}, where $c_0(\kf)$ and
$c_2(\kf)$ appear much flatter than $c_3(\kf)$ (\NNLO) and $c_4(\kf)$ (\NNNLO). An investigation in this direction is presented in Appendix~\ref{sec:model_checking_infinite_matter}. There we have attempted to isolate the strongly
repulsive 3N contributions that change the correlation structure by splitting the coefficients into NN-only and residual 3N coefficients with each having different $\kf$ dependence in $\genobsref(x)$.
%$c_n^{(2)}(\kf)$ and $c_n^{(3)}(\kf)$, 
This succeeds in making the coefficients more uniform and improves the diagnostics for PNM, but does not improve SNM significantly. Crucially, 
the order-by-order uncertainty bands for PNM and SNM presented in the next section are almost unchanged when  this alternative model is used; the saturation ellipses do become slightly larger though.
We provide these
details, along with annotated Jupyter notebooks~\cite{BUQEYEgithub} that generate them, to promote further
investigation, possibly with other EFT implementations, into the systematic
convergence of infinite matter.

%%%%%%%%%%%%%%%%%%%%%%%%%%%%%%%%%%%%%%%%%%%%%%%%%%%%%%%%%%%%%%%
\subsection{Quantified uncertainties for PNM and SNM}
\label{sec:uncertainties_for_pnm_snm}

%%%%%%%%%%%%%%%%%%%%%%%%%%%%%%%%%%%%%%%%%%%%%%%%%%%%%%%%%%%%%%%
\begin{figure}[tb]
    \centering
    \includegraphics{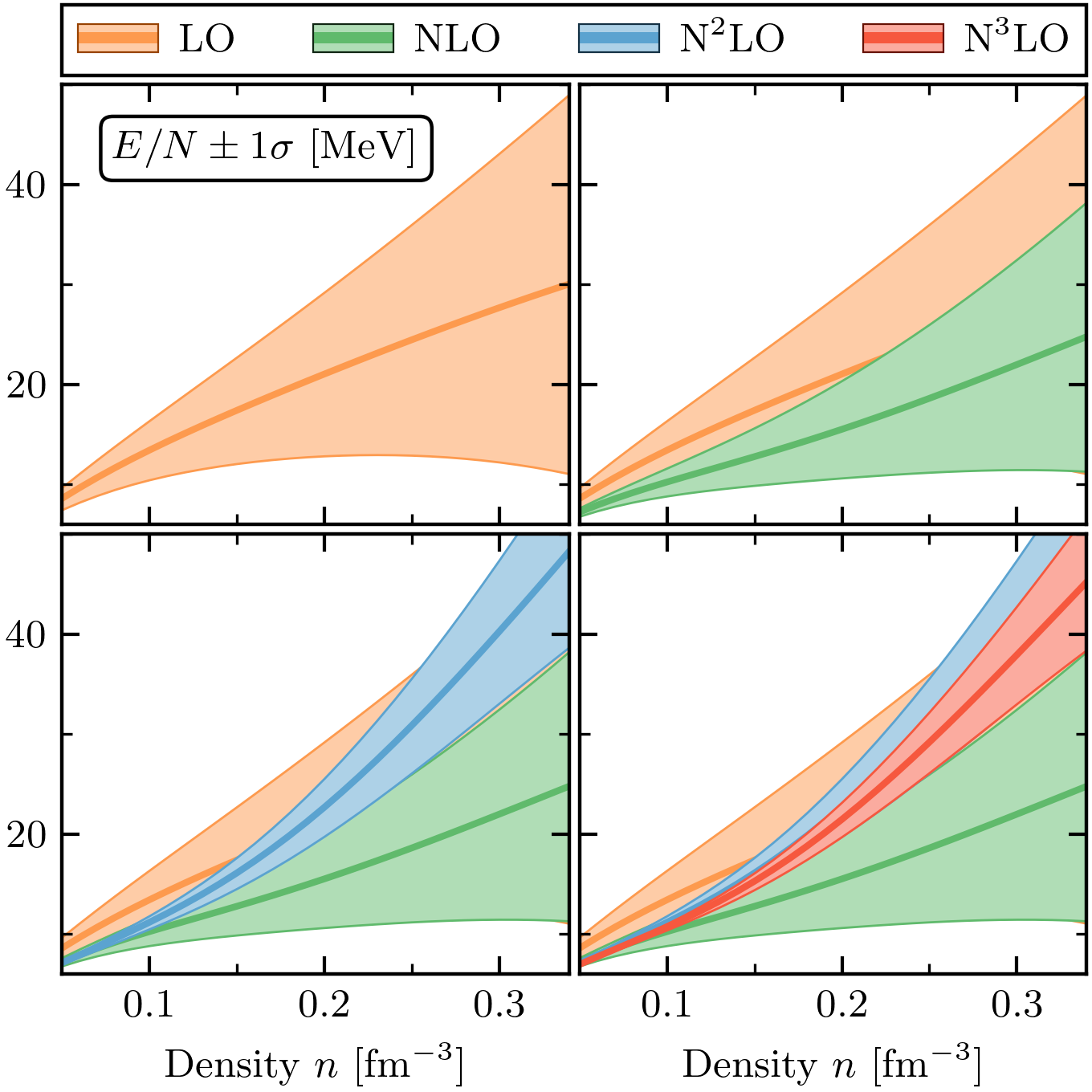}
    \caption{
    Energy per particle in PNM with truncation errors using the
    $\Lambda=500\MeV$ interactions in Table~\ref{tab:drischler_potential_fits}. From left to right, top to bottom, the panels show the order-by-order progression of EFT uncertainties as the \chiEFT\ order increases.
    The bands indicate
    68\% credible intervals.
    } \label{fig:neutron_matter_predictions_lambda-500}
\end{figure}
%%%%%%%%%%%%%%%%%%%%%%%%%%%%%%%%%%%%%%%%%%%%%%%%%%%%%%%%%%%%%%%

%%%%%%%%%%%%%%%%%%%%%%%%%%%%%%%%%%%%%%%%%%%%%%%%%%%%%%%%%%%%%%%
\begin{figure}[tb]
    \centering
    \includegraphics{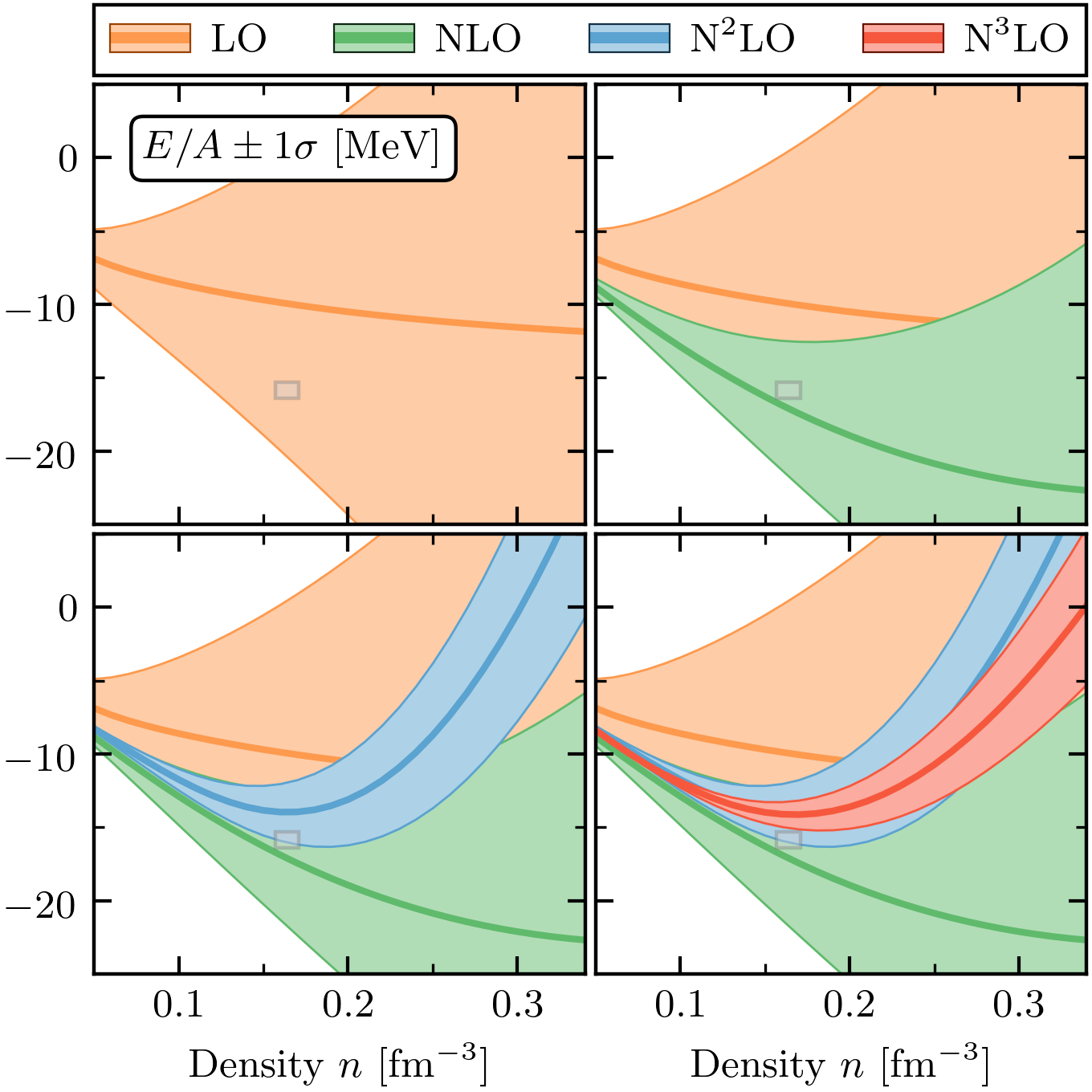}
    \caption{
    Similar to Fig.~\ref{fig:neutron_matter_predictions_lambda-500} but for SNM\@. The gray box depicts the empirical saturation point, $n_0 = 0.164 \pm 0.007 \fmiq$ with
$E/A(n_0) = -15.86 \pm 0.57 \MeV$, obtained from a set of energy-density functionals~\cite{Drischler:2015eba,Drischler:2017wtt} (see the main text for details).
    } \label{fig:nuclear_matter_predictions_lambda-500}
\end{figure}
%%%%%%%%%%%%%%%%%%%%%%%%%%%%%%%%%%%%%%%%%%%%%%%%%%%%%%%%%%%%%%%

%%%%%%%%%%%%%%%%%%%%%%%%%%%%%%%%%%%%%%%%%%%%%%%%%%%%%%%%%%%%%%%
\begin{figure}[!htb]
    \centering
    \includegraphics{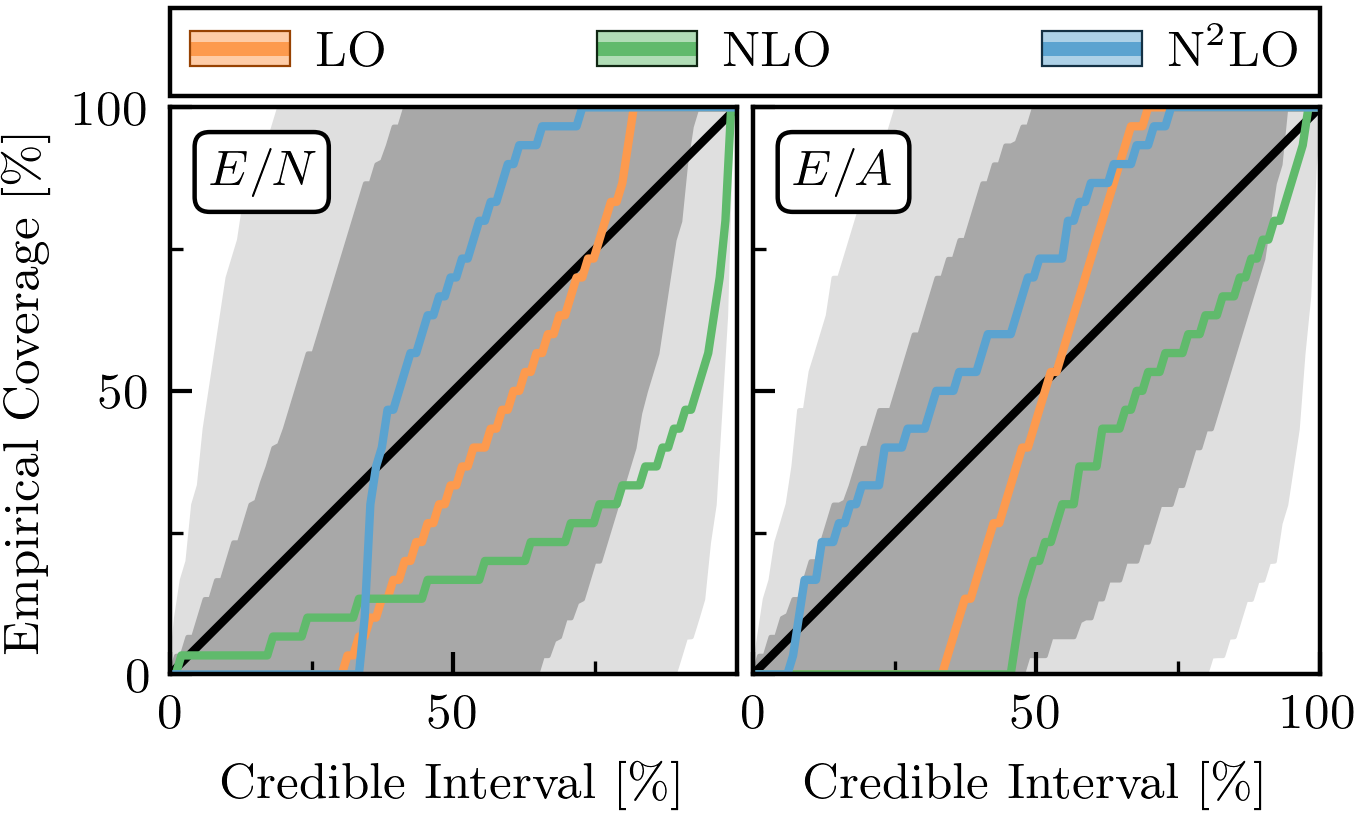}
    \caption{
    Credible-interval diagnostics for the $E/N(n)$ (left-hand side) and $E/A(n)$ uncertainty
    bands (right-hand side) for the $\Lambda = 500\MeV$ interactions in Table~\ref{tab:drischler_potential_fits}; for details see Ref.~\cite{Melendez:2019izc}. At each order we construct an uncertainty
    band for the upcoming correction (not the full truncation error) and test
    whether the next order is contained within it at  a specific credible
    interval. The expected size of fluctuations due to the finite effective
    sample size of the curves is depicted using dark (light) gray bands for the
    68\% (95\%) interval. Both bands are quite large, which shows that
    correlations are crucial to assess whether truncation errors have been properly assigned.
    }\label{fig:credible_interval_diagnostic_pnm_and_snm_lambda-500}
\end{figure}
%%%%%%%%%%%%%%%%%%%%%%%%%%%%%%%%%%%%%%%%%%%%%%%%%%%%%%%%%%%%%%%

The GP truncation error model described in Sec.~\ref{sec:the_model_nuclear_matter} combined
with the hyperparameter estimates now permit the first statistically
rigorous \chiEFT\ uncertainty bands in infinite matter.
Figures~\ref{fig:neutron_matter_predictions_lambda-500}
and~\ref{fig:nuclear_matter_predictions_lambda-500} depict predictions of $E/N(n)$ and $E/A(n)$, respectively, up to \NNNLO with 68\% credible intervals (colored bands). From left to right, top to bottom, the panels show the order-by-order progression of EFT uncertainties as the \chiEFT\ order increases.
The gray box in Fig.~\ref{fig:nuclear_matter_predictions_lambda-500}
represents the empirical saturation point, $n_0 = 0.164 \pm 0.007 \fmiq$ with
$E/A(n_0) = -15.86 \pm 0.57 \MeV$, obtained from a set of energy-density functionals in Ref.~\cite{Drischler:2015eba,Drischler:2017wtt}. We stress, however, that the quoted uncertainty in the empirical saturation point does not permit a statistical
interpretation (\eg, $1\sigma$ credibility interval), in contrast to the results discussed this work.

In these figures, one might be tempted to count how frequently $1\sigma$
bands from one order contain the prediction of the subsequent orders, and to
compare the frequency to the nominal value of 68\%. If the bands are too
conservative (aggressive), more (less) than 68\% of the points will lie within the bands. But the highly
correlated nature of the truncation error renders such an assessment difficult: if
any point along a given curve is contained (not contained) within an uncertainty band, nearby points will also likely be inside (outside) the band. Long correlation lengths indicate that the effective sample size, $N_{\rm eff}$, is much smaller then the number of data points. $\sqrt{N_{\rm eff}}$ fluctuations mean that the bands, within which statistically consistent truncation-error prescriptions fall, are wider.

This is not a failure of our convergence model; these correlations are real and
must be dealt with one way or another. On the contrary, we can provide
estimates of exactly how correlated the truncation error is, and use that to
inform us how perturbed we should be by any perceived ``failure'' of the
uncertainty bands to align with our intuitions. We show a useful diagnostic
tool in Fig.~\ref{fig:credible_interval_diagnostic_pnm_and_snm_lambda-500},
which plots the empirical coverage of the credible intervals versus the choice
of credible interval~\cite{Melendez:2017phj,Melendez:2019izc}. Importantly, we provide gray bands which account for
random fluctuations due to the finite sample size $N_{\rm eff}$. 
As anticipated, the bands are very large due to
the presence of correlations, much larger than the binomial bands one would
obtain by na\"ive counting (compare to\ Figs.~11 to 20 in Ref.~\cite{Melendez:2017phj}).
%The upshot is that in this case the empirical coverage is statistically consistent for all credible intervals.
Correctly accounting for these correlations shows that the empirical coverage is statistically consistent for all credible intervals.

%%%%%%%%%%%%%%%%%%%%%%%%%%%%%%%%%%%%%%%%%%%%%%%%%%%%%%%%%%%%%%%
\begin{figure*}[tb]
\subfloat[]{
  \includegraphics{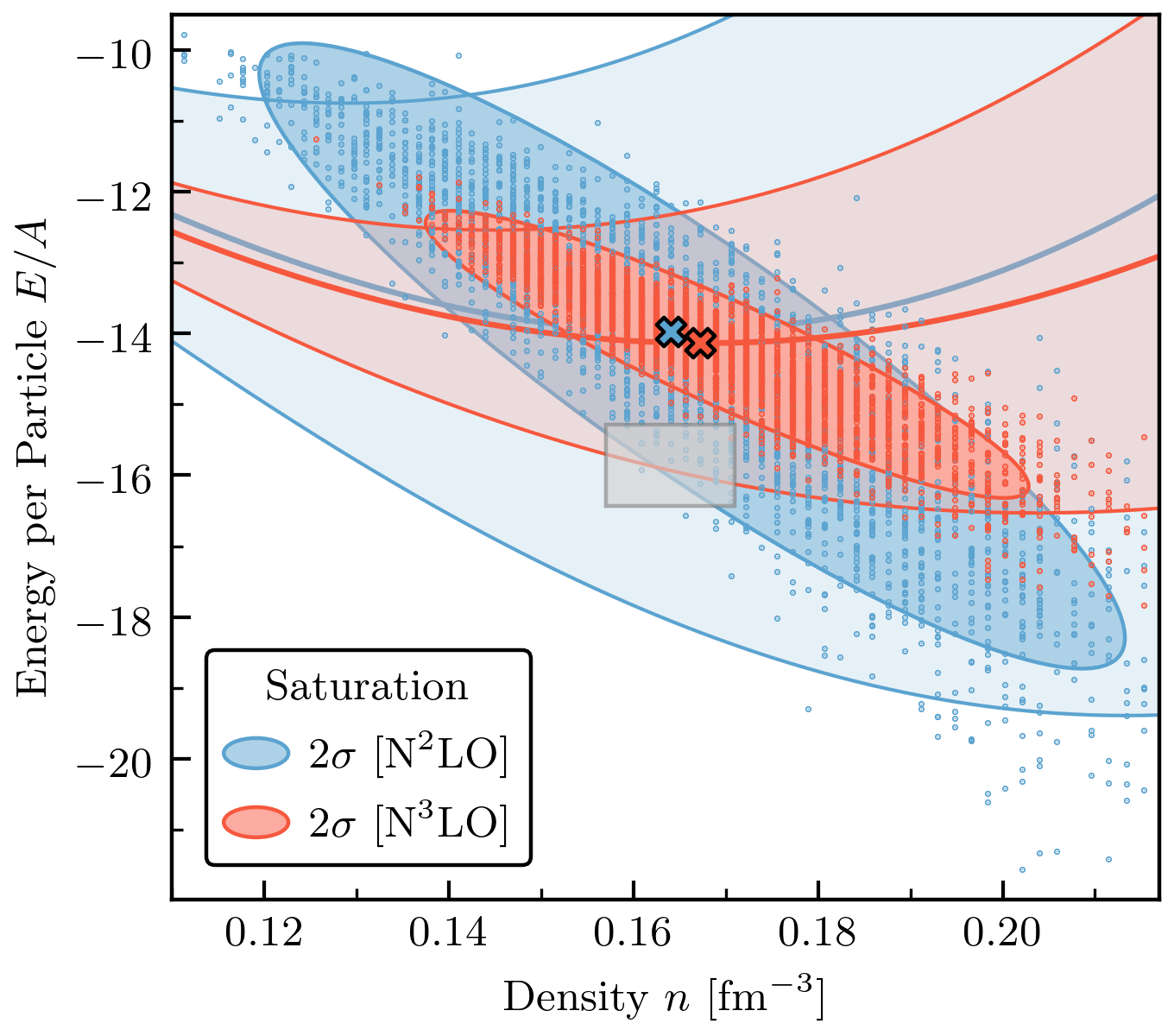}
  \label{fig:coester_ellipses_500MeV_fit_4_10}
  }
 \subfloat[]{
    \includegraphics{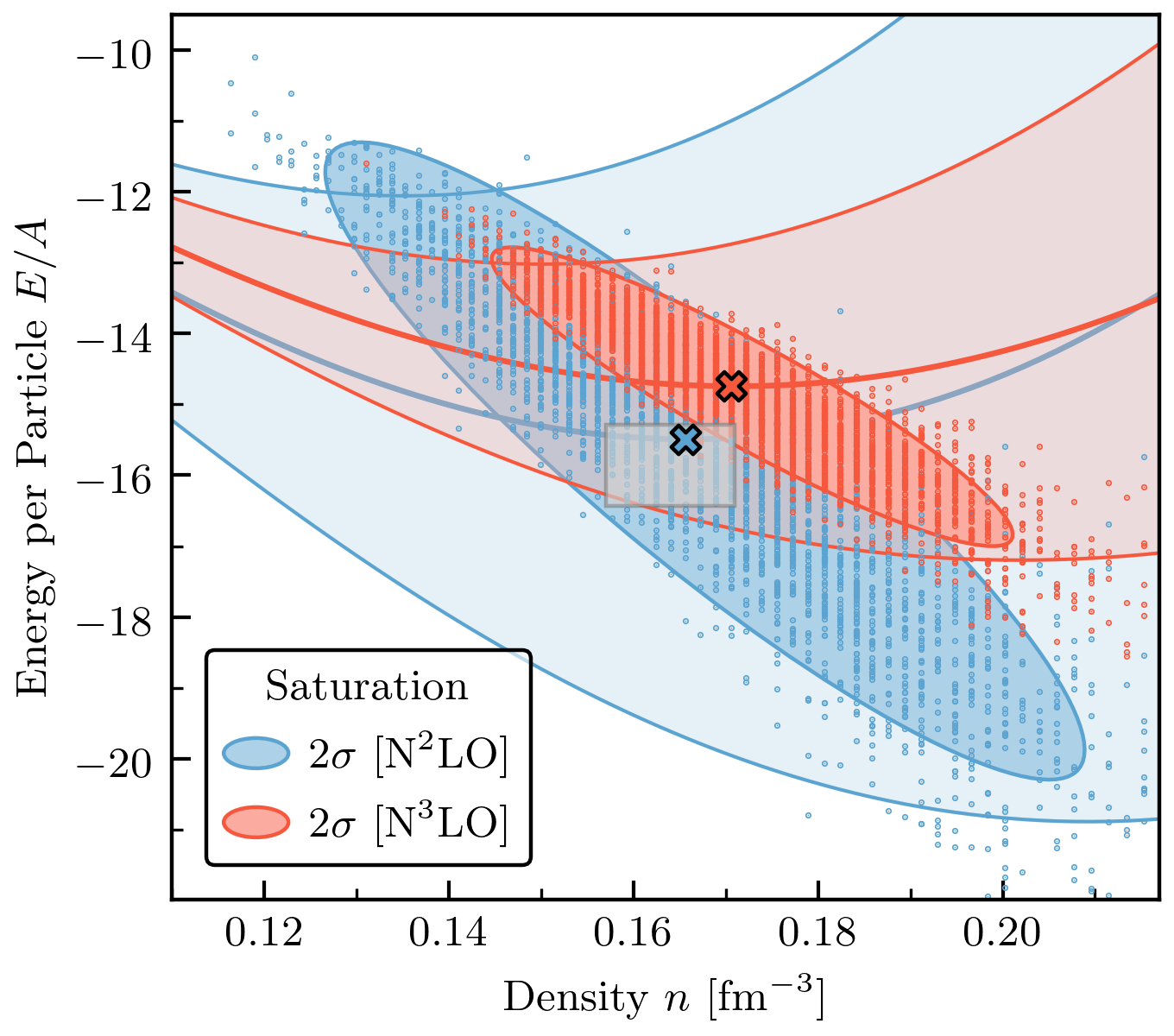}
    \label{fig:coester_ellipses_450MeV_fit_1_7}
    }
  \caption{
 Predicted nuclear saturation point of SNM at \NNLO (blue) and \NNNLO (red) with 95\% ($2\sigma$) credible interval ellipses including correlated truncation
  errors. Panels \protect\subref{fig:coester_ellipses_500MeV_fit_4_10} and \protect\subref{fig:coester_ellipses_450MeV_fit_1_7} show the $\Lambda=500\MeV$ and $\Lambda=450\MeV$ interactions described in Table~\ref{tab:drischler_potential_fits}, respectively. The colored crosses depict the minimum of each EOS obtained at that order without accounting for truncation errors. In contrast, each dot
  is the
  minimum of a curve sampled from the corresponding $E/A(n)$ GP in
  Fig.~\ref{fig:nuclear_matter_predictions_lambda-500}, which are used to fit the ellipses. These should be compared to the $2\sigma$ uncertainty bands in the same color, which are estimated using $\Lambda_b =
  600\MeV$. The gray box again depicts the empirical saturation point.
  }
  \label{fig:coester_ellipses_both_lambda}
\end{figure*}
%%%%%%%%%%%%%%%%%%%%%%%%%%%%%%%%%%%%%%%%%%%%%%%%%%%%%%%%%%%%%%%

%Theories with correlated error bands can provide realistic estimates for EOS properties of the EOS\@.
Next, we determine the location of the nuclear saturation point [\ie, the minimum of $E/A(n)$]. Ours is the first analysis to do this with fully correlated truncation errors. To assess saturation properties of nuclear interactions, one could follow a ``Coester plot'' approach (see, \eg, Refs.~\cite{Drischler:2016djf, Drischler:2017wtt}) and ask  whether the uncertainty bands in Fig.~\ref{fig:nuclear_matter_predictions_lambda-500} overlap with the empirical saturation point (gray box) at a certain credibility level and how close the mean values are to that region. For
example, while the \NkLO{2} band completely overlaps, the \NkLO{3} does so only partially.

Instead, we create here the foundation for a statistical analysis of nuclear saturation properties and obtain the joint posterior $\pr(E/A(n_0),
n_0 \given \data)$ given $\data$, the order-by-order predictions of $E/A(n)$ up to $2n_0$. We compute this distribution by sampling thousands of curves from the GP interpolant of $E/A(n)$ and extracting for each of them the minimum.
The resulting posteriors at \NkLO{2} (blue bands) and \NkLO{3} (red bands) are depicted in Fig.~\ref{fig:coester_ellipses_both_lambda} with $2\sigma$ ellipses. They are well-approximated at
\NkLO{2} and \NkLO{3} by a two-dimensional Gaussian. The mean and covariance of the highest-order prediction, is given by
\begin{align}
% x: 0.17014726962696783 +/- 0.01633229947392557
% y: -14.295079499264855 +/- 1.012689546110345
% mean:
%  [  0.17014727 -14.2950795 ]
% cov:
%  [[ 2.66832950e-04 -1.51811388e-02]
%  [-1.51811388e-02  1.02588208e+00]]
    \begin{bmatrix}
        n_0 \\ \frac{E}{A}(n_0)
    \end{bmatrix}
    \approx
    \begin{bmatrix}
        0.170 \\ -14.3
    \end{bmatrix}
    \quad \text{and} \quad
    \Sigma \approx
    \begin{bmatrix}
        0.016^2 & -0.015 \\
        -0.015 & 1.0^2
    \end{bmatrix}
    \label{eq:n0_EA_n0_mean_cov_500 MeV}
\end{align}
for the $\Lambda=500\MeV$ potentials [Fig.~\subref*{fig:coester_ellipses_500MeV_fit_4_10}] and by
\begin{align}
% x: 0.17286690634955418 +/- 0.01410628073764282
% y: -14.892831696254024 +/- 1.053694276861793
% mean:
%  [  0.17286691 -14.8928317 ]
% cov:
%  [[ 1.99053530e-04 -1.37172988e-02]
%  [-1.37172988e-02  1.11064197e+00]]
    \begin{bmatrix}
        n_0 \\ \frac{E}{A}(n_0)
    \end{bmatrix}
    \approx
    \begin{bmatrix}
        0.173 \\ -14.9
    \end{bmatrix} 
    \quad \text{and} \quad
    \Sigma \approx
    \begin{bmatrix}
        0.014^2 & -0.014 \\
        -0.014 & 1.1^2
    \end{bmatrix}
    \label{eq:n0_EA_n0_mean_cov_450 MeV}
\end{align}
for the $\Lambda=450\MeV$ potentials [Fig.~\subref*{fig:coester_ellipses_450MeV_fit_1_7}] in Table~\ref{tab:drischler_potential_fits}.
The off-diagonal terms in the covariance matrices render the
posteriors elliptical, rotated at an angle such that the \NkLO{3} bands barely (if at all) overlap with the empirical saturation point. These findings
are consistent with the conclusions in Ref.~\cite{Drischler:2017wtt}, albeit that work did not employ the statistical tools presented here.

Reference~\cite{Hoppe:2019uyw} studied binding energies and charge radii of medium-mass to heavy nuclei based on these \chiEFT\ NN and 3N interaction constrained by empirical saturation properties (see Figs.~6 and~7 in that reference).
While the selected closed-shell oxygen, calcium, and nickel isotopes are underbound (as expected from the findings in infinite matter), the charge radii are too large---opposite to the expectation from infinite matter.
Furthermore, the sensitivity of the observables to the 3N low-energy coupling $c_D$ is significantly less than that to the infinite-matter properties is (see Figs.~8 and 9 of Ref.~\cite{Hoppe:2019uyw}).
The link between finite nuclei in this mass range and infinite matter thus seems to be more intricate than one might na{\"i}vely expect~\cite{Hoppe:2019uyw}.

%%%%%%%%%%%%%%%%%%%%%%%%%%%%%%%%%%%%%%%%%%%%%%%%%%%%%%%%%%%%%%%
\section{Results for Derived Quantities} \label{sec:derivative_quantities}

This section describes the second set of correlations addressed in
this work: ``type-$y$'' correlations between observables.
Section~\ref{sec:symmetry_energy} proposes a novel correlation structure between
$E/N(n)$ and $E/A(n)$, prescribing how they combine to yield the symmetry
energy, whereas Sec.~\ref{sec:derivative_quantities_snm} discusses how the
energy per particle is correlated with its derivatives and related quantities.

%%%%%%%%%%%%%%%%%%%%%%%%%%%%%%%%%%%%%%%%%%%%%%%%%%%%%%%%%%%%%%%
\subsection{Nuclear symmetry energy} \label{sec:symmetry_energy}

%%%%%%%%%%%%%%%%%%%%%%%%%%%%%%%%%%%%%%%%%%%%%%%%%%%%%%%%%%%%%%%
\begin{figure*}[tb]
    \centering
    \includegraphics[width=\textwidth]{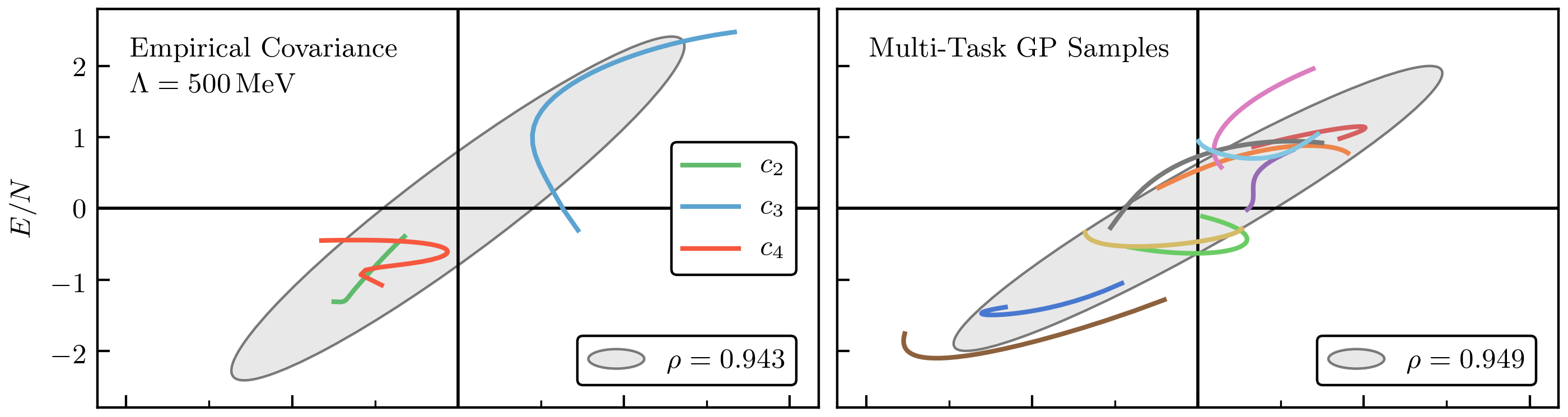}
    \phantomsublabel{-6.45}{0.11}{fig:empirical_correlation_ellipses_EN_vs_EA_true_coefficients_lambda-500}
    \phantomsublabel{-3.15}{0.11}{fig:empirical_correlation_ellipses_EN_vs_EA_sampled_coefficients_lambda-500}\\
    \includegraphics[width=\textwidth]{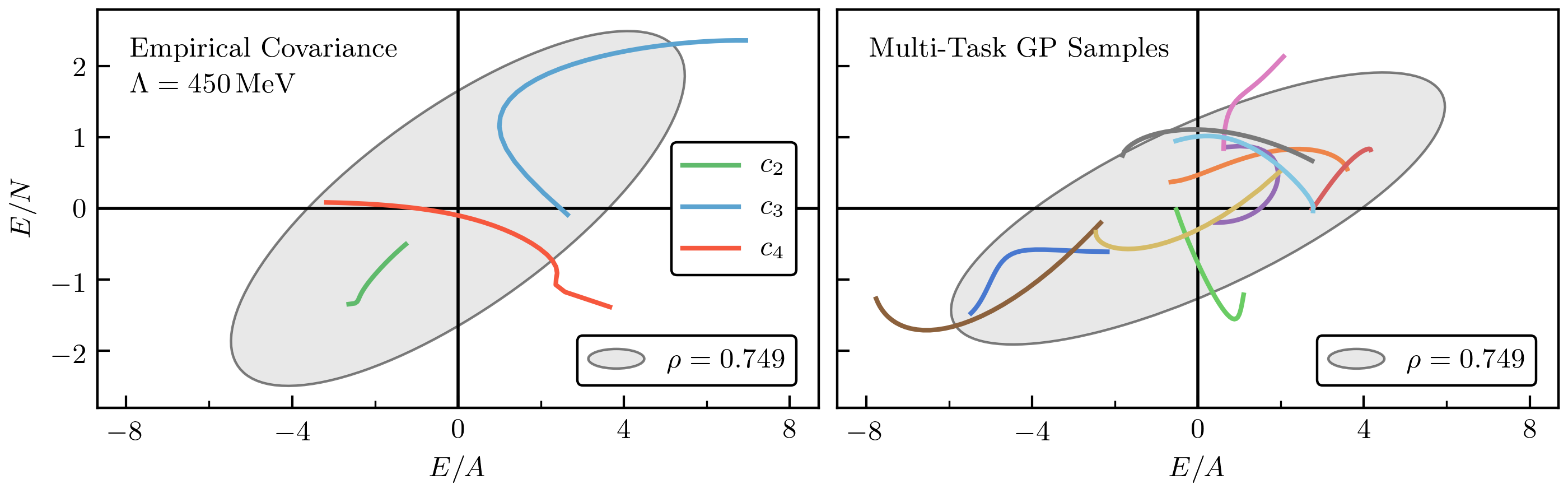}
    \phantomsublabel{-6.45}{0.45}{fig:empirical_correlation_ellipses_EN_vs_EA_true_coefficients_lambda-450}
    \phantomsublabel{-3.15}{0.45}{fig:empirical_correlation_ellipses_EN_vs_EA_sampled_coefficients_lambda-450}
    \caption{
    Correlations between the observable coefficients $c_n(\kf)$ from $E/N(n)$
    and $E/A(n)$.
    The points on each coefficient curve are at the same density.
    \protect\subref{fig:empirical_correlation_ellipses_EN_vs_EA_true_coefficients_lambda-500} The coefficients from the $\Lambda = 500\MeV$ interactions.
    \protect\subref{fig:empirical_correlation_ellipses_EN_vs_EA_sampled_coefficients_lambda-500} Toy coefficients created using a correlated multi-output GP trained on the coefficients shown in \protect\subref{fig:empirical_correlation_ellipses_EN_vs_EA_true_coefficients_lambda-500}. Each
    dimension of the multi-output GP ellipse takes into account the
    associated $\cbar$ from
    Figs.~\ref{fig:energy_per_neutron_coefficients_lambda-500}
    and~\ref{fig:energy_per_particle_coefficients_lambda-500}, which need not be
    the same as the empirical variances of the ellipse in \protect\subref{fig:empirical_correlation_ellipses_EN_vs_EA_true_coefficients_lambda-500}.
    For the $\Lambda = 450\MeV$ interactions, the corresponding plots are shown in \protect\subref{fig:empirical_correlation_ellipses_EN_vs_EA_true_coefficients_lambda-450} and \protect\subref{fig:empirical_correlation_ellipses_EN_vs_EA_sampled_coefficients_lambda-450}.
    }
    % \label{fig:empirical_correlation_ellipses_EN_vs_EA_lambda-500}
    \label{fig:empirical_correlation_ellipses_EN_vs_EA_both_lambdas}
\end{figure*}
%%%%%%%%%%%%%%%%%%%%%%%%%%%%%%%%%%%%%%%%%%%%%%%%%%%%%%%%%%%%%%%

The nuclear-matter EOS at zero temperature as a function of the total nucleon density $n = n_n +
n_p$ and isospin asymmetry $\beta = (n_n-n_p)/n$ can be expanded about SNM
($\beta = 0$),
\begin{equation} \label{eq:EA_quadratic_expansion}
 \frac{E}{A}(n, \, \beta) \approx \frac{E}{A}(n,\, \beta = 0) + \beta^2 \, S_2(n) \,,
\end{equation}
with the neutron (proton) density given by $n_n$ ($n_p$). Microscopic
calculations based on \chiEFT\ NN and 3N interactions up to $\lesssim n_0$ have
shown that the standard (quadratic) expansion~\eqref{eq:EA_quadratic_expansion}
works reasonably well~\cite{Drischler:2013iza,Drischler:2015eba} (see also
Refs.~\cite{Kaiser:2015qia,Wellenhofer:2016lnl}). The density-dependent nuclear
symmetry energy $S_2(n)$ is then given by the difference,
\begin{equation} \label{eq:S2_difference}
	S_2(n) \approx \frac{E}{N}(n) - \frac{E}{A}(n) \,.
\end{equation}
That is, $S_2(n)$ is determined by the energy per
particle in PNM ($\beta = 1$) and SNM. Hence, treating the truncation error of $S_2(n)$ completely uncorrelated with the errors of $E/N(n)$ and $E/A(n)$ is questionable. But how can
the correlations between the convergence pattern of $E/N(n)$ and
$E/A(n)$ be estimated and incorporated into our UQ model?

The contribution to the energy per particle from each \chiEFT\
order is dictated by observable coefficients $c_n(\kf)$. We take a first step
towards quantifying the extent to which the $c_n(\kf)$ in PNM are correlated to those of SNM. This enables a diagnosis of the correlation between their EFT truncation errors. 
%and hence, quantifying the correlations between their truncation errors. 
To this end, we calculate the empirical Pearson correlation coefficient $\rho$
between the coefficients evaluated at the same density, assuming they have mean zero.
% For the $\Lambda = 500\MeV$ interactions,
% the correlation $\rho \approx \verifyvalue{0.94}$ is called \emph{very strong} and visualized in
% Fig.~\subref*{fig:empirical_correlation_ellipses_EN_vs_EA_true_coefficients_lambda-500}.
% For the $\Lambda = 450\MeV$ interactions, we also find with $\rho \approx \verifyvalue{0.75}$ a \emph{strong correlation} in Fig.~\subref*{fig:empirical_correlation_ellipses_EN_vs_EA_true_coefficients_lambda-450}.
For the $\Lambda = 500\MeV$ interactions,
we find a \emph{very strong}~\cite{Evans1996,Asuero:2006abc} correlation $\rho \approx 0.94$ that is visualized in
Fig.~\subref*{fig:empirical_correlation_ellipses_EN_vs_EA_true_coefficients_lambda-500}.
For the $\Lambda = 450\MeV$ interactions, there exists a \emph{strong}~\cite{Evans1996,Asuero:2006abc} correlation $\rho \approx 0.75$; see Fig.~\subref*{fig:empirical_correlation_ellipses_EN_vs_EA_true_coefficients_lambda-450}. Note that $\rho$ is not an observable, since it is the correlation of the $E/A$ and $E/N$ EFT coefficients at the EFT orders which have been computed. As such it depends on the choice of $\genobsref$ and $Q$. $\rho$ will also be a function of the EFT cutoff, since RG running shuffles contributions between different orders. 

But we cannot simply propose an arbitrary correlation structure between $E/N(n)$
and $E/A(n')$ at any given $n$ and $n'$ because the covariance must remain positive semi-definite. As described in Secs.~\ref{sec:the_model_nuclear_matter}
and~\ref{sec:results_for_pnm_and_snm}, the truncation errors of each EOS are treated as an output from a GP\@.
Correlations between discrete outputs of various GPs have been addressed in the literature by
\emph{multitask} GPs, also known as co-kriging~\cite{alvarez2012kernels, Melkumyan:2011multikernel}.
% An area of active research in
% multitask GPs regards finding the \emph{cross-correlation} kernels between
% outputs in such a way that the total kernel remains valid.
% The problem of choosing the \emph{cross-correlation} between generic outputs
% in a way that is sure to yield positive semidefinite kernels is an area of
% open research.

Since we use the RBF kernel~\eqref{eq:rbf_kernel_nuclear_matter} for each individual observable, the correlations between $E/N(n)$ and $E/A(n')$ can be directly modeled. Furthermore, if
the correlation lengths in PNM and SNM are permitted to differ between the observables, we can
model this dependence using an RBF cross covariance kernel with a length scale
and correlation coefficient $\rho$ that are determined by each individual
correlation length~\cite{Melkumyan:2011multikernel}; see Eq.~\eqref{eq:cross_correlation_two_different_rbfs}.
We choose to employ this model for the $\Lambda = 500\MeV$ interactions, where the predicted $\rho$ accurately matches the empirical correlation.
Appendix~\ref{sec:multitask_gps} describes this multitask model and also proposes an alternative where $\rho$ can be tuned to the data. This modified model is applied to the
$\Lambda = 450\MeV$ interactions, since it can better fit the empirical correlation in that case.

We make use of the scale independence of the GP input in $\kf$, as discussed
in Sec.~\ref{sec:extracting_coefficients_pnm_snm}, so that we can define a space
where $E/N(n)$ and $E/A(n)$ at the same $\kf$ are also at the same density.
Concretely, we simply use $\kf^\text{PNM}$ as the input space for
$E/A(n)$. This only involves scaling the $c_n(\kf)$ length scale to leave the RBF
kernel invariant, but the $Q$ and $\genobsref$ defined in
Sec.~\ref{sec:extracting_coefficients_pnm_snm} for $E/A(n)$ still use $\kf^\text{SNM}$, and hence remain the same.
This is a necessary step in correlating $E/N(n)$ with $E/A(n')$ such that points
with $n=n'$ are the most highly correlated, while the correlations drop as
distance $\abs{n-n'}$ grows.

%%%%%%%%%%%%%%%%%%%%%%%%%%%%%%%%%%%%%%%%%%%%%%%%%%%%%%%%%%%%%%%
\begin{figure}[tb]
    \centering
    \includegraphics{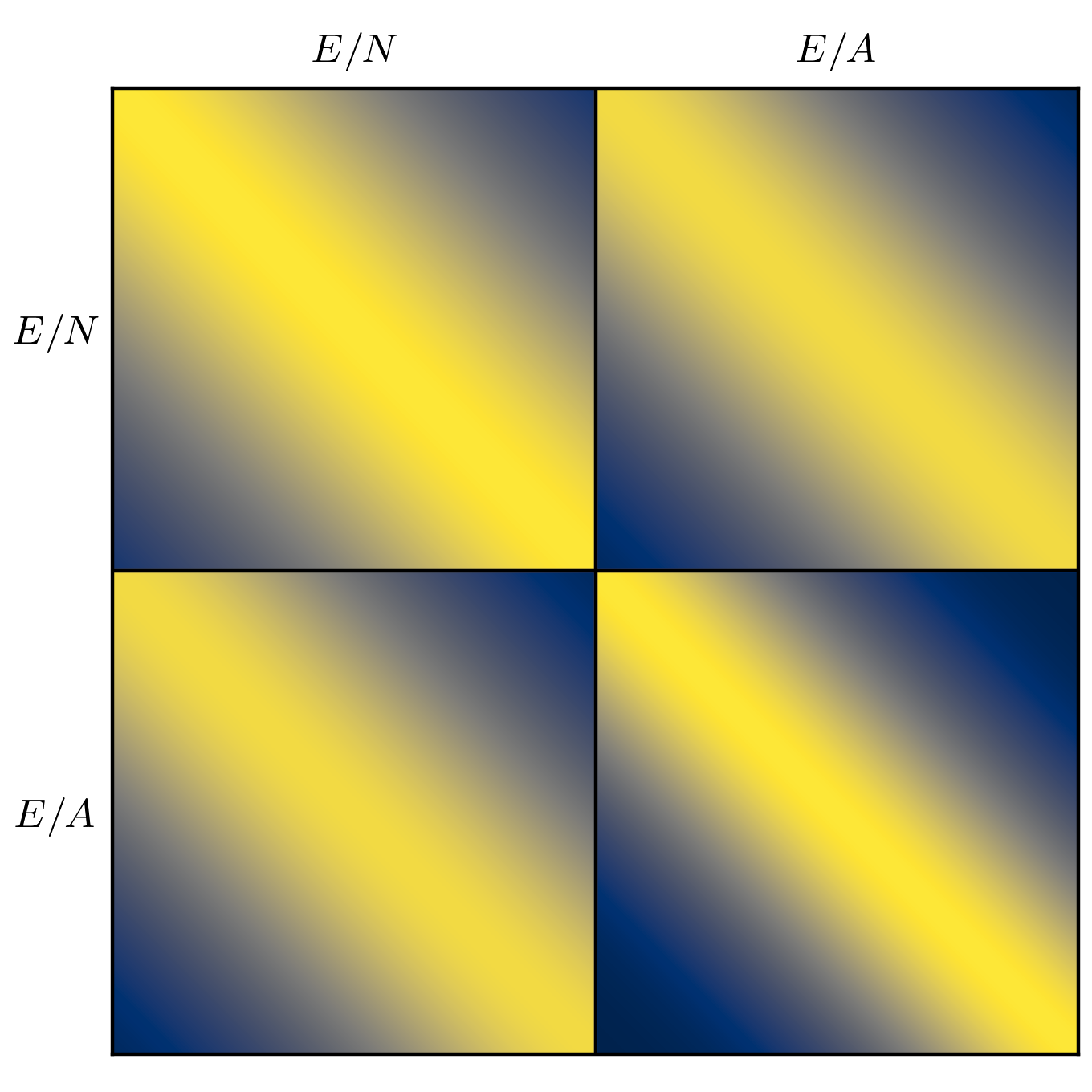}
    \caption{
    Total correlation matrix of $E/N(n)$ and $E/A(n)$ assuming a multitask GP
    model that was trained to order-by-order results of the $\Lambda=500\MeV$ interactions.
    Each submatrix uses the same grid spaced linearly in $\kf$.
    The diagonal blocks show the autocorrelation, and the off-diagonal block is known as the cross correlation. All use RBF kernels [see Eqs.~\eqref{eq:rbf_kernel_nuclear_matter},~\eqref{eq:joint_gaussian_nuclear_matter}, and~\eqref{eq:cross_correlation_two_different_rbfs}].
    The length scale of the $E/A(n)$ blocks has been transformed
    to $\kf^\text{PNM}$ as discussed in the text.
    Hence, points of equal density between $E/N(n)$
    and $E/A(n)$ lie on the diagonal band of the off-diagonal blocks, making them the most highly correlated, but the $E/A(n)$ autocorrelation is unchanged.
    The length scales were determined by fitting to the $E/N(n)$ and
    $E/A(n)$ coefficients independently as shown in Sec.~\ref{sec:extracting_coefficients_pnm_snm}.
    The correlation $\rho = 0.95$ is a prediction (which agrees with the empirical correlation) given these length scales.
    } \label{fig:correlation_matrix_EN_and_EA_lambda-500}
\end{figure}
%%%%%%%%%%%%%%%%%%%%%%%%%%%%%%%%%%%%%%%%%%%%%%%%%%%%%%%%%%%%%%%

We take the point estimates of $\cbar_i$ and $\ell_i$ in PNM and SNM, as determined in Sec.~\ref{sec:extracting_coefficients_pnm_snm}, and
use them to construct the total covariance matrix within and between PNM and
SNM\@.
Remarkably, the predicted correlation coefficient for the $\Lambda = 500\MeV$ interactions is
$\rho=0.949$, which agrees with the empirical value to two
digits.\footnote{This should likely not be taken as anything other than
coincidence, as $\rho$ was not tuned to the empirical correlation. Rather,
$\rho$ is a prediction based on the individual length scales of each marginal
process. Tuning of $\rho$ can be done if the length scales of PNM and SNM are
the same; see Appendix~\ref{sec:GPderivatives}.} We can then emulate $c_n(\kf)$
from our correlated PNM--SNM system as shown in
Fig.~\subref*{fig:empirical_correlation_ellipses_EN_vs_EA_sampled_coefficients_lambda-500}.
The emulated coefficients appear quite similar to the actual $c_n(\kf)$, in that we could not
tell them apart if they were not already distinguished. This gives us confidence
in our approach.
The total correlation matrix describing correlations within and
between PNM and SNM is given in
Fig.~\ref{fig:correlation_matrix_EN_and_EA_lambda-500}.

The correlation between PNM and SNM is somewhat weaker (but still strong) for the  $\Lambda = 450\MeV$ interactions, see Fig.~\subref*{fig:empirical_correlation_ellipses_EN_vs_EA_true_coefficients_lambda-450}.
% The coefficients and samples from the multitask GP for these potentials are shown in Figs.~\subref*{fig:empirical_correlation_ellipses_EN_vs_EA_true_coefficients_lambda-450} and~\subref*{fig:empirical_correlation_ellipses_EN_vs_EA_sampled_coefficients_lambda-450}.
This cross correlation is not well predicted by Eq.~\eqref{eq:cross_correlation_two_different_rbfs}, so we present a different approach to modeling $\rho$ in this case.
One can tune $\rho$, which we take to be equal to the empirical correlation $\rho = 0.75$, so long as we make the reasonable approximation $\ell_{\text{PNM}} \approx \ell_{\text{SNM}}$  (see Appendix~\ref{sec:multitask_gps} for details).
The samples from this tuned multitask GP look similar to the actual $c_n(\kf)$, see Fig.~\subref*{fig:empirical_correlation_ellipses_EN_vs_EA_sampled_coefficients_lambda-450}.
This shows that our multitask kernels are a flexible way to introduce correlations between observables.
% ; the best choice, however, may depend on the scenario of the modeler.

%%%%%%%%%%%%%%%%%%%%%%%%%%%%%%%%%%%%%%%%%%%%%%%%%%%%%%%%%%%%%%%
\begin{figure}[tb]
    \centering
    \includegraphics{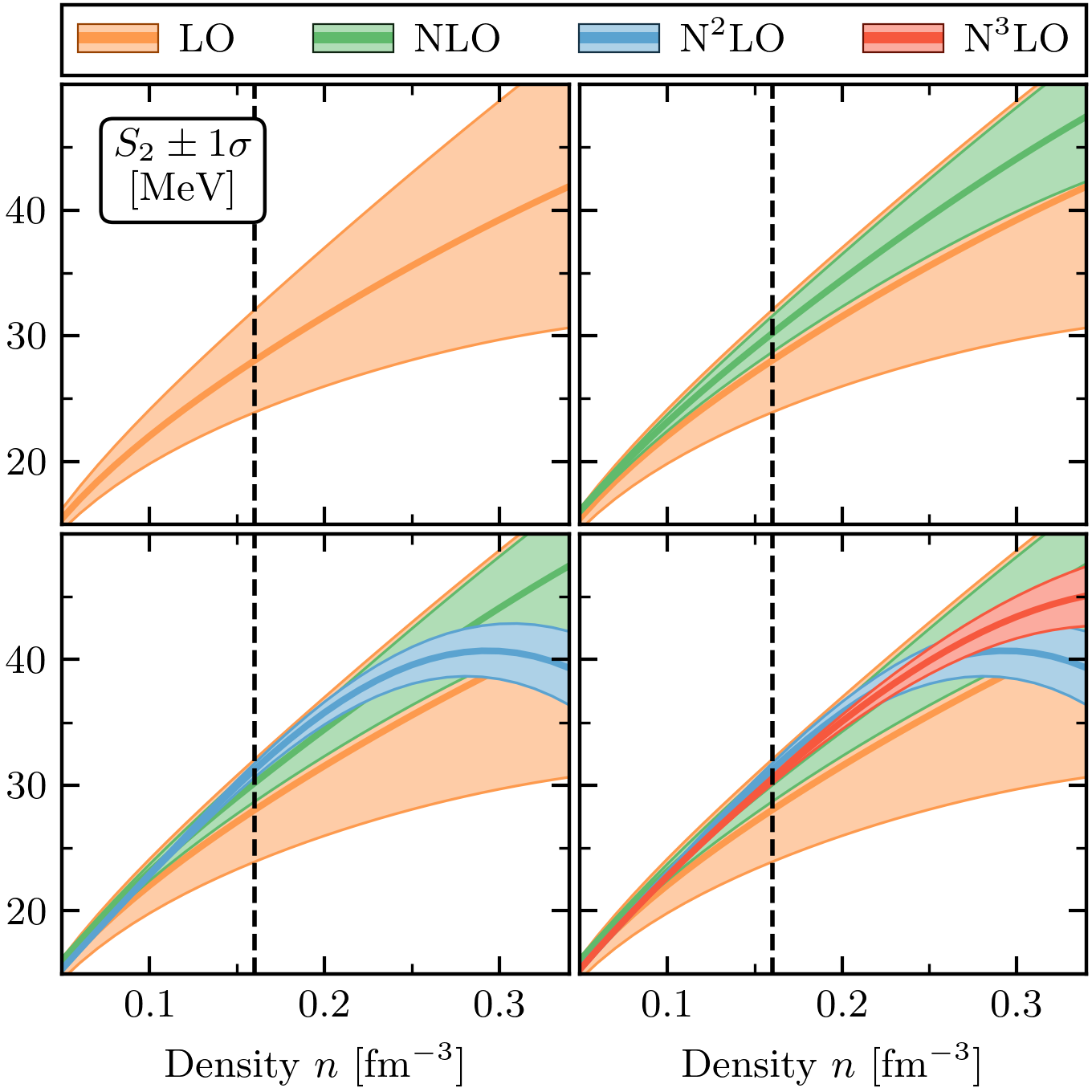}
    \caption{Similar to Fig.~\ref{fig:neutron_matter_predictions_lambda-500}, but these are the
    order-by-order predictions of the symmetry energy $S_2(n)$ for the $\Lambda = 500\MeV$ interactions. 
    } \label{fig:S2_order_by_order_predictions_lambda-500}
\end{figure}
%%%%%%%%%%%%%%%%%%%%%%%%%%%%%%%%%%%%%%%%%%%%%%%%%%%%%%%%%%%%%%%

With the full correlation structure of PNM and SNM in hand, we can compute
$S_2(n)$ with truncation errors. The results for the $\Lambda = 500\MeV$ interactions are shown in
Fig.~\ref{fig:S2_order_by_order_predictions_lambda-500}. If correlations between
PNM and SNM had instead been neglected, \ie, truncation errors simply added in quadrature, the size of the truncation uncertainty
would be $>5$ times larger ($\approx 2$ times larger for $\Lambda=450\MeV$). This factor is particularly
important for constraining $S_2(n)$ and its (rescaled) density dependence,
\begin{equation}
    L(n) = 3n\,\dv{n}\,S_2(n)\,,
\end{equation}
as discussed in the companion paper~\cite{Drischler:2020preparationAstro}. Computing $L(n)$ with
full uncertainty propagation requires a discussion of how to take derivatives of
GPs, the topic of the next subsection and Appendix~\ref{sec:GPderivatives}.

%%%%%%%%%%%%%%%%%%%%%%%%%%%%%%%%%%%%%%%%%%%%%%%%%%%%%%%%%%%%%%%
\subsection{Derivatives and related observables}
\label{sec:derivative_quantities_snm}

%If a function is completely known, then its derivative, in principle, has no
%uncertainty associated with it. So it makes sense that if a function is treated
%as a random variable, it would be highly correlated with its derivative.

An important feature of GP interpolants is that they allow straightforward
computation of derivatives that are smooth and have theoretical uncertainties that are fully propagated from their anti-derivatives. Finite differencing and parametric fits do not achieve this. In the following, we discuss how to evaluate first- and second-order derivatives in SNM using GPs  and refer to the companion publication~\cite{Drischler:2020preparationAstro} for a detailed discussion of PNM\@.  Appendix~\ref{sec:GPderivatives}
gives details on how to compute the GP for derivative quantities.

Specifically, we consider here the pressure,
\begin{align}
    P(n) = n^2 \, \dv{n} \frac{E}{A}(n) \,,
\end{align}
and the incompressibility,
\begin{equation}
    K = 9 n_0^2 \, \dv[2]{n} \frac{E}{A}(n) \bigg|_{n=n_0} \,.
\end{equation}
Notice that $K$ is evaluated at the predicted (not at the empirical)
saturation density, $n_0$. We thus need to account for the uncertainties in
$n_0$ in addition to the ones in $E/A(n)$ and its derivatives. The \chiEFT\
truncation errors in $E/A(n)$ lead to a distribution for $n_0$.
In Sec.~\ref{sec:uncertainties_for_pnm_snm} we derived the posterior $\pr(n_0 \given \data)$, which is approximately Gaussian.
Hence, we can estimate $K$ and
its full uncertainty at any given $n_0$ by computing $\pr(K \given \data, n_0)$,
and subsequently sum over all plausible $n_0$ values via
\begin{equation} \label{eq:incomp_sat_marg}
    \pr(K \given \data) = \int \pr(K \given \data, n_0) \pr(n_0 \given \data) \dd{n_0} \,.
\end{equation}
We perform a similar summation in Ref.~\cite{Drischler:2020preparationAstro} to
compute the posterior for the symmetry energy [$S_v = S_2(n_0)$] and its slope parameter
[$L = L(n_0)$] evaluated at the predicted range for $n_0$.

GPs have advantages over parametrizations of the EOS (\eg, series expansions) when derivatives up to high densities are desired. In particular, second-order
(and higher) derivatives tend to magnify numerical instabilities in these
parametrizations. We have verified this instability by performing a global
Bayesian fit of a power series in $\kf$~\cite{Drischler:2015eba},
\begin{equation} \label{eq:model_EA}
  \frac{E}{A} \left( n;\, \left \lbrace d_\nu \right \rbrace_{\nu = 2}^M \right)
  = \sum_{\nu= 2}^M d_\nu \left(\frac{n}{n_0}\right)^\frac{\nu}{3} \,,
\end{equation}
to $E/A(n)$, with $M \geq 2$ assigned by maximizing the Bayesian evidence in an attempt to
prevent overfitting.
Despite using the evidence as a safeguard, we obtained fit coefficients $d_\nu$ that are unnaturally large in magnitude and that alternated in sign---a classic symptom of over-fitting.
(Similar symptoms are seen in the fit coefficients presented in
Table~II of Ref.~\cite{Drischler:2015eba}.) Consequently, our parametric fits of
Eq.~\eqref{eq:model_EA} were unable to predict, \eg, $K$ reliably.
In contrast, GPs are like splines~\cite{Mackay:1998introduction} in being more sensitive to local information than a global parametric fit like Eq.~\eqref{eq:model_EA}.
Our GP model therefore yields more reliable derivatives. Its locality does mean, though, that there are possible edge effects when computing derivatives---especially higher derivatives---near the edge of the region where there is data.

Honest UQ for these observables is only possible with a correlated model of
uncertainty. If correlations are neglected, as in the ``standard EFT'' error
prescription~\cite{Epelbaum:2014sza, Epelbaum:2014efa}, then derivatives can be
arbitrarily uncertain and thus unrealistic. 

%%%%%%%%%%%%%%%%%%%%%%%%%%%%%%%%%%%%%%%%%%%%%%%%%%%%%%%%%%%%%%%
\begin{figure}[tb]
    \centering
    \includegraphics{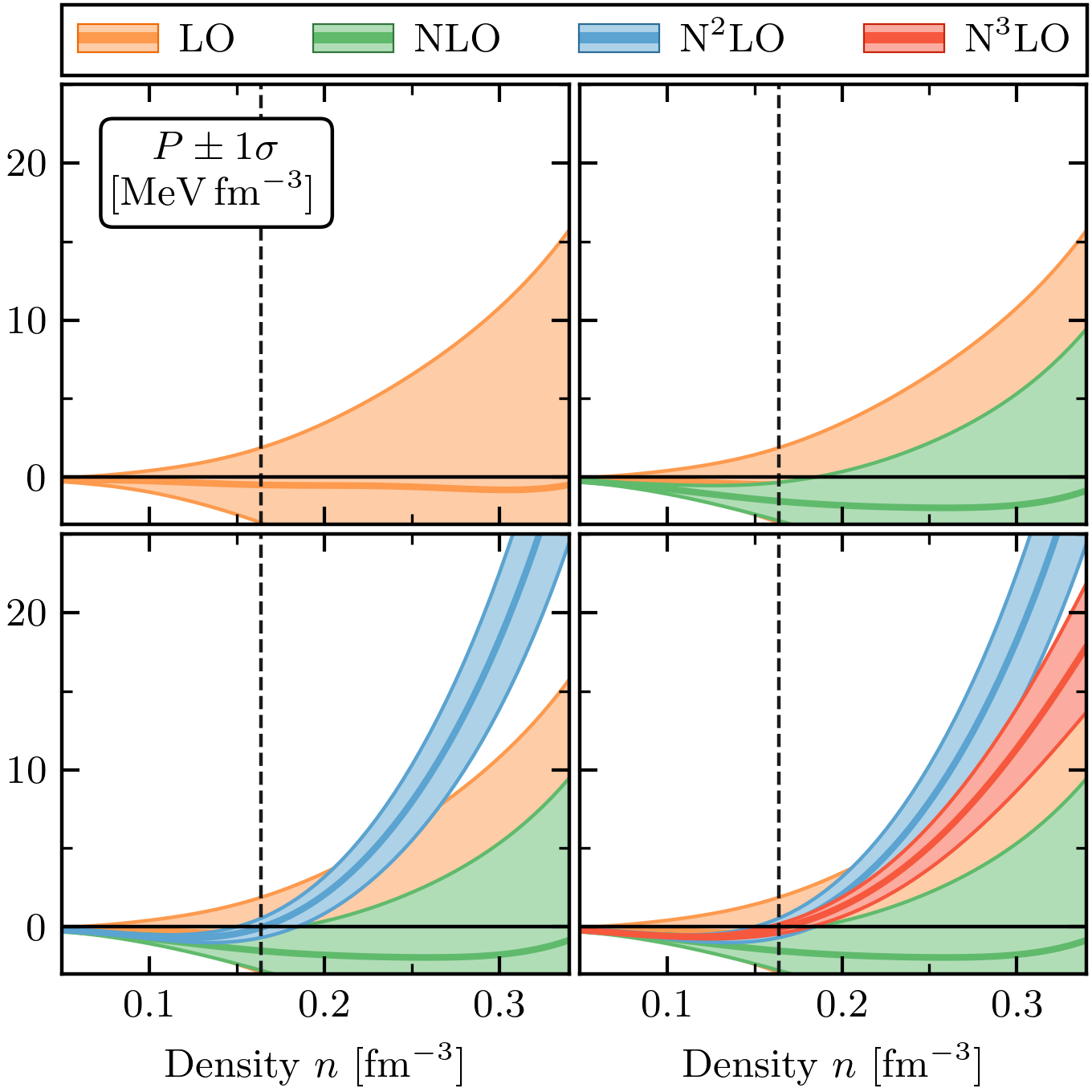}
    \caption{
    Order-by-order predictions of the pressure $P(n)$ of SNM, including
    differentiation and truncation uncertainty, for the $\Lambda = 500\MeV$ interactions. See the main text for details.
    } \label{fig:pressure_lambda-500}
\end{figure}
%%%%%%%%%%%%%%%%%%%%%%%%%%%%%%%%%%%%%%%%%%%%%%%%%%%%%%%%%%%%%%%

Figure~\ref{fig:pressure_lambda-500} shows the pressure of SNM with 68\%
credible intervals. Although LO and NLO have negative mean values across all
shown densities, nuclear saturation (\ie, $P = 0$) near $n_0$ could be achieved
within the large uncertainties, even for these NN-only interactions. Indeed, the wide range of densities at which the pressure \emph{could} cross zero at LO and NLO suggests that the nuclear saturation point is somewhat fine tuned. Perhaps this is not surprising: in EFT, if an observable that is not zero at LO has a zero crossing, the position of that crossing is, by definition, sensitive to higher-order corrections because lower orders must cancel there.
From a \chiEFT\ consistency point of view it is reassuring that \NkLO{2}
and \NkLO{3} are consistently within the bands of the previous orders at
$\lesssim n_0$. They do, however, begin to diverge at higher densities. As with $E/N(n)$
and $E/A(n)$ we stress that the consistency of the uncertainty bands
is difficult to gauge due to the long correlation length of the truncation error, and
hence small effective sample size of data.

%%%%%%%%%%%%%%%%%%%%%%%%%%%%%%%%%%%%%%%%%%%%%%%%%%%%%%%%%%%%%%%
\begin{figure}[tb]
    \centering
    \includegraphics{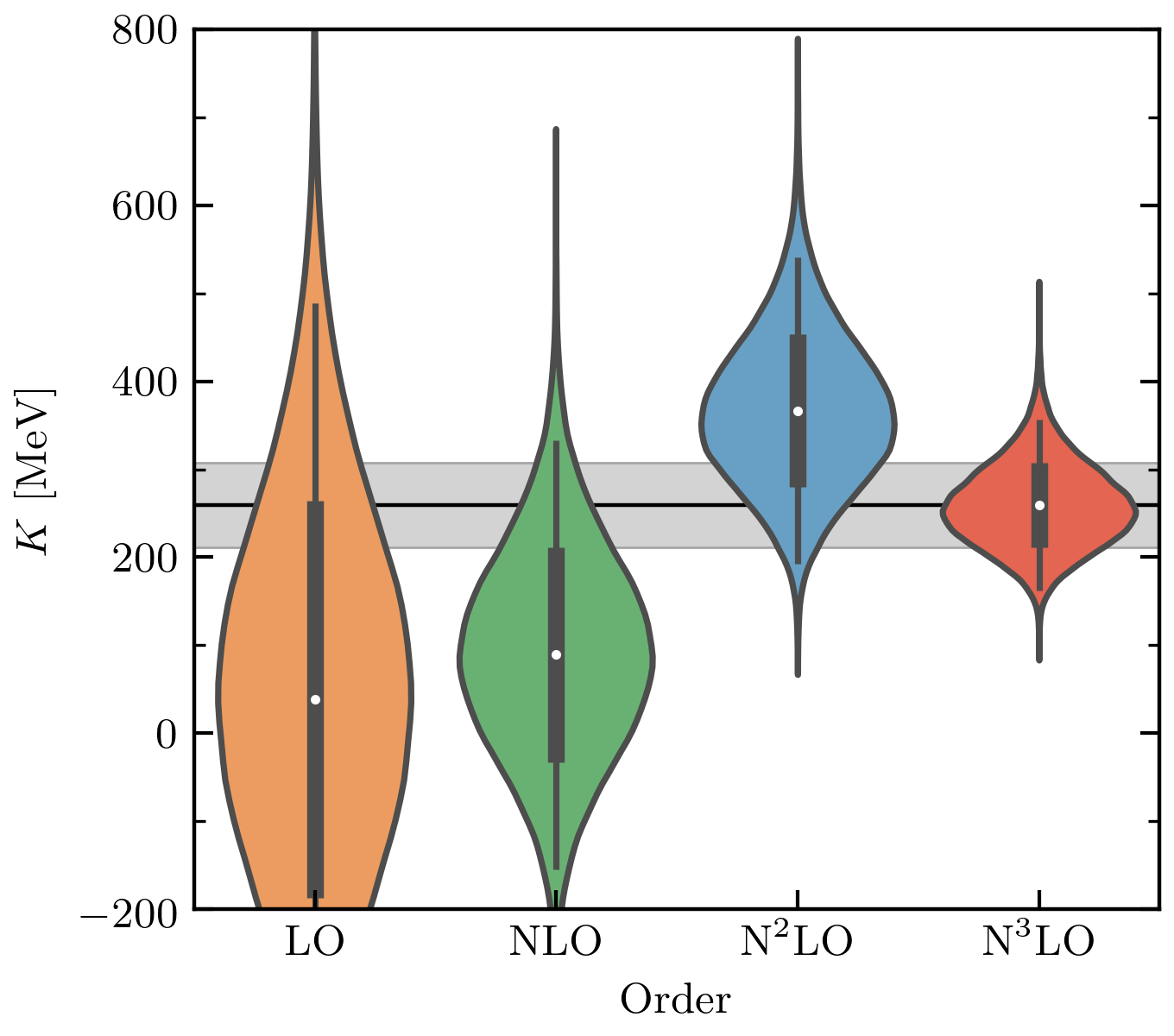}
    \caption{
    Violin plots of the incompressibility $K$ of SNM, shown order-by-order for
    the $\Lambda = 500 \MeV$ interactions.
    Curves show the entire smoothed posterior (and its reflection). Each
    posterior includes differentiation and truncation uncertainty, and are
    marginalized over all plausible saturation densities $n_0 = 0.17 \pm 0.01\fmiq$; see the main text. Dots and bars
    indicate the mean value, along with the $1\sigma$ and $2\sigma$
    uncertainties. The black line and gray band extends the \NkLO{3} mean and
    $1\sigma$ uncertainty to more easily compare \chiEFT\ orders.
    } \label{fig:K_sat_marginalized-500}
\end{figure}
%%%%%%%%%%%%%%%%%%%%%%%%%%%%%%%%%%%%%%%%%%%%%%%%%%%%%%%%%%%%%%%

Figure~\ref{fig:K_sat_marginalized-500} shows our order-by-order results for $K$
based on the $\Lambda = 500 \MeV$ interactions.
These predictions use our best estimate for $n_0$ in
the integral~\eqref{eq:incomp_sat_marg}: the Gaussian posterior $n_0 = 0.17 \pm
0.01 \fmiq$ at \NkLO{3} determined in Sec.~\ref{sec:uncertainties_for_pnm_snm}.
At LO and NLO, where the empirical saturation point is typically not well
reproduced, this choice leads to wide-spread distributions whose $1\sigma$
regions reach $K < 0$, even though nuclear saturation requires $K > 0$. In
general, the uncertainty bands are consistent across \chiEFT\ orders and settle at
$260 \pm 54\MeV$ ($292 \pm 54\MeV$) for the \NkLO{3}
interaction with $\Lambda = 500 \MeV$ ($450 \MeV$).

%%%%%%%%%%%%%%%%%%%%%%%%%%%%%%%%%%%%%%%%%%%%%%%%%%%%%%%%%%%%%%%
\section{Summary and Outlook} \label{sec:summary_matter_uq}

\begin{table}[tbp]
    \centering
    \caption{Our $1\sigma$-level constraints on the density (in $\fmiq$), energy per particle, and incompressibility of SNM at saturation as well as the symmetry energy and its derivative (all in $\MeV$). These are given for two different \NNNLO \chiEFT\ Hamiltonians. We emphasize that $n_0$ and $\frac{E}{A}(n_0)$ as well as $S_2(n_0)$ and $L(n_0)$ are correlated (\ie, the covariance matrix is not diagonal). We also remind the reader that the uncertainties quoted here are solely due to truncation of the \chiEFT\ expansion: they do not account for the uncertainty in the \chiEFT\ LECs. A full assessment of the uncertainties in \chiEFT's predictions for nuclear matter will require more work on \chiEFT\ NN and 3N interactions up to \NNNLO, as discussed in the main text.}
    \label{tab:summary_table}
        \begin{ruledtabular}
    \begin{tabular}{Scd{2.2}d{2.2}Sl}
         & \multicolumn{1}{Sc}{$\Lambda = 450 \MeV$} &	\multicolumn{1}{Sc}{$\Lambda =500 \MeV$} & See also\\
        \hline
        $n_0$ & 0.173(14) & 0.170(16) & Fig.~\ref{fig:coester_ellipses_both_lambda}\footnotemark[1]\\
        $\frac{E}{A}(n_0)$ & -14.9(1.1) & -14.3(1.0)   & Fig.~\ref{fig:coester_ellipses_both_lambda}\footnotemark[1]\\
        $K$ & 292.0(54.0) & 260.0(54.0) & Fig.~\ref{fig:K_sat_marginalized-500} \\
        $S_2(n_0)$ & 33.5(1.3) & 31.7(1.1)  & Fig.~\ref{fig:S2_order_by_order_predictions_lambda-500}\footnotemark[2]\\
        $L(n_0)$ & 67.8(4.0) & 59.8(4.1)  & Ref.~\cite{Drischler:2020preparationAstro}\footnotemark[2]\\
    \end{tabular}
     \end{ruledtabular}
     \footnotetext[1]{Mean and covariance matrix are given in Eqs.~\eqref{eq:n0_EA_n0_mean_cov_500 MeV} and~\eqref{eq:n0_EA_n0_mean_cov_450 MeV}.}
      \footnotetext[2]{Figure~2 of Ref.~\cite{Drischler:2020preparationAstro} shows the constraints in the $S_2(n_0)$--$L(n_0)$ plane for the $\Lambda = 500\MeV$ interaction; for the $\Lambda = 450\MeV$ interaction see the Supplemental Material of Ref.~\cite{Drischler:2020preparationAstro}.}
\end{table}

Order-by-order predictions of a well-behaved EFT should converge regularly
towards the all-orders value. We have formalized this idea into a falsifiable EFT
convergence model using Bayesian statistics and presented the first application
to infinite matter up to $2n_0$. The EOS is based on
order-by-order calculations in MBPT with NN and 3N interactions up to
\NNNLO~\cite{Drischler:2017wtt, Leonhardt:2019fua, Drischler:2020preparationAstro}. While this work focuses on
key properties of SNM, our companion
publication~\cite{Drischler:2020preparationAstro} is dedicated to PNM and its
astrophysical applications. Together, they set a new
standard~\cite{BUQEYEgithub} for UQ in infinite matter calculations.

Section~\ref{sec:results_for_pnm_and_snm} provides the first truncation error
bands for infinite matter that account for correlations in density. Our
findings indicate that the truncation errors are highly correlated, rendering
the qualitative judgment of credible intervals more difficult. A full
understanding, therefore, requires the diagnostic tools discussed in
Ref.~\cite{Melendez:2019izc}. Specifically, we have verified the importance of
truncation error correlations between
\begin{enumerate}
    \item[(i)] the EOS at different densities,

    \item[(ii)] different observables, such as $E/N(n)$ and $E/A(n)$ to determine $S_2(n)$, and

    \item[(iii)] the EOS and its derivatives.
\end{enumerate}

Our truncation error model also allows us to do the first efficient and accurate
propagation of EOS theoretical uncertainties (including \chiEFT\ truncation errors) to derived quantities. We have pointed out the advantages of our
approach over global parametrizations of the EOS such as fitted series expansions: even
maximizing the Bayesian evidence could not prevent overfitting of the series
expansion~\eqref{eq:model_EA}. Such numerical instabilities are
magnified when computing derivatives: especially second and higher derivatives.
Our nonparametric GP approach does not suffer from
these instabilities.

We then studied nuclear saturation properties, including the incompressibility
and pressure of SNM, and symmetry energy, with theoretical uncertainties fully
quantified. We have also provided the first probability distributions for the
predicted saturation region by sampling from our correlated error model. Table~\ref{tab:summary_table} summarizes our constraints for the two momentum cutoffs at the $1\sigma$ level. The results all agree at the level of $2\sigma$, and for the saturation properties the agreement is better than that. This indicates only a mild cutoff dependence for these observables. Our
findings indicate that taking into account the correlations in density is
necessary to quantify these properties. The methods
developed here are publicly available as annotated Jupyter notebooks.~\cite{BUQEYEgithub}.

In Sec.~\ref{sec:introduction_matter_uq} we raised several points about the 
in-medium convergence of \chiEFT\@. Here are some of our conclusions:
\begin{itemize}
    \item[(1)] The convergence plots for PNM and SNM show regular convergence with
    increasing order, as seen for NN observables.

    \item[(2)] The statistical model formulated with $Q(\kf) = \kf/\Lambda_b$ and
    $\genobsref(\kf) = 16\MeV\times(\kf / \kfzero)^2$ provides a reasonable
    characterization of the convergence pattern in infinite matter. This yields
    consistent uncertainty bands for both $E/N(n)$ and $E/A(n)$, within
    statistical fluctuations.
    % The coefficient functions $c_n(\kf)$ and
    % the model checking diagnostics support the BUQEYE statistical model

    \item[(3)] With our model checking diagnostics given in
    Appendix~\ref{sec:model_checking_infinite_matter}, we have found evidence
    that 3N interactions show somewhat different characteristics (\eg, different
    length scales), and the \NNLO coefficient $c_3(\kf)$ may be an outlier due
    to the first nonvanishing 3N contributions at this order. This points to the
    possibility that one should not use $c_3(\kf)$ to infer truncation error
    properties, though more work along these lines with different interactions is needed.
    % Contributions from 3N interactions exhibit the same natural size as NN
    % terms, but have a noticeably different correlation length in \kf, which
    % might be interpreted as a new scale and traced back to Pauli blocking.

    \item[(4)] The posterior for $\Lambda_{b}$ calculated using $Q(\kf) = \kf/\Lambda_b$
    is consistent with determinations from NN observables.
\end{itemize}

Calculations of finite nuclei with \chiEFT\ potentials have often been found to predict too-small radii~\cite{Binder:2013xaa, Lapoux:2016exf, Epelbaum:2019kcf}.
Since the link between infinite matter, heavy to medium-mass nuclei, and few-body systems has yet to be fully understood~\cite{Hoppe:2019uyw,Huth19int},
examining the statistical correlation of radius systematics with the predicted saturation density in SNM could clarify the origin of these deficiencies.
The methods developed here could facilitate the statistically consistent inclusion of empirical saturation properties in fits of \chiEFT\ potentials (cf. Refs.~\cite{Drischler:2017wtt, Ekstrom:2015rta}).
  
Future studies will extend our analysis to asymmetric matter with
arbitrary proton fractions (\eg, neutron-star matter) and finite temperature.
In this case, the discrete correlations between $E/N(n)$ and $E/A(n)$ found here could be naturally handled by $c_n(\kf)$ that are correlated in both density and proton fraction.
To elucidate the full dependence of the EOS on the nuclear interactions, however, improved order-by-order NN and 3N interactions need to be developed up to
\NNNLO~\cite{Hoppe:2019uyw, Huth19int, Epelbaum:2019kcf}. Our physically motivated GP model can also be applied to
efficiently compute nuclear saturation properties using Bayesian optimization
frameworks~\cite{frazier2018tutorial, Picheny:2013BayesianOpt} and
to account for uncertainties in the fits of the LECs using Monte Carlo
sampling~\cite{Carlsson:2015vda, Wesolowski:2015fqa, Wesolowski:2018lzj}. This last task is particularly important. It will presumably expand the error bars presented here, which only account for truncation error, and not for uncertainties in the LECs.

% Challenging nuclear theory with observation and experiment is one of the primary
% goals in the era of multi-messenger astronomy. The presented methods will
% be key for achieving this goal in terms of statistically meaningful comparisons
% of microscopic EOS calculations (based on different EFT prescriptions) with
% empirical constraints. This will provide important insights into the structure
% of neutron stars and supernova explosions. The recent joint mass-radius
% measurement of a millisecond pulsar by NASA's Neutron star Interior Composition
% ExploreR (NICER)~\cite{Bogd19NICER2, Riley19NICER, Mill19NICER, Raai19NICER},
% for instance, is an exciting prospect in this direction.

%%%%%%%%%%%%%%%%%%%%%%%%%%%%%%%%%%%%%%%%%%%%%%%%%%%%%%%%%%%%%%%
% \clearpage

\begin{acknowledgments}
We thank M.~Grosskopf and S.~Reddy for fruitful discussions. We are also grateful to
the organizers of ``Bayesian Inference in Subatomic Physics---A Marcus Wallenberg Symposium'' at Chalmers University of Technology, Gothenburg, for creating a stimulating environment to learn and discuss the use of statistical methods in nuclear physics.
C.D. acknowledges support by
the Alexander von Humboldt Foundation through a Feodor-Lynen Fellowship and the
U.S. Department of Energy, the Office of Science, the Office of Nuclear Physics,
and SciDAC under awards DE-SC00046548 and DE-AC02-05CH11231. The work of R.J.F. and
J.A.M. was supported in part by the National Science Foundation under Grants
No.~PHY--1614460 and No. PHY--1913069, and the NUCLEI SciDAC Collaboration under U.S.
Department of Energy MSU subcontract RC107839-OSU\@. The work of D.R.P. was
supported by the U.S. Department of Energy under contract DE-FG02-93ER-40756 and
by the National Science Foundation under PHY-1630782, N3AS FRHTP\@. C.D. thanks
the Physics Departments of The Ohio State University and Ohio University for
their warm hospitality during extended stays in the BUQEYE state.
\end{acknowledgments}

%%%%%%%%%%%%%%%%%%%%%%%%%%%%%%%%%%%%%%%%%%%%%%%%%%%%%%%%%%%%%%%

%%%%%%%%%%%%%%%%%%%%%%%%%%%%%%%%%%%%%%%%%%%%%%%%%%%%%%%%%%%%%%%
\appendix

\input{appendix_model_checking}
\input{appendix_tabulated_EOS}
\input{appendix_details_GPs}

%%%%%%%%%%%%%%%%%%%%%%%%%%%%%%%%%%%%%%%%%%%%%%%%%%%%%%%%%%%%%%%
%\clearpage
\bibliographystyle{apsrev4-1}
\bibliography{bayesian_refs,additional,misc}

\end{document}

%% file: buqeye_macros.tex
\makeatletter
\newcommand\newsubcommand[3]{\newcommand#1{#2\sc@sub{#3}}}
\def\sc@sub#1{\def\sc@thesub{#1}\@ifnextchar_{\sc@mergesubs}{_{\sc@thesub}}}
\def\sc@mergesubs_#1{_{\sc@thesub#1}}

\newcommand\newsupcommand[3]{\newcommand#1{#2\sc@sup{#3}}}
\def\sc@sup#1{\def\sc@thesup{#1}\@ifnextchar^{\sc@mergesups}{^{\sc@thesup}}}
\def\sc@mergesups^#1{^{\sc@thesup#1}}
\makeatother

\DeclareMathAlphabet{\mathbcal}{OMS}{cmsy}{b}{n}

\newcommand{\kernel}{\kappa}

\newcommand{\cbar}{\bar c}
\newcommand{\sdth}{\cbar}

\newcommand{\DVAR}[1]{\inputvec{D}_{\textup{#1}}}

\newcommand{\iid}{\text{i.i.d.}}

\newcommand{\discrcorr}[1]{R_{\delta #1}}

\newcommand{\fmi}{\, \text{fm}^{-1}}
\newcommand{\fmiq}{\, \text{fm}^{-3}}
\newcommand{\keV}{\, \text{keV}}
\newcommand{\MeV}{\, \text{MeV}}

\newcommand{\NNLO}{\ensuremath{{\rm N}{}^2{\rm LO}}\xspace}
\newcommand{\NNNLO}{\ensuremath{{\rm N}{}^3{\rm LO}}\xspace}

\newcommand{\NkLO}[1]{\ensuremath{\mathrm{N}^{#1}\mathrm{LO}}\xspace}

\newcommand{\data}{\mathcal{D}}

\DeclareMathOperator{\GP}{\mathcal{GP}}

\newcommand{\npr}{\ensuremath{np}}

\newcommand{\ordervec}{\vec}
\newcommand{\inputdimvec}{}
\newcommand{\inputvec}{\mathbf}

\newsubcommand{\ckvec}{\ordervec{c}}{k}

\newsubcommand{\bkvec}{\ordervec{b}}{k}
\newcommand{\kinparvec}{\inputdimvec{x}}

\newsubcommand{\ckvecset}{\ordervec{\inputvec{c}}}{k}

\newsubcommand{\ckvecapprox}{\mathbf{c}'}{k}
\newsubcommand{\ckvecapproxset}{\mathbf{C}'}{k}

\newsubcommand{\bkvecapprox}{\mathbf{b}'}{k}
\newsubcommand{\bkvecset}{\mathbf{B}}{k}
\newsubcommand{\bkvecapproxset}{\mathbf{B}'}{k}

\newcommand{\genobs}{y}

\newsubcommand{\genobsvec}{\ordervec{\genobs}}{k}
\newsubcommand{\genobsvecset}{\ordervec{\inputvec{\genobs}}}{k}

\newsubcommand{\akvec}{\mathbf{a}}{k}

\newsubcommand{\akvecapprox}{\mathbf{a}'}{k}
\newsubcommand{\akvecset}{\mathbf{A}}{k}
\newsubcommand{\akvecapproxset}{\mathbf{A}'}{k}

{}  % Remove the definition from the Physics package

\DeclareMathOperator{\pr}{pr} % Good, but want to handle sizing and | spacing?
\newcommand{\given}{\,|\,}  % Use for | in \pr

\newcommand{\normal}{\mathcal{N}}

\newcommand{\trans}{\intercal}

\newcommand{\chiEFT}{$\chi$EFT}

\newcommand{\genobsref}{\ensuremath{y_{\mathrm{ref}}}}

\newcommand{\kf}{\ensuremath{k_{\scriptscriptstyle\textrm{F}}}}
\newcommand{\kfzero}{\ensuremath{k_{\scriptscriptstyle\textrm{F,0}}}}

\def\diffd{\mathrm{d}}  % Upright differentials
\DeclareDocumentCommand\differential{ o g d() }{ % Differential 'd'
    \IfNoValueTF{#2}{
        \IfNoValueTF{#3}
            {\diffd\IfNoValueTF{#1}{}{^{#1}}}
            {\mathinner{\diffd\IfNoValueTF{#1}{}{^{#1}}\argopen(#3\argclose)}}
        }
        {\mathinner{\diffd\IfNoValueTF{#1}{}{^{#1}}#2} \IfNoValueTF{#3}{}{(#3)}}
    }
\DeclareDocumentCommand\dd{}{\differential} % Shorthand for \differential

\newcommand{\pathd}{\mathcal{D}}  % differential symbol for path integrals

\DeclareDocumentCommand\pathdifferential{ o g d() }{ % Path 'D'
    \IfNoValueTF{#2}{
        \IfNoValueTF{#3}
            {\pathd\IfNoValueTF{#1}{}{^{#1}}}
            {\mathinner{\pathd\IfNoValueTF{#1}{}{^{#1}}\argopen(#3\argclose)}}
        }
        {\mathinner{\pathd\IfNoValueTF{#1}{}{^{#1}}#2} \IfNoValueTF{#3}{}{(#3)}}
    }

%% file: appendix_model_checking.tex
%%%%%%%%%%%%%%%%%%%%%%%%%%%%%%%%%%%%%%%%%%%%%%%%%%%%%%%%%%%%%%%
\section{Model-checking diagnostics}
\label{sec:model_checking_infinite_matter}

In this appendix we provide more details on our model-checking diagnostics.

There are three underlying assumptions in our statistical model for EFT
truncation errors that we seek to validate~\cite{Melendez:2019izc}. First, the
coefficients $c_n(x)$ at each order are well-characterized as independent
draws (\iid\ realizations) from a single GP. Second, the parametrized mean and
covariance functions that characterize this GP have been correctly estimated
(\ie, we have found a consistent mean, variance, and correlation length). Third, the GP
learned from the known coefficients predicts a statistically meaningful
distribution for the truncation error.

For this validation we adopt here the preferred menu of model-checking
diagnostics advocated in Ref.~\cite{Melendez:2019izc}. These are:
\begin{itemize}
    \item[(i)] Distribution of Mahalanobis distance (MD) for the order-by-order
    coefficients.
    To check if pointwise data from a coefficient $\{c_n(x_i)\}_i$ followed an
    uncorrelated normal distribution, we would calculate the sum of squares of
    the scaled residuals and compare to the $\chi^2$ distribution with the appropriate number of degrees of freedom.
    The generalization for a multivariate correlated normal distribution---that is, when $c_n(x)$ is correlated in $x$, as in our case---is to
    calculate the MD for the extracted $c_n(\kf)$ at a specific order $n$ at $M$ validation $\kf$'s and compare to a reference distribution. For a GP that is a
    $\chi^2$ distribution with $M$ degrees of freedom.

    \item[(ii)] Pivoted Cholesky (PC) decomposition of the MD plotted graphically
    against the index and compared to a standard normal distribution. This
    provides specific information about mis-estimated variance (too large or
    small values across all indices) or correlation structure (failing distribution at
    large index).

    \item[(iii)] Credible interval diagnostic (CID). A plot of the CID for
    truncation error shows whether a $100\alpha\%$ credible interval learned up
    to a given order contains approximately $100\alpha\%$ of a set of validation points representing the next order result. This test can be carried out at $k-1$ orders, where $k$ is the number of \chiEFT\ coefficients in hand, since it requires knowledge of the result one order beyond that at which validation is being carried out.
\end{itemize}
Examples and associated Python code for carrying out these diagnostics are given in
Refs.~\cite{Melendez:2019izc,gsum}.
% We will apply each of them in Sec.~\ref{sec:model_checking_infinite_matter}.

%%%%%%%%%%%%%%%%%%%%%%%%%%%%%%%%%%%%%%%%%%%%%%%%%%%%%%%%%%%%%%%
\begin{figure*}[tbp]
	\centering
	\includegraphics[trim= 0 0 270 0, clip]{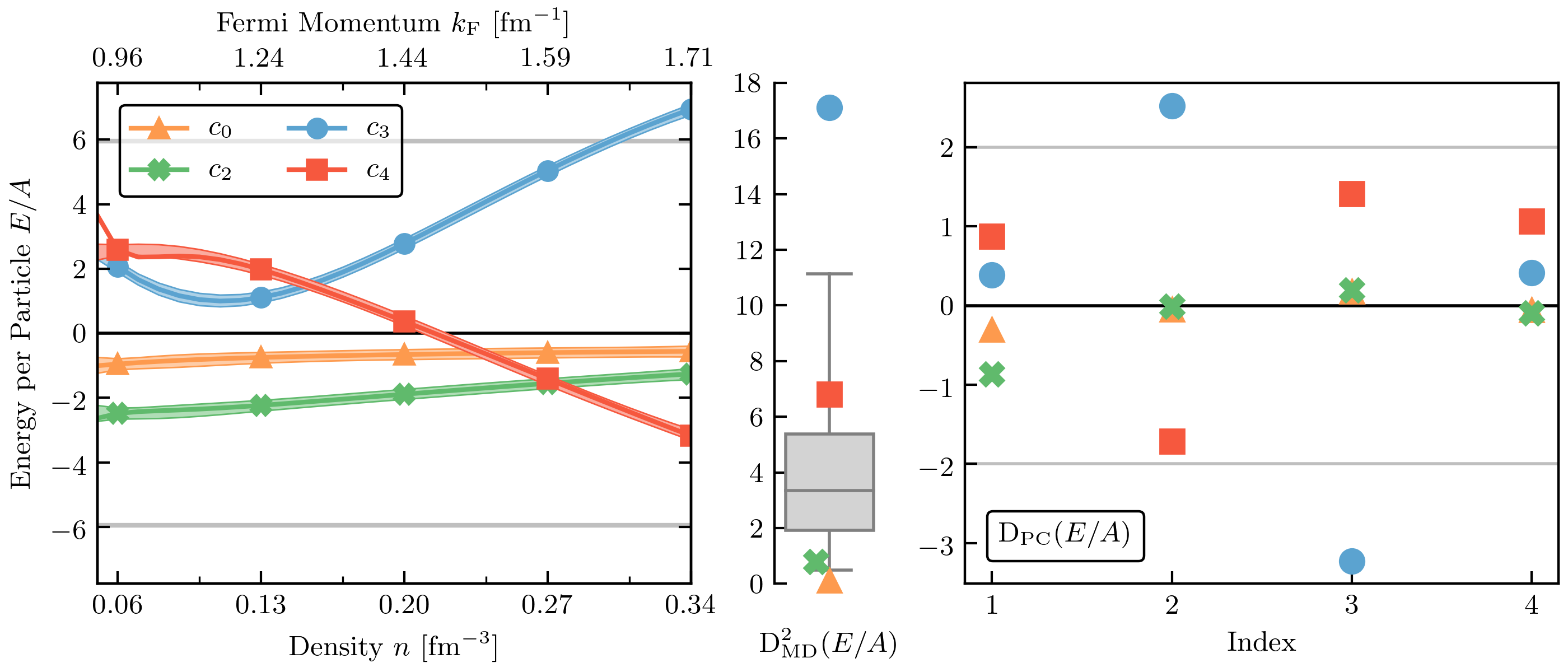}
	\includegraphics[trim= 0 0 270 0, clip]{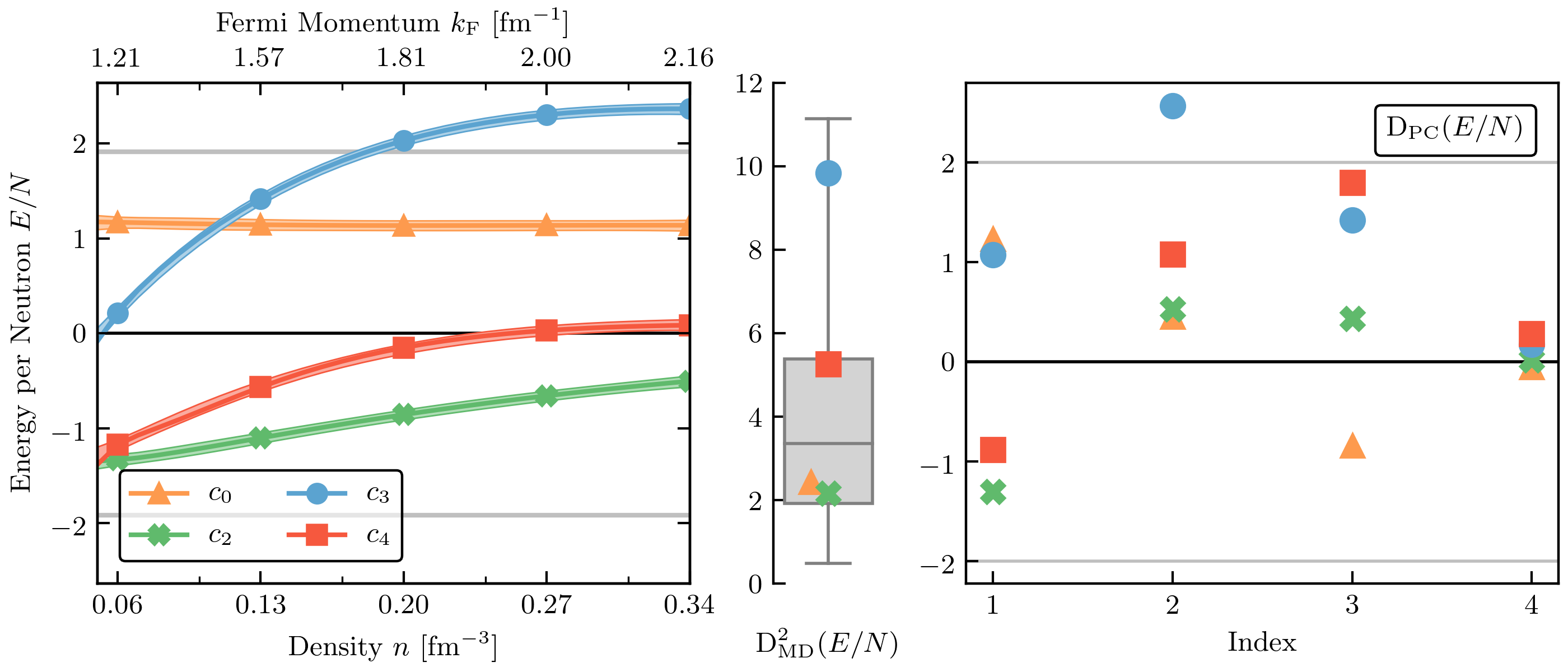}
	    \phantomsublabel{-6.45}{0.11}{fig:cn_coefficients_lambda-450_snm}
	        \phantomsublabel{-3.1}{0.11}{fig:cn_coefficients_lambda-450_pnm}
	\caption{
	    Observable coefficients for \protect\subref{fig:cn_coefficients_lambda-450_snm} $E/A(n)$ (SNM) and \protect\subref{fig:cn_coefficients_lambda-450_pnm} $E/N(n)$ (PNM) up to \NNNLO using the $\Lambda = 450\MeV$ interactions in Table~\ref{tab:drischler_potential_fits}.
    Markers indicate training points, gray bands
    indicate $2\cbar$ and colored bands are 68\% credible intervals of the
    interpolating GPs.
    The estimated hyperparameters are given by $\cbar = 3.0$ and $\ell = 0.50\fmi$ for SNM and $\cbar = 0.96$ and $\ell = 0.81\fmi$ for PNM.
}
	%\label{fig:pnm_diagnostics_lambda-450}
	\label{fig:cn_coefficients_lambda-450}
\end{figure*}
%%%%%%%%%%%%%%%%%%%%%%%%%%%%%%%%%%%%%%%%%%%%%%%%%%%%%%%%%%%%%%%

%%%%%%%%%%%%%%%%%%%%%%%%%%%%%%%%%%%%%%%%%%%%%%%%%%%%%%%%%%%%%%%
%\begin{figure*}[tbhp]
%    \centering
%    \includegraphics{cn_diags_sys-s_NN+3N_fit-1-7_Lamb-450_Q-kf_Lb-600_ls-x_midx-x_hyp-0-0-10-0p8.png}
%    \caption{
%    Model checking diagnostics for MD and PC applied to SNM coefficients from the $\Lambda = 450\MeV$ potential.
%    See Fig.~\ref{fig:pnm_diagnostics_lambda-500} for the notation.
%    }
%    \label{fig:snm_diagnostics_lambda-450}
%\end{figure*}
%%%%%%%%%%%%%%%%%%%%%%%%%%%%%%%%%%%%%%%%%%%%%%%%%%%%%%%%%%%%%%%

%%%%%%%%%%%%%%%%%%%%%%%%%%%%%%%%%%%%%%%%%%%%%%%%%%%%%%%%%%%%%%%
\begin{figure*}[tbp]
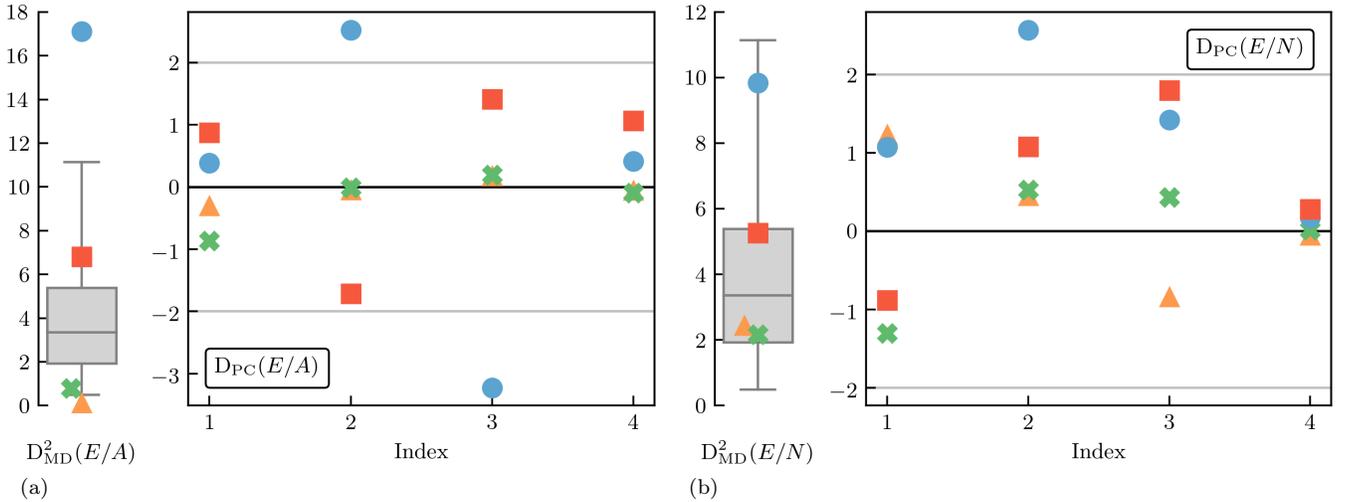

	\centering
	\includegraphics[scale=0.92, trim= 230 0 0 0, clip]{cn_diags_sys-s_NN+3N_fit-1-7_Lamb-450_Q-kf_Lb-600_ls-x_midx-x_hyp-0-0-10-0p8.png}
	\includegraphics[scale=0.92, trim= 230 0 0 0, clip]{cn_diags_sys-n_NN+3N_fit-1-7_Lamb-450_Q-kf_Lb-600_ls-x_midx-x_hyp-0-0-10-0p8.png}
		    \phantomsublabel{-6.9}{-0.18}{fig:model_diagnostics_lambda-450_snm}
	        \phantomsublabel{-3.4}{-0.18}{fig:model_diagnostics_lambda-450_pnm}
	\caption{
		Model checking diagnostics applied to \protect\subref{fig:model_diagnostics_lambda-450_snm} SNM and \protect\subref{fig:model_diagnostics_lambda-450_pnm} PNM coefficients from the $\Lambda = 450\MeV$ interactions in Fig.~\ref{fig:cn_coefficients_lambda-450}.
		The MD computed against the underlying process is shown in the left panel of each subplot.
		The interior line, box end caps, and whiskers on the box plot show the median, 50\% credible intervals, and 95\% credible intervals, respectively.
		The right panel shows the PC diagnostic $\DVAR{PC}$ vs index, with gray lines that represent its $2\sigma$ error bands.
		Both diagnostics point to the $c_3(\kf)$ coefficient as a possible outlier.
		See Ref.~\cite{Melendez:2019izc} for more details about analyzing these plots.}
	\label{fig:model_diagnostics_lambda-450}
\end{figure*}
%%%%%%%%%%%%%%%%%%%%%%%%%%%%%%%%%%%%%%%%%%%%%%%%%%%%%%%%%%%%%%%

%%%%%%%%%%%%%%%%%%%%%%%%%%%%%%%%%%%%%%%%%%%%%%%%%%%%%%%%%%%%%%%
\begin{figure*}[tbp]
    \centering
        \includegraphics[scale=0.92, trim= 230 0 0 0, clip]{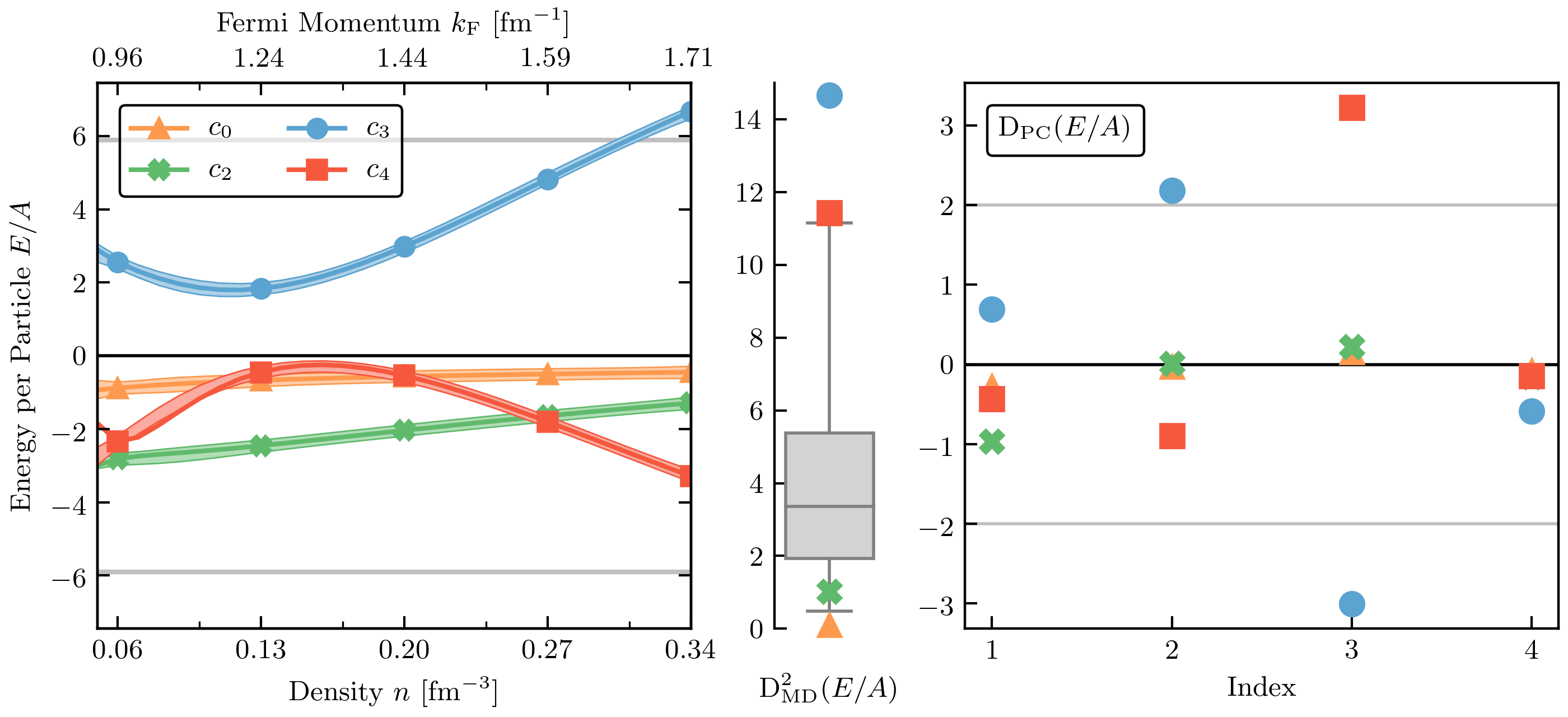}
    \includegraphics[scale=0.92, trim= 230 0 0 0, clip]{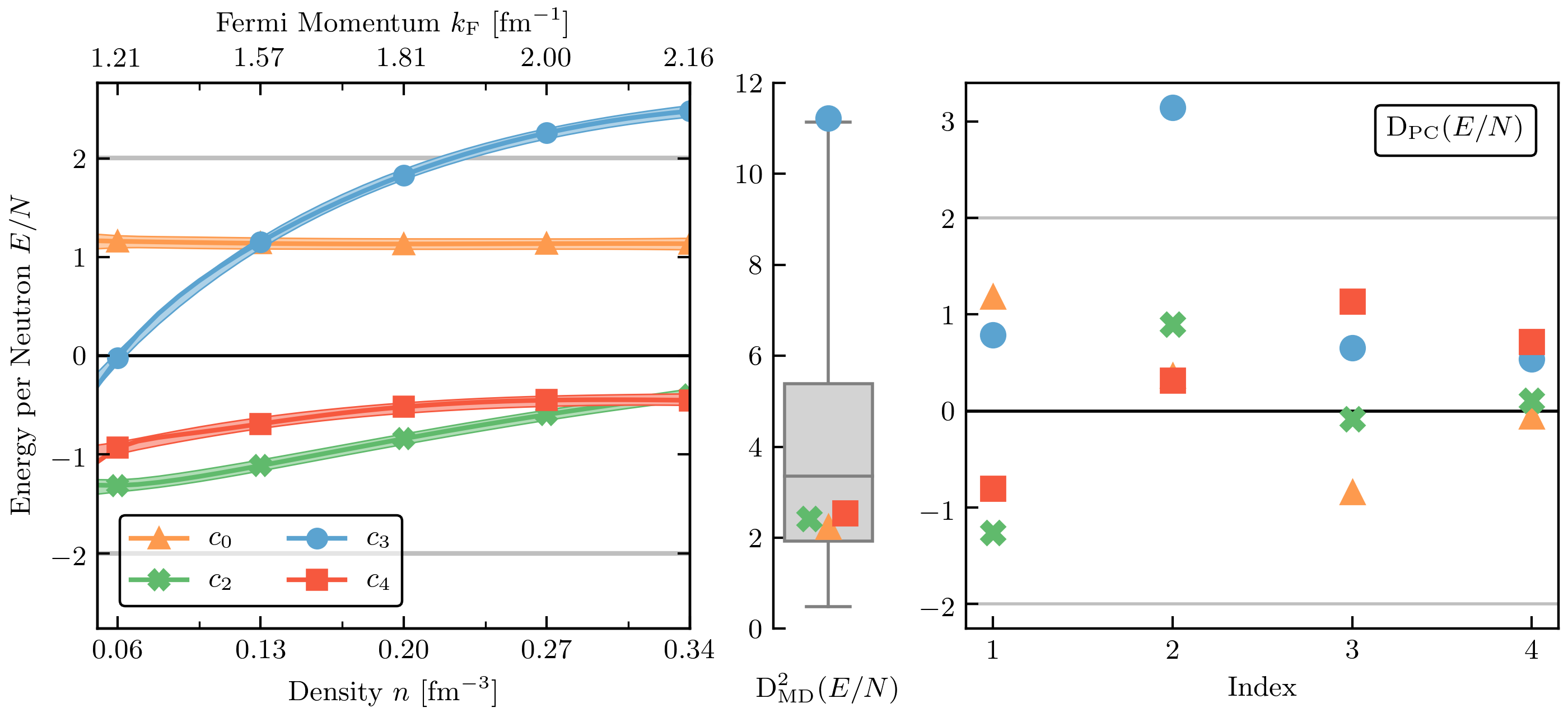}
    		    \phantomsublabel{-6.9}{-0.18}{fig:model_diagnostics_lambda-500_snm}
	        \phantomsublabel{-3.4}{-0.18}{fig:model_diagnostics_lambda-500_pnm}
    \caption{
    Similar to Fig.~\ref{fig:model_diagnostics_lambda-450} but for the $\Lambda = 500\MeV$ interactions in Figs.~\ref{fig:energy_per_neutron_coefficients_lambda-500} and \ref{fig:energy_per_particle_coefficients_lambda-500}.
    \label{fig:model_diagnostics_lambda-500}}
\end{figure*}
%%%%%%%%%%%%%%%%%%%%%%%%%%%%%%%%%%%%%%%%%%%%%%%%%%%%%%%%%%%%%%%

%%%%%%%%%%%%%%%%%%%%%%%%%%%%%%%%%%%%%%%%%%%%%%%%%%%%%%%%%%%%%%%
%\begin{figure*}
%    \centering
%    \includegraphics{cn_diags_sys-s_NN+3N_fit-4-10_Lamb-500_Q-kf_Lb-600_ls-x_midx-x_hyp-0-0-10-0p8.png}
%    \caption{
%    Model checking diagnostics for MD and PC applied to SNM coefficients from the $\Lambda = 500\MeV$ potential.
%    See Fig.~\ref{fig:pnm_diagnostics_lambda-500} for the notation.
%    }
%    \label{fig:snm_diagnostics_lambda-500}
%\end{figure*}
%%%%%%%%%%%%%%%%%%%%%%%%%%%%%%%%%%%%%%%%%%%%%%%%%%%%%%%%%%%%%%%

%%%%%%%%%%%%%%%%%%%%%%%%%%%%%%%%%%%%%%%%%%%%%%%%%%%%%%%%%%%%%%%
\begin{figure*}[tbp]
    \centering
        \includegraphics{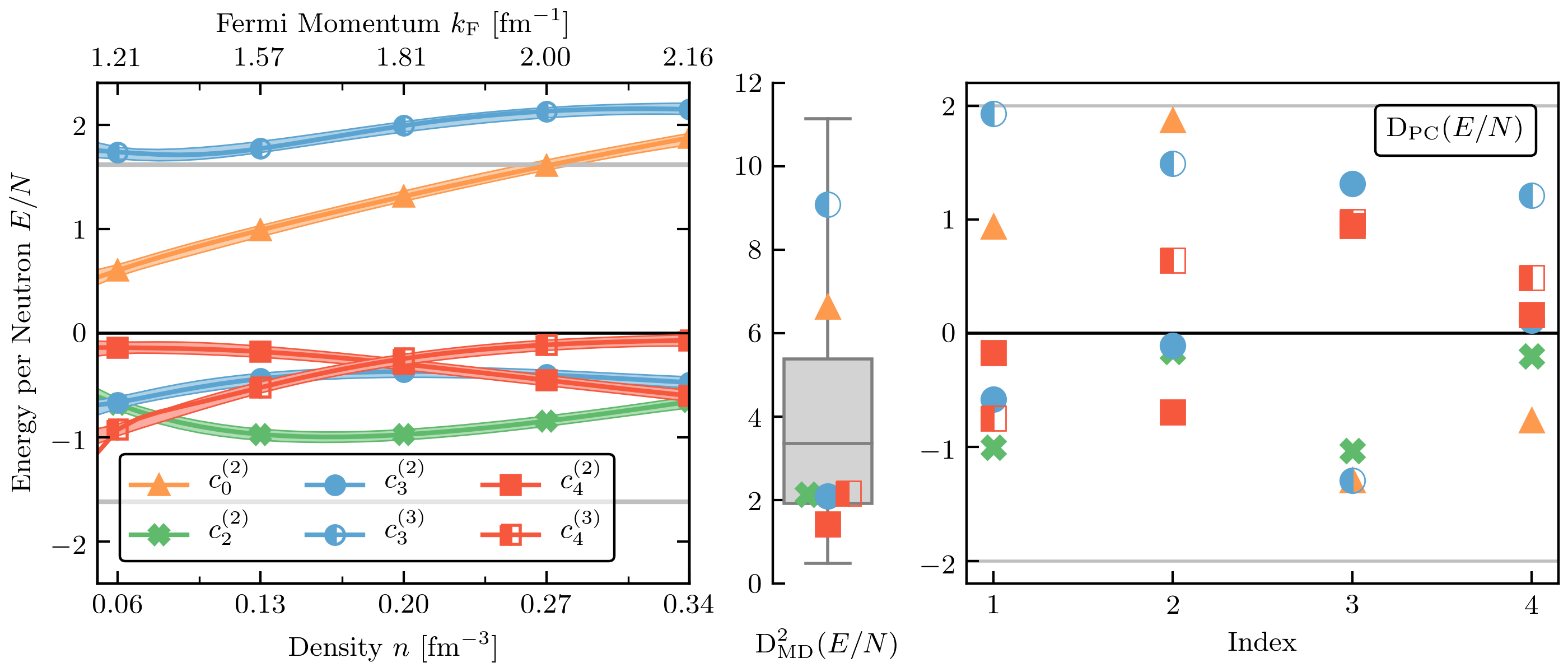}
	        \phantomsublabel{-6.9}{-0.18}{fig:cn_coefficients_lambda-500_pnm_Appended}
    		\phantomsublabel{-3.7}{-0.18}{fig:model_diagnostics_lambda-500_pnm_Appended}
    \caption{
    Observables coefficients and model diagnostics for $E/N(n)$ (PNM) up to \NNNLO for the $\Lambda = 500\MeV$ interactions in Table~\ref{tab:drischler_potential_fits} using  separate reference scales for NN-only (denoted with a superscript ``$(2)$'') and 3N contributions (denoted with a superscript ``$(3)$''). The model for $\genobsref(\kf)$ is explained in the text [see Eq.~\eqref{eq:y_ref_alt}].
    \label{fig:coeffs_and_diagnostics_lambda-500_pnm_Appended}}
\end{figure*}
%%%%%%%%%%%%%%%%%%%%%%%%%%%%%%%%%%%%%%%%%%%%%%%%%%%%%%%%%%%%%%%

%%%%%%%%%%%%%%%%%%%%%%%%%%%%%%%%%%%%%%%%%%%%%%%%%%%%%%%%%%%%%%%
\begin{figure}[tb]
    \centering
    \includegraphics{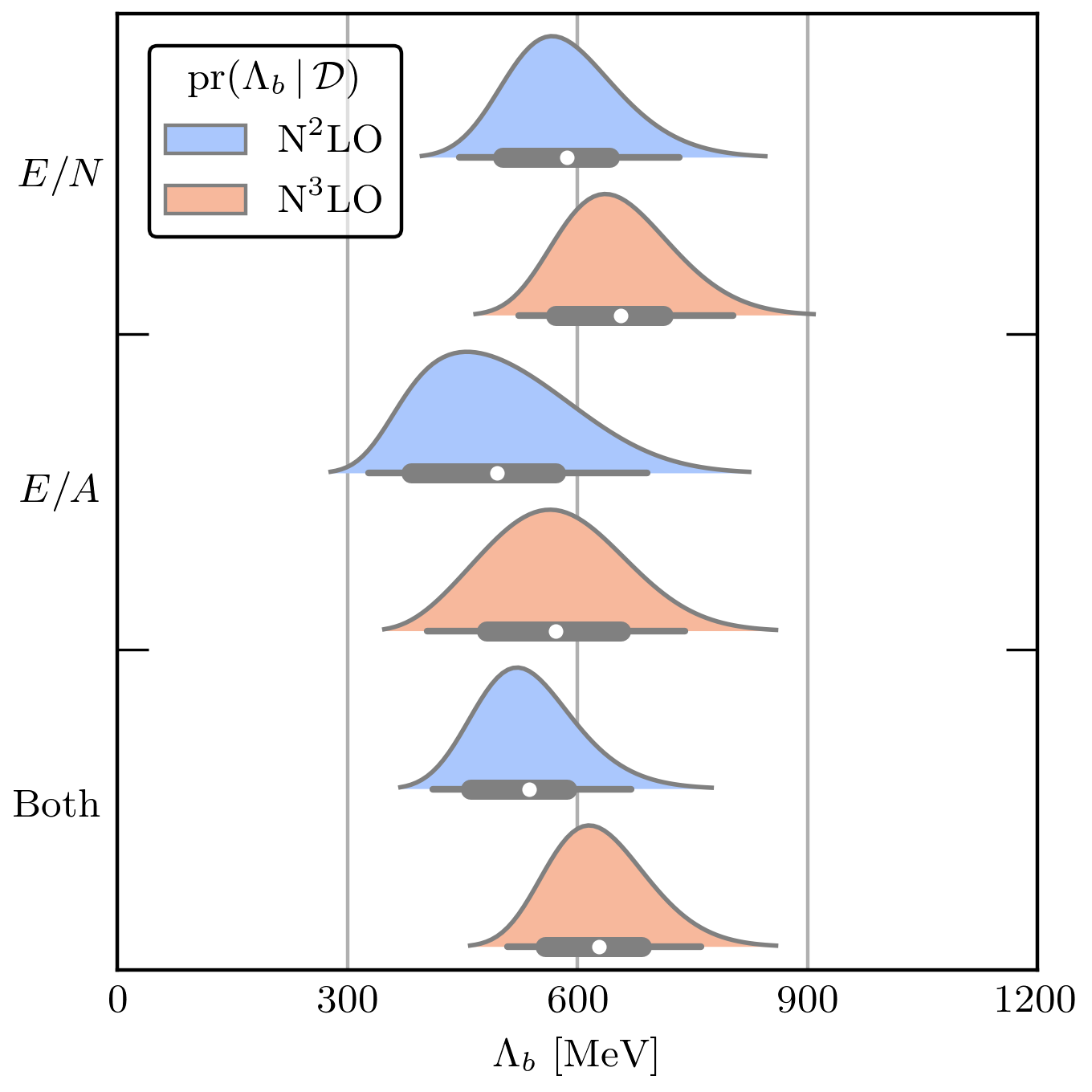}
    \caption{
    The posteriors for the EFT breakdown scale $\Lambda_b$ as in Fig.~\ref{fig:breakdown_posteriors_lambda-500}
    but using the alternative model for $\genobsref(\kf)$ as explained in the text [see Eq.~\eqref{eq:y_ref_alt}].
    %using orders through \NkLO{2}(blue bands) and \NkLO{3} (red bands). The upper pair of posteriors comes from analyzing $E/N$, the middle pair from $E/A$, and the bottom from a combined analysis. In all cases a Gaussian prior
    %centered at $\Lambda_b = 600\pm 150\MeV$ is used.
    %The combined \NkLO{3} posterior is consistent with the $\Lambda_b \approx 600\MeV$ found when considering free-space NN scattering observables~\cite{Melendez:2017phj}.
    } \label{fig:breakdown_posteriors_lambda-500_Appended}
\end{figure}

%%%%%%%%%%%%%%%%%%%%%%%%%%%%%%%%%%%%%%%%%%%%%%%%%%%%%%%%%%%%%%%

%%%%%%%%%%%%%%%%%%%%%%%%%%%%%%%%%%%%%%%%%%%%%%%%%%%%%%%%%%%%%%%
\begin{figure}[tb]
    \centering
    \includegraphics{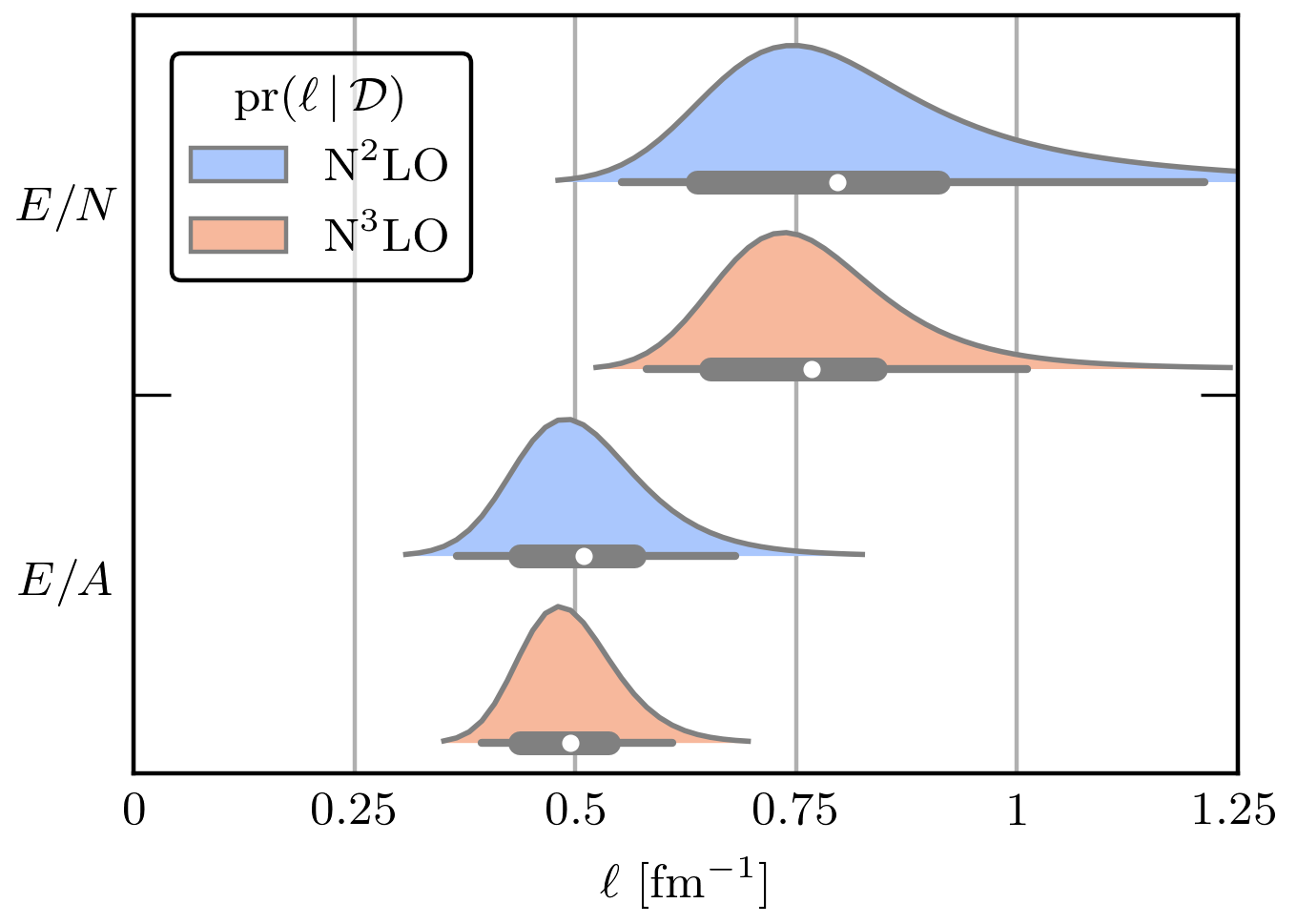}
    \caption{
    Length-scale posteriors as in Fig.~\ref{fig:length_scale_posteriors_lambda-500}
    but using the alternative model for $\genobsref(\kf)$ as explained in the text [see Eq.~\eqref{eq:y_ref_alt}].
    } \label{fig:length_scale_posteriors_lambda-500_Appended}
\end{figure}
%%%%%%%%%%%%%%%%%%%%%%%%%%%%%%%%%%%%%%%%%%%%%%%%%%%%%%%%%%%%%%%

%%%%%%%%%%%%%%%%%%%%%%%%%%%%%%%%%%%%%%%%%%%%%%%%%%%%%%%%%%%%%%%
\begin{figure*}[tb]
    \centering
        \includegraphics{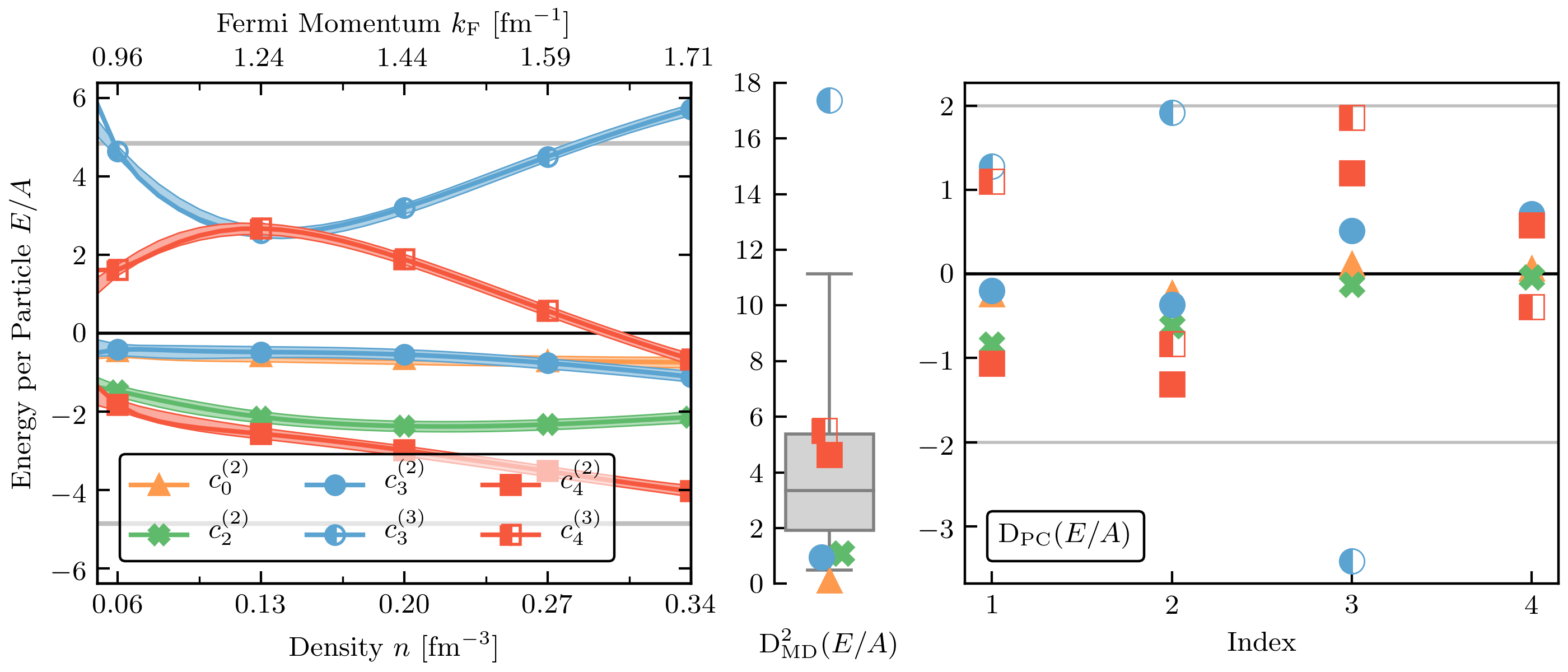}
	        \phantomsublabel{-6.9}{-0.18}{fig:cn_coefficients_lambda-500_snm_Appended}
    		\phantomsublabel{-3.7}{-0.18}{fig:model_diagnostics_lambda-500_snm_Appended}
    \caption{
   Same as Fig.~\ref{fig:coeffs_and_diagnostics_lambda-500_pnm_Appended} but for $E/A(n)$ (SNM).
    \label{fig:coeffs_and_diagnostics_lambda-500_snm_Appended}}
\end{figure*}
%%%%%%%%%%%%%%%%%%%%%%%%%%%%%%%%%%%%%%%%%%%%%%%%%%%%%%%%%%%%%%%

The observable coefficients for the
$\Lambda=450\MeV$ interactions of Table~\ref{tab:drischler_potential_fits} are shown in Fig.~\ref{fig:cn_coefficients_lambda-450}.
The two panels are the analog of  those shown for the $\Lambda=500\MeV$
interactions in
Figs.~\ref{fig:energy_per_neutron_coefficients_lambda-500} and \ref{fig:energy_per_particle_coefficients_lambda-500}.
The MD and PC diagnostics are applied to the $\Lambda=450\MeV$ and $\Lambda=500\MeV$ observable coefficients
in Figs.~\ref{fig:model_diagnostics_lambda-450} and
\ref{fig:model_diagnostics_lambda-500} respectively.
The reference scale~\eqref{eq:y_ref} and expansion parameter~\eqref{eq:Qkf} are used in all of these figures.

The CID for $\Lambda=450\MeV$ is very similar to $\Lambda=500\MeV$, which was shown in Fig.~\ref{fig:credible_interval_diagnostic_pnm_and_snm_lambda-500}.
The MD diagnostics for the two PNM cases are also similar. Both show that the $c_3(\kf)$ coefficient may be an outlier.
This supports the qualitative observation, apparent in Fig.~\ref{fig:energy_per_neutron_coefficients_lambda-500} and the right-hand panel of Fig.~\ref{fig:cn_coefficients_lambda-450}, and discussed in Sec.~\ref{sec:extracting_coefficients_pnm_snm}, that $c_3(\kf)$ has a different shape to $c_0(\kf)$ and $c_2(\kf)$.
We reiterate that this presumably happens because of the 3N
contributions that enter \chiEFT\ at \NNLO.
The PC diagnostic, while behaving well at small index, shows a decreased range of points at the highest index for both PNM and SNM. This happens because of the size of the white noise term $\sigma^2 = 5 \times 10^{-4}$ that was used for numerical stability (see Sec.~\ref{sec:extracting_coefficients_pnm_snm}).
The MD diagnostics for the two SNM cases show larger discrepancies with the reference distributions, with no points lying within the 50\% credible intervals.

As pointed out in Sec.~\ref{sec:extracting_coefficients_pnm_snm}, this problem reflects a mismatch of the assumed NN-only and 3N correlation structures.
To explore this further, we consider an alternative \chiEFT\ convergence model.
In particular, we split the coefficients into NN-only and residual 3N coefficients,
$c_n^{(2)}(\kf)$ and $c_n^{(3)}(\kf)$,
and assign different $\kf$ dependencies to the $\genobsref(\kf)$ associated with each. In this variant of the truncation-error model
we use a constant $\genobsref(\kf) = 16\MeV$ for the $c_n^{(2)}(\kf)$ coefficients, while for the $c_n^{(3)}(\kf)$ coefficients we use
\begin{align} \label{eq:y_ref_alt}
    \genobsref(\kf) = 16 \MeV \times \left(\frac{\kf}{\kfzero}\right)^3 \,.
\end{align}
Here, as in Eq.~\eqref{eq:y_ref}, $\kfzero$ is the Fermi momentum associated with $n_0 =0.16 \fmiq$, namely ${\kfzero^\text{PNM}} = 1.680 \fmi$ and $\kfzero^\text{SNM} = 1.333 \fmi$.
This form is chosen to roughly capture the extra $\kf$ dependence of 3N contributions relative to leading NN contributions.

The observable coefficients for $E/N(n)$ (PNM) for the alternative model are given in Fig.~\ref{fig:coeffs_and_diagnostics_lambda-500_pnm_Appended}
along with the MD and PC diagnostics.
Comparing to Fig.~\ref{fig:energy_per_neutron_coefficients_lambda-500}, we see that the observable coefficients have less variability with the alternative reference scale.
This is verified by the diagnostic plots, which show reasonable distributions for MD and the PC at each index, with the $c_3^{(3)}(\kf)$ coefficient being less of an outlier than the combined $c_3(\kf)$ coefficient in Fig.~\ref{fig:model_diagnostics_lambda-500}.
The order-by-order credible intervals for $E/N(n)$ obtained with this alternative truncation-error model are not shown, but are very close to those from using Eq.~\eqref{eq:y_ref}.
The $\Lambda_b$ posterior (Fig.~\ref{fig:breakdown_posteriors_lambda-500_Appended}) is compatible with that shown in the main text (Fig.~\ref{fig:breakdown_posteriors_lambda-500}) although the maximum a posteriori (MAP) value is somewhat larger here. The posterior for $\ell$, Fig.~\ref{fig:length_scale_posteriors_lambda-500_Appended}, has a smaller MAP value than the one shown in Fig.~\ref{fig:length_scale_posteriors_lambda-500}, but is consistent
with Fig.~\protect\subref*{fig:cn_coefficients_lambda-500_pnm_Appended}.

The observable coefficients and diagnostics for $E/A(n)$ (SNM) for the alternative model are given in Fig.~\ref{fig:coeffs_and_diagnostics_lambda-500_snm_Appended}.
Here we see the $c_3^{(3)}(\kf)$ coefficient has become even more of an outlier than it was in the approach used in the main text, so the model with split $\genobsref(\kf)$'s has not succeeded.

We invite the reader to take advantage of the freely available Jupyter notebooks~\cite{BUQEYEgithub} to further
investigate these issues.

%% file: appendix_tabulated_EOS.tex
%%%%%%%%%%%%%%%%%%%%%%%%%%%%%%%%%%%%%%%%%%%%%%%%%%%%%%%%%%%%%%%
\section{Tabulated values for the EOS} \label{sec:tables_pnm_snm}

Tables~\ref{tab:energies_per_nucleon_450MeV} and~\ref{tab:energies_per_nucleon_500MeV} give numerical values for the EOS in the limit of SNM
(left-hand side) and PNM (right-hand side) up to \NNNLO. The NN and
3N interactions in Table~\ref{tab:drischler_potential_fits} are used. See the captions for more details. Our GitHub repository provides all data sets in a machine-readable format along with annotated Jupyter notebooks~\cite{BUQEYEgithub}.

%%%%%%%%%%%%%%%%%%%%%%%%%%%%%%%%%%%%%%%%%%%%%%%%%%%%%%%%%%%%%%%
\begin{table*}[p]
    \caption{
    Energy per particle in MeV for SNM (left-hand side) and PNM (right-hand
    side) at four orders in the \chiEFT\ expansion for the interactions with $\Lambda=450\MeV$ in Table~\ref{tab:drischler_potential_fits}. The density $n$ is given in units of $\fmiq$ and the Fermi momentum
    $\kf$ in $\fmi$. Notice that $\kf^\text{SNM}$ and $\kf^\text{PNM}$ at same
    density are different.
    }\label{tab:energies_per_nucleon_450MeV}
\begin{tabular}{Srrrrrr|}
\toprule
\multicolumn{6}{Sc}{Symmetric nuclear matter (SNM)} \\
\colrule
  \multicolumn{1}{Sc}{$n$}   &  \multicolumn{1}{Sc}{$\kf^\text{SNM}$} & \multicolumn{1}{Sc}{LO} & \multicolumn{1}{Sc}{
  NLO} & \multicolumn{1}{Sc}{N$^2$LO} & \multicolumn{1}{Sc|}{N$^3$LO} \\
\colrule
 0.05 & $0.90$ &  $-7.40$ &  $-9.12$ &  $-8.60$ &  $-8.41$ \\
 0.06 & $0.96$ &  $-7.93$ &  $-9.99$ &  $-9.46$ &  $-9.22$ \\
 0.07 & $1.01$ &  $-8.39$ & $-10.87$ & $-10.30$ & $-10.01$ \\
 0.08 & $1.06$ &  $-8.79$ & $-11.71$ & $-11.13$ & $-10.78$ \\
 0.09 & $1.10$ &  $-9.16$ & $-12.55$ & $-11.94$ & $-11.50$ \\
 0.10 & $1.14$ &  $-9.49$ & $-13.36$ & $-12.70$ & $-12.20$ \\
 0.11 & $1.18$ &  $-9.79$ & $-14.14$ & $-13.41$ & $-12.80$ \\
 0.12 & $1.21$ & $-10.12$ & $-14.90$ & $-14.04$ & $-13.36$ \\
 0.13 & $1.24$ & $-10.43$ & $-15.62$ & $-14.58$ & $-13.82$ \\
 0.14 & $1.27$ & $-10.71$ & $-16.34$ & $-14.98$ & $-14.19$ \\
 0.15 & $1.30$ & $-10.99$ & $-17.01$ & $-15.29$ & $-14.47$ \\
 0.16 & $1.33$ & $-11.27$ & $-17.69$ & $-15.46$ & $-14.66$ \\
 0.17 & $1.36$ & $-11.50$ & $-18.32$ & $-15.51$ & $-14.72$ \\
 0.18 & $1.39$ & $-11.79$ & $-18.92$ & $-15.39$ & $-14.65$ \\
 0.19 & $1.41$ & $-11.99$ & $-19.51$ & $-15.13$ & $-14.50$ \\
 0.20 & $1.44$ & $-12.29$ & $-20.08$ & $-14.69$ & $-14.23$ \\
 0.21 & $1.46$ & $-12.48$ & $-20.61$ & $-14.14$ & $-13.83$ \\
\toprule
\end{tabular}
\begin{tabular}{|Srrrrrr}
\toprule
\multicolumn{6}{Sc}{Pure neutron matter (PNM)} \\
\colrule
    \multicolumn{1}{|Sc}{$n$}   &  \multicolumn{1}{Sc}{$\kf^\text{PNM}$} & \multicolumn{1}{Sc}{LO} & \multicolumn{1}{Sc}{
  NLO} & \multicolumn{1}{Sc}{N$^2$LO} & \multicolumn{1}{Sc}{N$^3$LO} \\
\colrule
 0.05 & $1.14$ &  $8.64$ &  $7.25$ &  $7.22$ &  $7.02$ \\
 0.06 & $1.21$ &  $9.72$ &  $7.96$ &  $8.07$ &  $7.81$ \\
 0.07 & $1.27$ & $10.73$ &  $8.61$ &  $8.91$ &  $8.60$ \\
 0.08 & $1.33$ & $11.69$ &  $9.21$ &  $9.76$ &  $9.40$ \\
 0.09 & $1.39$ & $12.60$ &  $9.77$ & $10.64$ & $10.22$ \\
 0.10 & $1.44$ & $13.48$ & $10.32$ & $11.56$ & $11.09$ \\
 0.11 & $1.48$ & $14.32$ & $10.84$ & $12.51$ & $12.01$ \\
 0.12 & $1.53$ & $15.14$ & $11.35$ & $13.51$ & $12.97$ \\
 0.13 & $1.57$ & $15.94$ & $11.86$ & $14.56$ & $14.01$ \\
 0.14 & $1.61$ & $16.71$ & $12.36$ & $15.66$ & $15.10$ \\
 0.15 & $1.64$ & $17.46$ & $12.86$ & $16.81$ & $16.26$ \\
 0.16 & $1.68$ & $18.20$ & $13.36$ & $18.02$ & $17.48$ \\
 0.17 & $1.71$ & $18.93$ & $13.85$ & $19.27$ & $18.75$ \\
 0.18 & $1.75$ & $19.62$ & $14.35$ & $20.56$ & $20.10$ \\
 0.19 & $1.78$ & $20.32$ & $14.86$ & $21.90$ & $21.49$ \\
 0.20 & $1.81$ & $20.99$ & $15.36$ & $23.27$ & $22.94$ \\
 0.21 & $1.84$ & $21.67$ & $15.87$ & $24.70$ & $24.41$ \\
\toprule
\end{tabular}

\bigskip

    \caption{
    Same as Table~\ref{tab:energies_per_nucleon_450MeV} but for the interactions with $\Lambda=500\MeV$ in Table~\ref{tab:drischler_potential_fits}.
    }\label{tab:energies_per_nucleon_500MeV}
\begin{tabular}{Srrrrrr|}
\toprule
\multicolumn{6}{Sc}{Symmetric nuclear matter (SNM)} \\
\colrule
  \multicolumn{1}{Sc}{$n$}   &  \multicolumn{1}{Sc}{$\kf^\text{SNM}$} & \multicolumn{1}{Sc}{LO} & \multicolumn{1}{Sc}{
  NLO} & \multicolumn{1}{Sc}{N$^2$LO} & \multicolumn{1}{Sc|}{N$^3$LO} \\
\colrule
 0.05 & $0.90$ &  $-6.86$ &  $-8.82$ &  $-8.26$ &  $-8.37$ \\
 0.06 & $0.96$ &  $-7.34$ &  $-9.68$ &  $-9.00$ &  $-9.20$ \\
 0.07 & $1.01$ &  $-7.72$ & $-10.52$ &  $-9.73$ &  $-9.97$ \\
 0.08 & $1.06$ &  $-8.05$ & $-11.33$ & $-10.43$ & $-10.71$ \\
 0.09 & $1.10$ &  $-8.34$ & $-12.11$ & $-11.10$ & $-11.40$ \\
 0.10 & $1.14$ &  $-8.60$ & $-12.89$ & $-11.74$ & $-12.01$ \\
 0.11 & $1.18$ &  $-8.84$ & $-13.63$ & $-12.32$ & $-12.57$ \\
 0.12 & $1.21$ &  $-9.06$ & $-14.33$ & $-12.83$ & $-13.05$ \\
 0.13 & $1.24$ &  $-9.28$ & $-15.00$ & $-13.25$ & $-13.46$ \\
 0.14 & $1.27$ &  $-9.46$ & $-15.65$ & $-13.60$ & $-13.76$ \\
 0.15 & $1.30$ &  $-9.67$ & $-16.28$ & $-13.84$ & $-13.97$ \\
 0.16 & $1.33$ &  $-9.88$ & $-16.89$ & $-13.95$ & $-14.10$ \\
 0.17 & $1.36$ & $-10.06$ & $-17.46$ & $-13.96$ & $-14.14$ \\
 0.18 & $1.39$ & $-10.24$ & $-17.99$ & $-13.86$ & $-14.04$ \\
 0.19 & $1.41$ & $-10.46$ & $-18.52$ & $-13.62$ & $-13.88$ \\
 0.20 & $1.44$ & $-10.60$ & $-19.00$ & $-13.24$ & $-13.59$ \\
 0.21 & $1.46$ & $-10.74$ & $-19.48$ & $-12.67$ & $-13.23$ \\
\toprule
\end{tabular}
\begin{tabular}{|Srrrrrr}
\toprule
\multicolumn{6}{Sc}{Pure neutron matter (PNM)} \\
\colrule
    \multicolumn{1}{|Sc}{$n$}   &  \multicolumn{1}{Sc}{$\kf^\text{PNM}$} & \multicolumn{1}{Sc}{LO} & \multicolumn{1}{Sc}{
  NLO} & \multicolumn{1}{Sc}{N$^2$LO} & \multicolumn{1}{Sc}{N$^3$LO} \\
\colrule
 0.05 & $1.14$ &  $8.59$ &  $7.23$ &  $7.11$ &  $6.96$ \\
 0.06 & $1.21$ &  $9.66$ &  $7.92$ &  $7.91$ &  $7.71$ \\
 0.07 & $1.27$ & $10.66$ &  $8.55$ &  $8.70$ &  $8.45$ \\
 0.08 & $1.33$ & $11.62$ &  $9.14$ &  $9.50$ &  $9.19$ \\
 0.09 & $1.39$ & $12.54$ &  $9.69$ & $10.31$ &  $9.94$ \\
 0.10 & $1.44$ & $13.42$ & $10.22$ & $11.16$ & $10.72$ \\
 0.11 & $1.48$ & $14.26$ & $10.75$ & $12.05$ & $11.53$ \\
 0.12 & $1.53$ & $15.09$ & $11.26$ & $12.99$ & $12.39$ \\
 0.13 & $1.57$ & $15.88$ & $11.77$ & $13.97$ & $13.30$ \\
 0.14 & $1.61$ & $16.66$ & $12.28$ & $15.02$ & $14.27$ \\
 0.15 & $1.64$ & $17.41$ & $12.78$ & $16.12$ & $15.29$ \\
 0.16 & $1.68$ & $18.15$ & $13.29$ & $17.29$ & $16.38$ \\
 0.17 & $1.71$ & $18.88$ & $13.82$ & $18.52$ & $17.53$ \\
 0.18 & $1.75$ & $19.58$ & $14.35$ & $19.81$ & $18.74$ \\
 0.19 & $1.78$ & $20.27$ & $14.89$ & $21.17$ & $20.01$ \\
 0.20 & $1.81$ & $20.97$ & $15.43$ & $22.59$ & $21.37$ \\
 0.21 & $1.84$ & $21.63$ & $16.00$ & $24.06$ & $22.78$ \\
\toprule
\end{tabular}
\end{table*}
%%%%%%%%%%%%%%%%%%%%%%%%%%%%%%%%%%%%%%%%%%%%%%%%%%%%%%%%%%%%%%%

%% file: appendix_details_GPs.tex
%%%%%%%%%%%%%%%%%%%%%%%%%%%%%%%%%%%%%%%%%%%%%%%%%%%%%%%%%%%%%%%
\section{Multitask Gaussian processes} \label{sec:gp_details_infinite_matter}

In this appendix, we provide more technical details on modeling with \emph{multitask} GPs, with particular emphasis on two cases: (1) a function and its derivatives, and (2) multiple generic functions.
multitask GPs, also known as multi-output GPs, are used to model multiple
curves $y_i(x)$ simultaneously, while possibly learning about their
interdependencies to improve predictions.

%%%%%%%%%%%%%%%%%%%%%%%%%%%%%%%%%%%%%%%%%%%%%%%%%%%%%%%%%%%%%%%
\subsection{A function and its derivatives} \label{sec:GPderivatives}

The derivative is a linear operator, so a Gaussian random variable remains
closed under this operation. Assume that $f(x)$ is distributed as $\GP[m(x),
\kernel(x,\,x')]$. The joint distribution of a function $f(x)$ and
its derivative $\partial_x f(x)$ is then
\begin{align}
    \begin{bmatrix}
        f(x) \\ \partial_x f(x)
    \end{bmatrix}
    & \sim
    \GP\left[\meangrad(x),\, \kernelgrad(x, x')\right] \,,\\
    \text{with} \quad \meangrad(x) & =
    \begin{bmatrix}
        m(x) \\ \partial_x m(x)
    \end{bmatrix} \,, \text{and}\\
    \kernelgrad(x, x') & =
    \begin{bmatrix}
    \kernel(x, x') & \partial_{x'}^\trans\kernel(x,x') \\
    \partial_{x}\kernel(x,x') &  \partial_x \partial_{x'}^\trans \kernel(x,x')
    \end{bmatrix} \,, \label{eq:joint_gradient_kernel}
\end{align}
where $\partial_x$ is a
$d$-dimensional vector if $x \in \mathbb{R}^d$, making $\meangrad(x)$ and
$\kernelgrad(x, x')$ then $(d+1)$- and $(d+1)\times (d+1)$-dimensional,
respectively. For example, if $f(x)$ has a prior mean of $0$ and has been
estimated by fitting to a set of training points ($\mathbf{x},\, \mathbf{y})$,
then the conditional mean and variance are given by
\begin{align}
    \tilde m(x) & = \kernel(x,\, \mathbf{x}) K^{-1} \mathbf{y} \,, \\
    \tilde \kernel(x,\, x') & = \kernel(x,\, x') - \kernel(x,\, \mathbf{x}) K^{-1} \kernel(\mathbf{x},\, x')\,,
\end{align}
where $K = \kernel(\mathbf{x},\, \mathbf{x})$ (see also
Refs.~\cite{Melendez:2019izc,rasmussen2006gaussian}). Then the
distribution of $f(x)$ with any of its derivatives would involve differentiating
$\tilde m$ and $\tilde \kernel$, which includes at least two derivatives of the
kernel $\kernel$.
% Thus,
% \begin{align}
%     \partial_x \tilde m(x) & = \partial_x \kernel(x, \mathbf{x}) K^{-1} \mathbf{y} \\
%     \partial^2 \tilde \kernel(x, x') & = \partial^2 \kernel(x, x') - \partial_x \kernel(x, \mathbf{x}) K^{-1} \partial_{x'}^\trans\kernel(\mathbf{x}, x').
% \end{align}
The generalization to higher derivatives follows straightforwardly.

% To evaluate such formulas, we must be able to take at least 2 derivatives of
% the kernel $\kernel$.
For the squared exponential kernel (RBF) employed here, an analytic
expression for an arbitrary number of derivatives exists. Consider the $n$th order
(scalar) mixed partial derivative,
\begin{align}
    \partial_{n_1,n_2,\dotsc,n_d}^n \equiv \frac{\partial^n}{\partial x_1^{n_1} \partial x_2^{n_2} \dotsm \partial x_d^{n_d}}\,,
\end{align}
with $\partial_{n_1',n_2',\dotsc,n_d'}^{n'}$ defined similarly for $x'$.
Stationary kernels like the RBF kernel obey $\partial_{x'}\kernel(x,\,x') = -\partial_x \kernel(x,\,x')$, which implies
\begin{equation}
    \partial_{n_1,\dotsc,n_d}^n \partial_{n_1',\dotsc,n_d'}^{n'}\kernel(x,x') = (-1)^{n'} \partial_{N_1,\dotsc,N_d}^{N} \kernel(x,x')\,,
\end{equation}
where $N_i = n_i + n_i'$ and all derivatives act on $x$. To compute this $N$th
order derivative, we make use of the (physicists') Hermite polynomial relation
\begin{align} \label{eq:hermite_definition}
    H_n(z) = (-1)^n e^{z^2} \frac{\dd^n}{\dd z^n} e^{-z^2} \,.
\end{align}
The squared exponential covariance function is given by
\begin{align}
    \kernel(x,\,x') = \cbar^2 e^{-\frac{1}{2} (x-x')^\trans L^{-1} (x-x')}\,,
\end{align}
where we assume a diagonal correlation length matrix $L =
\mathrm{diag}(\ell_1^2,\, \ell_2^2,\, \dotsc,\, \ell_d^2)$. Now it is useful to make a
change of variables $z = L^{-1/2}(x-x')/\sqrt{2}$, from which it follows that
\begin{align}
    \frac{\dd}{\dd x_i} = \frac{1}{\sqrt{2}\ell_i}\frac{\dd}{\dd z_i} \,.
\end{align}
This transformation allows the kernel to be separable in $z$, \ie,
$\kernel(x,\,x') = \cbar^2 \prod_{i=1}^d e^{-z_i^2}$. Using
Eq.~\eqref{eq:hermite_definition} the desired derivative follows as
\begin{align}
    \partial_{n_1,n_2,\dotsc,n_d}^n & \partial_{n_1',n_2',\dotsc,n_d'}^{n'}\kernel(x,x') \notag \\
    & = (-1)^{n'} \cbar^2 {\left[\prod_{i=1}^d \left(\frac{1}{\sqrt{2}\ell_i} \right)^{\!N_i} \frac{\partial^{N_i}}{\partial z_i^{N_i}} e^{-z_i^2}\right]} \notag \\
    & = (-1)^n \left[\prod_{i=1}^d \left(\frac{1}{\sqrt{2}\ell_i} \right)^{\!N_i}\! H_{N_i}(z_i)\right] \kernel(x,\, x') \,.
\end{align}

Certain observables require the sum of a function with one or more of its
derivatives. The distribution of such a sum follows straightforwardly as the sum
of correlated Gaussians.  If $X$ and $Y$ are distributed
jointly as
\begin{align} \label{eq:joint_gaussian_nuclear_matter}
    \begin{bmatrix}
        X \\ Y
    \end{bmatrix}
    &\sim
    \normal\left(
    \begin{bmatrix}
        \mu_X \\ \mu_Y
    \end{bmatrix} ,
    \begin{bmatrix}
        K_{XX} & K_{XY} \\
        K_{YX} & K_{YY}
    \end{bmatrix}
    \right)\,,
\intertext{then}
    AX + BY &\sim \normal\left(\mu, \Sigma\right)\,, \label{eq:sum_of_gaussians_matter}\\
    \mu & = A\mu_X + B\mu_Y \,, ~~\text{and} \\
    \begin{split}
    \Sigma & = K_{XX} + K_{YY} \\
    & ~~ + B K_{YX} A^\trans + A K_{XY} B^\trans\,. \label{eq:sum_of_gaussians_cov_matter}
    \end{split}
\end{align}

%%%%%%%%%%%%%%%%%%%%%%%%%%%%%%%%%%%%%%%%%%%%%%%%%%%%%%%%%%%%%%%
\subsection{Two generic functions}
\label{sec:multitask_gps}

Equations~\eqref{eq:sum_of_gaussians_matter}--\eqref{eq:sum_of_gaussians_cov_matter} appear simple enough, and for derivatives  the cross covariance $K_{XY} = K_{YX}^\trans$ is given by Eq.~\eqref{eq:joint_gradient_kernel}.
However, the form this cross-covariance takes is less clear in the case of generic \emph{multitask} Gaussian processes.
One of the main difficulties with
using GPs in the generic multitask setting is coming up with valid covariance functions
that accurately model the relationships in the data.

In this work, we are interested in a multitask GP that describes $E/N(n)$ and $E/A(n)$.
This means we have two processes, each individually distributed as a GP with an RBF kernel, and whose
outputs are correlated with one another.
Reference~\cite{Melkumyan:2011multikernel} showed that if each autocovariance,
$K_{XX}$ and $K_{YY}$, is generated from RBF kernels $\kernel_1(x,\, x'; \sigma_1,\,\ell_1)$ and $\kernel_2(x,\, x';\, \sigma_2,\, \ell_2)$,
then their cross covariance $K_{XY}$ can be written as another RBF kernel with a correlation length
$\ell = \sqrt{(\ell_1^2 + \ell_2^2) / 2}$ and correlation coefficient $\rho$:
\begin{align}
    \kernel(x,\,x';\, \sigma_1,\,\sigma_2,\, \ell_1,\, \ell_2) & = \sigma_1\sigma_2\rho \exp[- \frac{(x-x')^2}{\ell_1^2 + \ell_2^2}] \,, \label{eq:cross_correlation_two_different_rbfs} \\
    \text{with} \quad \rho & = \sqrt{\frac{2\ell_1 \ell_2}{\ell_1^2 + \ell_2^2}} \,. \label{eq:rho_cross_correlation}
\end{align}
Note that $\rho$ is uniquely determined by $\ell_1$ and $\ell_2$, and, as $\ell_1
\to \ell_2$, the two outputs become 100\% correlated.
% This can be a
% desirable feature, depending on if this predicted $\rho$ aligns with the
% empirical correlations found in the data.
For the $\Lambda = 500\MeV$ interactions this model accurately reproduces the
correlations found between the observable coefficients of SNM and PNM.

If one instead makes the constraint $\ell_1 = \ell_2$, or, more generally, that
the same correlation kernel is used for each output, then
Eq.~\eqref{eq:rho_cross_correlation} need not be enforced. Rather, in this case,
one can use the intrinsic coregionalization model~\cite{alvarez2012kernels}
\begin{align} \label{eq:lmc_kernel}
    \kernel_{\text{joint}}(x,\, x') = C \otimes \kernel(x,\, x') \,,
\end{align}
where $C$ is a positive semi-definite matrix called the coregionalization matrix
and $\otimes$ denotes the Kronecker product. $C$ imposes the correlation
structure \emph{between} curves, while $\kernel(x,\,x')$ imposes a correlation
structure \emph{within} each curve. The off-diagonal components of $C$ can then
be tuned to data if desired, as long as $C$ remains positive semidefinite.

We use the intrinsic coregionalization model, Eq.~\eqref{eq:lmc_kernel}, for the
$\Lambda = 450\MeV$ interactions. We choose the diagonal components to be $\cbar^2$
of $E/N(n)$ and $E/A(n)$, and the off-diagonal component of $C$ to be $\cbar_\text{PNM}
\cbar_\text{SNM} \rho$, where $\rho$ is the empirical correlation of the $c_n$. For
Eq.~\eqref{eq:lmc_kernel}, which is combined with
Eqs.~\eqref{eq:joint_gaussian_nuclear_matter}--\eqref{eq:sum_of_gaussians_cov_matter}
to compute $S_2(n)$, we must choose a common length scale for $\kernel$. We take
this to be $(\ell_\text{PNM} + \ell_\text{SNM}) / 2$, with $\ell_\text{SNM}$ on the $\kf^\text{PNM}$ scale (\ie, it has the appropriate factor of $\sqrt[3]{2}$).
Note, however, that the individual length scales are still used for all predictions
requiring only PNM or SNM.

The total covariance of the truncation error model then follows directly
from the total covariance of the coefficients $\kappa$:
\begin{equation} \label{eq:cross_truncation_corrfunc_matter}
\begin{split}
    \Sigma_{ij}(x,\, x')  &\equiv {\genobsref}_i(\kinparvec){\genobsref}_j(\kinparvec') \\
    & \quad \times \frac{[Q_i(\kinparvec)Q_j(\kinparvec')]^{k+1}}{1 - Q_i(\kinparvec)Q_j(\kinparvec')} \kappa_{ij}(\kinparvec, \kinparvec')\,,
    \end{split}
\end{equation}
where $i$ and $j$ denote the observable, here $E/N(n)$ or $E/A(n)$.
For $i = j$, this reduces to Eq.~\eqref{eq:truncation_corrfunc_matter}, but this extension describes correlations between the SNM and PNM truncation errors when $i \neq j$.